\documentclass{IEEEtran}
\usepackage{mathrsfs}
\usepackage{algorithm}
\usepackage{algorithmic}
\usepackage{cite}
\usepackage{psfrag}
\usepackage{amssymb}
\usepackage{epsfig}
\usepackage{pifont}
\usepackage{amsmath}
\usepackage{array}
\usepackage{pifont}
\usepackage{amsfonts}
\usepackage{fancyhdr}
\usepackage{graphics,eepic,epic}
\usepackage{latexsym}
\usepackage{euscript}
\usepackage{graphics,eepic,epic,psfrag}
\usepackage{graphicx,bm,dsfont}
\usepackage{xcolor}
\usepackage{pbox}

\usepackage{float}
\usepackage{booktabs}
\usepackage{multirow}
\usepackage{tikz}
\usetikzlibrary{shapes,snakes,plotmarks}

\usepackage{enumerate}

\usepackage{hyperref}
\hypersetup{
	colorlinks=true,
	linkcolor=blue,
	citecolor=blue
}

\usepackage{subcaption}
\newtheorem{theorem}{Theorem}
\newtheorem{proposition}{Proposition}
\newtheorem{definition}{Definition}

\newtheorem{lemma}{Lemma}

\newtheorem{assumption}{Assumption}

\newcommand{\vecn}{\op{vec}_{\operatorname{non-diag}}}
\newcommand{\s}{\mathbf{s}}
\renewcommand{\S}{\mathbf{S}}
\newcommand{\op}[1]{\mathrm{#1}}
\newcommand{\qam}{\left\{ \pm \frac{\sqrt{2}}{2} \pm \imath \frac{\sqrt{2}}{2} \right\}}
\newcommand{\lip}{\eta}

\renewcommand{\algorithmicrepeat}{\textbf{Repeat}}

\begin{document}
\title{Covariance-Based Activity Detection in Cooperative Multi-Cell Massive MIMO: Scaling Law and Efficient Algorithms}
\author{Ziyue~Wang,~\IEEEmembership{Graduate Student~Member,~IEEE},
        Ya-Feng~Liu,~\IEEEmembership{Senior~Member,~IEEE},
        Zhaorui~Wang,~\IEEEmembership{Member,~IEEE},
        and~Wei~Yu,~\IEEEmembership{Fellow,~IEEE}
        \thanks{Received November 23, 2023; revised July 16, 2024; accepted September 23, 2024.
        The work of Ziyue Wang and Ya-Feng Liu was supported in part by the National Natural Science Foundation of China (NSFC) under Grant 12371314, Grant 12021001, and Grant 12288201.
        The work of Zhaorui Wang was supported by the Hetao Shenzhen-Hong Kong Science and Technology Cooperation Zone through the Basic Research Project under Grant HZQB-KCZYZ-2021067.
        The work of Wei Yu was supported by a Discovery Grant from the Natural Sciences and Engineering Research Council of Canada.
        An earlier version of this paper was presented in part at the 2021 IEEE International Workshop on Signal Processing Advances in Wireless Communications (SPAWC)~\cite{wang2021accelerating},
        and in part at the 2023 IEEE International Conference on Acoustics, Speech, and Signal Processing (ICASSP)~\cite{wang2023scaling}. \textit{(Corresponding author: Ya-Feng Liu.)}}
        \thanks{Ziyue Wang is with the State Key Laboratory of Scientific and Engineering Computing, Institute of Computational Mathematics and Scientific/Engineering Computing, Academy of Mathematics and Systems Science, Chinese Academy of Sciences, Beijing 100190, China, and also with the School of Mathematical Sciences, University of Chinese Academy of Sciences, Beijing 100049, China (e-mail: ziyuewang@lsec.cc.ac.cn).}
        \thanks{Ya-Feng Liu is with the State Key Laboratory of Scientific and Engineering Computing, Institute of Computational Mathematics and Scientific/Engineering Computing, Academy of Mathematics and Systems Science, Chinese Academy of Sciences, Beijing 100190, China (e-mail: yafliu@lsec.cc.ac.cn).}
        \thanks{Zhaorui Wang is with the Shenzhen Future Network of Intelligence Institute (FNii-Shenzhen), the School of Science and Engineering (SSE), and Guangdong Provincial Key Laboratory of Future Networks of Intelligence, The Chinese University of Hong Kong, Shenzhen 518172, China (e-mail: wangzhaorui@cuhk.edu.cn).}
        \thanks{Wei Yu is with the Edward S. Rogers Sr. Department of Electrical and Computer Engineering, University of Toronto, Toronto, ON M5S 3G4, Canada (e-mail: weiyu@comm.utoronto.ca).}
}

\maketitle

\begin{abstract}
This paper focuses on the covariance-based activity detection problem in a multi-cell massive multiple-input multiple-output (MIMO) system. In this system, active devices transmit their signature sequences to multiple base stations (BSs), and the BSs cooperatively detect the active devices based on the received signals. While the scaling law for the covariance-based activity detection in the \textit{single-cell} scenario has been extensively analyzed in the literature, this paper aims to analyze the scaling law for the covariance-based activity detection in the \textit{multi-cell} massive MIMO system. Specifically, this paper demonstrates a quadratic scaling law in the multi-cell system, under the assumption that the path-loss exponent of the fading channel $\bm{\gamma > 2.}$ This finding shows that, in the multi-cell massive MIMO system, the maximum number of active devices that can be correctly detected in each cell increases quadratically with the length of the signature sequence and decreases logarithmically with the number of cells (as the number of antennas tends to infinity). Moreover, in addition to analyzing the scaling law for the signature sequences randomly and uniformly distributed on a sphere, the paper also establishes the scaling law for signature sequences based on a finite alphabet, which are easier to generate and store.
Finally, this paper proposes two efficient accelerated coordinate descent (CD) algorithms with a convergence guarantee for solving the device activity detection problem.
The first algorithm reduces the complexity of CD by using an inexact coordinate update strategy. The second algorithm avoids unnecessary computations of CD by using an active set selection strategy.
Simulation results show that the proposed algorithms exhibit excellent performance in terms of computational efficiency and detection error probability.
\end{abstract}
\IEEEpeerreviewmaketitle

\section{Introduction}

\IEEEPARstart{M}{assive} machine-type communication (mMTC) is an important application scenario in fifth-generation (5G) cellular systems and beyond~\cite{bockelmann2016massive}.
One of the main challenges in mMTC is handling massive random access, where a large number of devices are connected to the network in the uplink, but their activities are sporadic~\cite{chen2021massive}.
To address this challenge, an \textit{activity detection} phase can be used,
during which active devices transmit their unique preassigned signature sequences.
The network then identifies the active devices by detecting which sequences are transmitted based on the received signals at the base stations (BSs)~\cite{liu2018sparse}.

In contrast to conventional cellular systems that provide devices with orthogonal sequences, the signature sequences preassigned to devices in mMTC have to be non-orthogonal due to the large number of devices and limited coherence time.
The non-orthogonality of the signature sequences inevitably causes both intra-cell and inter-cell interference, making the task of device activity detection more complicated.
In this paper, we investigate the problem of device activity detection in massive multiple-input multiple-output (MIMO) systems, which utilize spatial dimensions for intra-cell interference mitigation~\cite{liu2018massive}.
We also consider a cloud-radio access network (C-RAN) architecture for inter-cell interference cancellation, where the BSs are connected to a central unit (CU) via fronthaul links and cooperate in performing device activity detection.

There are two main mathematical optimization approaches used for device activity detection~\cite{liu2024survey}.
The first method takes advantage of the sporadic nature of device activities and jointly estimates the instantaneous channel state information (CSI) and device activities~\cite{liu2018massive}.
In this paper, we call this approach the compressed sensing (CS) technique.
The second method focuses on the statistical information of the channel only, instead of the instantaneous channel realizations. It estimates the device activities (along with the large-scale fading components of the channels) by solving a maximum likelihood estimation (MLE) problem~\cite{haghighatshoar2018improved}.
This method is called the covariance-based approach, because the MLE formulation depends on the received signals only through the sample covariance matrix.
Studies have shown that the covariance-based approach generally outperforms the CS-based approach, particularly in massive MIMO systems~\cite{chen2022phase}.
The covariance-based approach was first proposed for the single-cell scenario in the pioneering work~\cite{haghighatshoar2018improved}  and later extended to the multi-cell scenario in \cite{chen2021sparse} and \cite{ganesan2021clustering}.

One significant advantage of the covariance-based approach over the CS-based approach is its ability to detect a larger number of active devices, due to its quadratic scaling law~\cite{fengler2021non,chen2022phase}.
This scaling law represents the feasible set of system parameters under which the covariance-based approach can successfully recover device activities in a massive MIMO system, and is closely related to early works on the fundamental limits of sparse recovery, including~\cite{tang2010performance,pal2015pushing,koochakzadeh2018fundamental,balkan2014localization}.
Specifically, when applied to the activity detection problem in the massive random access setting, in the single-cell scenario, \cite{fengler2021non} demonstrates that given a set of signature sequences of length $L$ (randomly and uniformly) generated from the sphere of radius $\sqrt{L},$ the maximum number of active devices that can be correctly detected from $N$ potential devices scales as $K=\mathcal{O}(L^2/\log^2(N/L^2))$ by solving a restricted MLE problem.
The work \cite{chen2022phase} further shows that the above scaling law also holds true for the more practical unrestricted MLE model.
This paper aims to establish the scaling law for the multi-cell network. In this realm, to the best of our knowledge, \cite{chen2021sparse} is the first and only work to analyze the phase transition of the unrestricted MLE model/covariance-based approach in multi-cell massive MIMO.
In particular, \cite{chen2021sparse} conjectures and verifies through simulations that the scaling law for the covariance-based approach in the multi-cell scenario is approximately the same as that in the single-cell scenario.

This paper addresses the conjecture in \cite{chen2021sparse} by explicitly characterizing the scaling law for the covariance-based approach in the multi-cell massive MIMO scenario.
Moreover, we consider two types of signature sequences:
one where each element of the sequence is randomly and uniformly generated from the finite alphabet $\qam,$ where $\imath$ is the imaginary unit, called Type~\ref{item:qam} signature sequences in this paper; and another type where the sequence is randomly and uniformly generated from the sphere of radius $\sqrt{L}$~\cite{fengler2021non}, called Type~\ref{item:sphere} signature sequences in this paper.
We assume that the large-scale fading coefficients from every device to every BS in the network are known, and they satisfy the path-loss model in \cite{rappaport1996wireless}, with the path-loss exponent $\gamma > 2.$
We show that, as the number of antennas tends to infinity, the maximum number of active devices in each cell that can be correctly detected at the CU is
\begin{equation}\label{eq:introduction-scaling-law}
	K=\mathcal{O}(L^2/\log^2(BN/L^2)),
\end{equation}
where $B$ is the total number of cells.
This scaling law is approximately the same as that of the single-cell scenario~\cite{fengler2021non,chen2022phase}, indicating that the covariance-based approach can correctly detect almost as many active devices in each cell in the multi-cell scenario as it can in the single-cell scenario.
Moreover, the inter-cell interference is not a limiting factor of the detection performance because $B$ affects $K$ only through $\log B$ in \eqref{eq:introduction-scaling-law}.
This result also clarifies the condition under which the conjecture in \cite{chen2021sparse} holds true.

In addition to the theoretical analysis, this paper also investigates the algorithmic aspect of the covariance-based activity detection.
In this regard, the coordinate descent (CD) algorithm~\cite{haghighatshoar2018improved,chen2021sparse} is a commonly used algorithm, as it can achieve excellent detection performance by iteratively updating the activity indicator of each device.
In the single-cell scenario, the CD algorithm is computationally efficient because each subproblem (i.e., optimizing the original objective with respect to only one of the variables) has a closed-form solution~\cite{haghighatshoar2018improved}.
However, in the multi-cell scenario, the CD algorithm becomes less appealing because the subproblem for updating each coordinate does not have a closed-form solution.
In this case, the subproblem involves a polynomial root-finding problem whose degree depends on the number of cells~\cite{chen2021sparse,ganesan2021clustering}.
Moreover, the CD algorithm performs many unnecessary computations because many coordinates of the MLE solution are on the boundaries of the box constraints.
To address these issues, we propose computationally efficient inexact CD and active set CD algorithms for solving the covariance-based activity detection problem in the multi-cell scenario.
These two algorithms can be proved to converge, and they exhibit efficient numerical performance in simulations.

\subsection{Related Works}

The CS technique is a well-suited approach for detecting joint device activity and channel estimation.
To efficiently solve the CS-based activity detection problem, sparse recovery methods such as approximate message passing (AMP)~\cite{liu2018massive,senel2018grant,chen2018sparse,zhu2023message,rajoriya2023joint}, bilinear generalized AMP (BiGAMP)~\cite{iimori2021grant,zhang2023joint,bian2023joint}, group least absolute shrinkage and selection operator (LASSO)~\cite{li2022dynamic}, block coordinate descent (BCD)~\cite{liu2021efficient}, and approximation error method (AEM)~\cite{marata2023joint} have been developed.
As an alternative to joint activity detection and channel estimation, these two parts of the problem can also be addressed sequentially in a three-phase protocol~\cite{kang2021minimum,kang2022scheduling}, where the BSs use the covariance-based approach to identify the active devices and then schedule them in orthogonal transmission slots for channel estimation.
By scheduling devices in orthogonal channels, additional performance improvements in channel estimation and transmission rate can be obtained as compared to the CS-based grant-free protocol.

Recently, there has been significant research interest in the covariance-based activity detection due to its ability to detect a larger number of active devices and its superior performance compared to the CS-based approach in the massive MIMO system.
In particular, the pioneering work~\cite{haghighatshoar2018improved} introduces two formulations for the covariance-based activity detection: the non-negative least squares (NNLS) and the MLE.
While the NNLS formulation has a good recovery guarantee and a quadratic scaling law~\cite{fengler2021non}, the MLE formulation is more popular due to its superior detection performance, as demonstrated in \cite{chen2022phase}.
Several works have investigated the recovery guarantee of the MLE.
For instance, \cite{khanna2022support} considers the MLE problem with box constraints on the non-zero support set and provides the conditions for accurately recovering the true support set.
Meanwhile, \cite{fengler2021non} provides the quadratic scaling law for the restricted MLE problem, where the activity indicator to be estimated is binary.
However, the combinatorial constraints in the MLE models presented in \cite{khanna2022support} and \cite{fengler2021non} make them challenging to optimize.
To address this issue, \cite{chen2022phase} analyzes the more practical unrestricted MLE model, which provides a phase transition analysis, revealing that the unrestricted MLE shares the quadratic scaling law as the restricted MLE and the NNLS.
While the above theoretical results assume the single-cell scenario, \cite{chen2021sparse} extends the phase transition analysis of the unrestricted MLE to the multi-cell scenario.
Specifically, \cite{chen2021sparse} shows that we can determine the consistency of the MLE by numerically testing whether the intersection of two convex sets is a zero vector.
Nevertheless, the phase transition result in \cite{chen2021sparse} cannot directly determine the feasible region of the system parameters in which the consistency of the MLE holds.
This paper aims to address this issue by characterizing the scaling law for the MLE based on the results in \cite{chen2021sparse}.
The theoretical results for the covariance-based activity detection are summarized in Table~\ref{table:contribution}.

\begin{table*}[t]
	\centering
	\caption{A Summary of the Theoretical Results for the Covariance-Based Activity Detection.}
	\label{table:contribution}
	\resizebox{\linewidth}{!}{
	\begin{tabular}{|c|c|c|c|c|}
		\hline
		Scenario & Theoretical results & Types of signature sequences & Large-scale fading coefficients & References \\ \hline
		& Accurate recovery of the support set (for the   & Sequences with specific statistical & \multirow{2}{*}{Unknown} & \multirow{2}{*}{\cite{khanna2022support}} \\
		& MLE with box constraints on the support set) &  properties & & \\
		\cline{2-5}
		Single-cell & \multirow{2}{*}{Scaling law (for the MLE with binary constraints)}    &  \multirow{3}{*}{Randomly and uniformly generated} & \multirow{2}{*}{Known} & \multirow{2}{*}{\cite{fengler2021non}}  \\
		massive MIMO && \multirow{3}{*}{from the sphere of radius $\sqrt{L}$} &&\\
		\cline{2-2}\cline{4-5}
		& \multirow{2}{*}{Phase transition, scaling law, and estimation error} &  & \multirow{2}{*}{Unknown} &  \multirow{2}{*}{\cite{chen2022phase}}  \\
		&&&&\\
		\hline
		& \multirow{2}{*}{Phase transition}  & \multirow{2}{*}{Arbitrary}  & \multirow{2}{*}{Known} & \multirow{2}{*}{\cite{chen2021sparse}}  \\
		\multirow{3}{*}{Multi-cell} &&&&\\
		\cline{2-5}
		\multirow{3}{*}{massive MIMO} & \multirow{4}{*}{Scaling law and estimation error} & Randomly and uniformly generated & \multirow{3}{*}{Known and the path-loss} &    \multirow{4}{*}{This paper} \\ 
		&&   from the sphere of radius $\sqrt{L}$ or & \multirow{3}{*}{exponent $\gamma>2$}  & \\
		&&   the discrete set $\qam^L$ &  &\\
		\hline
	\end{tabular}}
\end{table*}

Efficient algorithm design for solving the covariance-based activity detection problem is an important research direction.
The CD algorithm is a commonly used approach for the covariance-based activity detection problem in both single-cell and multi-cell scenarios.
Specifically, it has been used in \cite{haghighatshoar2018improved} for the single-cell scenario and in \cite{chen2021sparse,jiang2022ml} for the multi-cell scenario.
To accelerate the convergence of the CD algorithm, variants such as \cite{dong2022faster,xie2020massive} use sampling strategies, and \cite{ganesan2021clustering} considers a clustering-based technique to reduce the complexity of coordinate updating.
Other algorithms for solving the detection problem in the single-cell scenario include the expectation maximization/minimization (EM) (i.e., sparse Bayesian learning)~\cite{wipf2007empirical}, the sparse iterative covariance-based estimation (SPICE)~\cite{stoica2011spice}, and gradient descent~\cite{wang2021efficient}.
In the multi-cell scenario, \cite{shao2020cooperative} proposes a decentralized approximate separating algorithm that exploits the similarity of activity indicators between neighboring BSs.
Another approach, presented in \cite{li2023asynchronous}, is a distributed algorithm that performs low-complexity computations in parallel at each BS.
Among the above algorithms, the CD algorithm and its variants are the most popular because they are computationally efficient and easy to implement~\cite{liu2024grant}.
However, there is still room for improving their computational efficiency because they do not fully exploit the special structure of the solution to the MLE problem, and the complexity of updating the coordinates is high.
In this paper, we aim to address these shortcomings by proposing new strategies to accelerate the CD algorithm.

The device activity detection and the covariance-based approach have been widely used for many related detection problems in practical applications.
For instance, in the joint activity and data detection problem, \cite{gao2023energy} analyzes the minimum transmission power required in a fading channel.
The covariance-based approach for this task has been studied in \cite{chen2019covariance,wang2021efficient}.
In asynchronous systems, where joint activity and delay detection are required, the BCD algorithm has been proposed for solving this problem in \cite{wang2022covariance}, while penalty-based algorithms with better detection performance have been proposed in \cite{lin2022sparsity,li2023asynchronous}.
In systems with frequency offsets, the covariance-based approach is studied in \cite{liu2022mle_icc}.
The covariance-based approach has also been extended to Rician fading channels in \cite{tian2022massive,liu2023mle}.
In the orthogonal frequency division multiplexing (OFDM)-based massive grant-free access scheme, the covariance-based approach is used in \cite{jia2023statistical}.
The problem formulation and algorithm for device activity detection using low-resolution analog-to-digital converters are studied in \cite{wang2023device}.
In irregular repetition slotted ALOHA systems, the covariance-based approach and algorithm are studied in \cite{srivatsa2022user}.
A comprehensive overview of the theoretical analysis, algorithm design, and practical issues of the covariance-based activity detection in both the single-cell and multi-cell scenarios is provided in~\cite{liu2024grant}.

In most of the works mentioned above, sourced random access is considered, which means that each device has a unique signature sequence for device identification.
This is in contrast to the unsourced random access scenario studied in \cite{polyanskiy2017perspective,amalladinne2020coded,fengler2021non,fengler2021sparcs}, where all devices use the same codebook, and the BSs decode the transmitted messages instead of decoding device activities.
This paradigm is suitable for situations where the transmitted messages are more important than the identities of the transmitters.

\subsection{Main Contributions}

This paper investigates the covariance-based activity detection problem in multi-cell massive MIMO systems, with a focus on the theoretical analysis and the algorithmic development.
On the theoretical side, we establish statistical properties of a more practical type of signature sequences and characterize the scaling law for system parameters of the multi-cell massive MIMO system for the first time.
In terms of algorithms, we propose two strategies to accelerate the CD algorithm in \cite{chen2021sparse} based on the special structure of the problem.
The main contributions of the paper are as follows:

\begin{itemize}
	\item \textit{Statistical properties.}
	We study the statistical properties of Type~\ref{item:qam} signature sequences randomly and uniformly generated from the finite alphabet $\qam^L,$  where $L$ is the length of the signature sequence.
	Compared to Type~\ref{item:sphere} signature sequences on the sphere of radius $\sqrt{L},$ which also exhibit similar statistical properties~\cite{fengler2021non}, Type~\ref{item:qam} signature sequences are easier to generate and store, resulting in significantly reduced hardware costs.
	To be specific, we establish the stable null space property (NSP) of order $\mathcal{O}(L^2)$ of the matrix whose each column is the Kronecker product of the conjugate of one signature sequence and itself. This property plays a crucial role in the scaling law analysis.
	\item \textit{Scaling law analysis.}
	We establish the scaling law in~\eqref{eq:introduction-scaling-law} for the system parameters $K, L, N,$ and $B,$ based on some mild assumptions of signature sequences and the large-scale fading coefficients.
	Under this scaling law, and as the number of antennas $M$ tends to infinity, the MLE can accurately recover active devices.
	Additionally, we characterize the asymptotic distribution of the MLE error, providing insight into the decay rate of the MLE error as $M$ tends to infinity.
	These results have important implications for understanding the performance of cooperative device activity detection in multi-cell massive MIMO systems.
	\item \textit{Efficient accelerated CD algorithms.}
	We propose two computationally efficient algorithms, namely the inexact CD algorithm and the active set CD algorithm, for solving the covariance-based activity detection problem.
	In particular, the inexact CD algorithm accelerates the CD algorithm by approximately solving the one-dimensional subproblems with low complexity;
	the active set CD algorithm fully exploits the special structure of the solution and accelerates the CD algorithm using the active set selection strategy.
	We provide convergence and iteration complexity\footnote{The iteration complexity of an algorithm is the maximum number of iterations that the algorithm needs to perform in order to return an approximate solution satisfying a given error tolerance.} guarantees for both algorithms.
	Furthermore, we demonstrate that the two acceleration strategies can be combined.
	In simulations, the combined algorithm, called active set inexact CD, not only achieves significantly better performance compared to the CD algorithm~\cite{chen2021sparse} and its accelerated version~\cite{ganesan2021clustering}, but also outperforms the proposed inexact CD and active set CD algorithms.
	Therefore, the combination of the two acceleration strategies yields the best algorithm among all the considered ones.
\end{itemize}

This paper extends the results presented in the two earlier conference publications~\cite{wang2021accelerating,wang2023scaling}. First, while the conference paper~\cite{wang2023scaling} focuses on the scaling law when using Type~\ref{item:sphere} signature sequences, this paper investigates the statistical properties of Type~\ref{item:qam} signature sequences and extends the scaling law result to this more practical type of sequences.
Second, this paper simplifies the active set selection strategy presented in \cite{wang2021accelerating}, and provides detailed proofs of its convergence and iteration complexity properties.
Finally, the paper proposes a new inexact coordinate update strategy that can be combined with the active set selection strategy to further accelerate the CD algorithm.

\subsection{Paper Organization and Notations}

The rest of the paper is organized as follows. Section~\ref{sec:system} introduces the system model and the covariance-based activity detection problem formulation. Section~\ref{sec:main-theory} presents the asymptotic detection performance analysis, which includes the statistical properties of signature sequences, the scaling law, and the distribution of the estimation error. In Section~\ref{sec:algorithm}, we propose two efficient accelerated CD algorithms. Simulation results are provided in Section~\ref{sec:simulation}. Finally we conclude the paper in Section~\ref{sec:conclusion}.

In this paper, lower-case, boldface lower-case, and boldface upper-case letters denote scalars, vectors, and matrices, respectively.
Calligraphic letters denote sets.
Superscripts $(\cdot)^{H},$ $(\cdot)^{T},$ $(\cdot)^{*},$ $(\cdot)^{-1},$ and $(\cdot)^{\dagger}$ denote conjugate transpose, transpose, conjugate, inverse, and Moore-Penrose inverse, respectively.
Additionally, $\mathbf{I}$ denotes the identity matrix with an appropriate dimension, $\mathbb{E}[\cdot]$ denotes expectation, $\op{tr}(\mathbf{X})$ denotes the trace of $\mathbf{X},$ $\op{diag}(x_1,x_2,\ldots,x_n)$ (or $\op{diag}(\mathbf{X}_1,\mathbf{X}_2,\ldots,\mathbf{X}_n)$) denotes a (block) diagonal matrix formed by $x_1,x_2,\ldots,x_n$ (or $\mathbf{X}_1,\mathbf{X}_2,\ldots,\mathbf{X}_n$), $\triangleq$ denotes definition, $|\cdot|$ denotes the determinant of a matrix, or the (coordinate-wise) absolute operator, or the cardinality of a set, $\|\mathbf{x}\|_p$ denotes the $\ell_p$-norm of $\mathbf{x},$ $\odot$ denotes element-wise product, and $\otimes$ denotes the Kronecker product. Finally, $\mathcal{N}(\boldsymbol\mu, \bm{\Sigma})$ (or $\mathcal{CN}(\boldsymbol\mu, \bm{\Sigma})$) denotes a (complex) Gaussian distribution with mean $\boldsymbol\mu$ and covariance $\bm{\Sigma}.$

\section{System Model and Problem Formulation}
\label{sec:system}

\subsection{System Model}

Consider an uplink multi-cell massive MIMO system comprising $B$ cells, where each cell includes one BS equipped with $M$ antennas and $N$ single-antenna devices. To mitigate inter-cell interference, we assume a C-RAN architecture, where all $B$ BSs are connected to a CU via fronthaul links, and the signals received at the BSs can be collected and jointly processed at the CU. During any coherence interval, only $K\ll N$ devices are active in each cell. Each device $n$ in cell $j$ is preassigned a unique signature sequence $\mathbf{s}_{jn}\in\mathbb{C}^{L}$ with sequence length $L.$ The signature sequence is generated from a specific distribution, such as the Type~\ref{item:qam} and \ref{item:sphere} signature sequences mentioned earlier.
We denote $a_{jn}$ as a binary variable with $a_{jn}=1$ for active and $a_{jn}=0$ for inactive devices. The channel between device $n$ in cell $j$ and BS $b$ is represented by $\sqrt{g_{bjn}}\mathbf{h}_{bjn},$ where $g_{bjn} \geq 0$ is the large-scale fading coefficient that depends on path-loss and shadowing, and $\mathbf{h}_{bjn}\in\mathbb{C}^{M}$ is the independent and identically distributed (i.i.d.) Rayleigh fading component following $\mathcal{CN}(\mathbf{0},\mathbf{I}).$

During the uplink pilot stage, all active devices synchronously transmit their preassigned signature sequences to the BSs as random access requests. Subsequently, the received signal at BS $b$ can be expressed as
\begin{align}\label{eq:sys}
	\mathbf{Y}_b&=\sum_{n=1}^{N}a_{bn}\mathbf{s}_{bn}g_{bbn}^{\frac{1}{2}}\mathbf{h}_{bbn}^T  + \sum_{j\neq b} \sum_{n=1}^N a_{jn}\mathbf{s}_{jn}g_{bjn}^{\frac{1}{2}}\mathbf{h}_{bjn}^T+\mathbf{W}_b\nonumber\\
	&=\mathbf{S}_b\mathbf{A}_b\mathbf{G}^{\frac{1}{2}}_{bb}\mathbf{H}_{bb}+\sum_{j\neq b}\mathbf{S}_j\mathbf{A}_j\mathbf{G}^{\frac{1}{2}}_{bj}\mathbf{H}_{bj}+\mathbf{W}_b,
\end{align}
where $\mathbf{S}_j=[\mathbf{s}_{j1},\mathbf{s}_{j2},\ldots,\mathbf{s}_{jN}]\in\mathbb{C}^{L\times N}$ is the signature sequence matrix of the devices in cell $j,$ $\mathbf{A}_{j}=\operatorname{diag}(a_{j1},a_{j2},\ldots,a_{jN})$ is a diagonal matrix that indicates the activity of the devices in cell $j,$ $\mathbf{G}_{bj}=\operatorname{diag}(g_{bj1},g_{bj2},\ldots,g_{bjN})$ contains the large-scale fading components between the devices in cell $j$ and BS $b,$ $\mathbf{H}_{bj}=[\mathbf{h}_{bj1},\mathbf{h}_{bj2},\ldots,\mathbf{h}_{bjN}]^T\in\mathbb{C}^{N\times M}$ is the Rayleigh fading channel between the devices in cell $j$ and BS $b,$ and $\mathbf{W}_b$ is the additive Gaussian noise that follows $\mathcal{CN}(\mathbf{0},\sigma_w^2\mathbf{I})$ with $\sigma_w^2$ being the variance of the background noise normalized by the transmit power.

For notational simplicity, we use $\mathbf{S}=[\mathbf{S}_{1},\mathbf{S}_{2},\ldots,\mathbf{S}_{B}]\in\mathbb{C}^{L\times BN}$ to denote the signature matrix of all devices, and use $\mathbf{G}_b=\operatorname{diag}(\mathbf{G}_{b1},\mathbf{G}_{b2}\ldots,\mathbf{G}_{bB})\in\mathbb{R}^{BN\times BN}$ to denote the matrix containing the large-scale fading components between all devices and BS $b.$
We also let $\mathbf{A} = \operatorname{diag}( \mathbf{A}_1, \mathbf{A}_2,\ldots, \mathbf{A}_B ) \in \mathbb{R}^{BN \times BN}$ be a diagonal matrix that indicates the activity of all devices, and use $\mathbf{a}\in\mathbb{R}^{BN}$ to denote its diagonal entries.
Throughout this paper we use two different notations to denote the components of $\mathbf{a}$ (or other vectors of length $BN$): (i) $a_i$ denotes the $i$-th component of $\mathbf{a},$ where $1 \le i \le BN;$ and (ii) $a_{bn}$ denotes the component of $\mathbf{a}$ that corresponds to device $n$ in cell $b,$ also termed as the $(b,n)$-th component of $\mathbf{a},$ where $1\le b \le B$ and $1\le n \le N.$

\subsection{Problem Formulation}

The task of device activity detection is to identify the active devices from the received signals $\mathbf{Y}_b,$ $b=1,2,\ldots,B.$
In this paper, we assume that the large-scale fading coefficients are known, i.e., the matrices $\mathbf{G}_{b}$ for all $b$ are known at the BSs.
Under this assumption, the activity detection problem can be formulated as the estimation of the activity indicator vector $\mathbf{a}.$

Notice that for each BS $b,$ the Rayleigh fading components and noises are both i.i.d.\ Gaussian over the antennas.
Therefore, given an activity indicator vector $\mathbf{a},$ the columns of the received signal $\mathbf{Y}_{b}$ in \eqref{eq:sys}, denoted by $\mathbf{y}_{bm},$ $m=1,2,\ldots,M,$ are i.i.d. Gaussian vectors,
i.e., $\mathbf{y}_{bm}\sim\mathcal{CN}(\mathbf{0},\boldsymbol\Sigma_b),$ where the covariance matrix $\boldsymbol\Sigma_b$ is given by
\begin{equation}
	\boldsymbol\Sigma_b = \frac{1}{M}\mathbb{E}\left[\mathbf{Y}_b\mathbf{Y}_b^H\right]=\mathbf{S} \mathbf{G}_b \mathbf{A} \mathbf{S}^H + \sigma_w^2 \mathbf{I}.
\end{equation}
Therefore, the likelihood function of $\mathbf{Y}_b$ can be expressed as
\begin{align}
	p(\mathbf{Y}_b\mid\mathbf{a}) & = \prod_{m=1}^{M} \frac{1}{| \pi \bm{\Sigma}_b |} \exp \left( - \mathbf{y}_{bm}^H \bm{\Sigma}_b^{-1} \mathbf{y}_{bm}  \right) \nonumber \\
	& = \frac{1}{| \pi \bm{\Sigma}_b |^M} \exp \left( - \op{tr} \left( \bm{\Sigma}_b^{-1} \mathbf{Y}_b \mathbf{Y}_b^H \right) \right) .
\end{align}
As the received signals $\mathbf{Y}_b,$ $b = 1,2,\ldots,B,$ are independent given $\mathbf{a}$ due to the i.i.d. Rayleigh fading channels, the likelihood function can be factorized as
\begin{equation}
	p(\mathbf{Y}_1,\mathbf{Y}_2,\ldots,\mathbf{Y}_B\mid\mathbf{a}) = \prod_{b=1}^{B} p(\mathbf{Y}_b\mid\mathbf{a}).
\end{equation}
Therefore, the MLE problem can be reformulated as the minimization of the negative log-likelihood function
$- \frac{1}{M} \sum_{b=1}^{B} \log p(\mathbf{Y}_b\,|\,\mathbf{a}),$ which can be expressed as
\begin{subequations}\label{eq:mle-binary}
	\begin{alignat}{2}
		&\underset{\mathbf{a}}{\operatorname{minimize}}    &\quad& \sum_{b=1}^B\left(\log\left|\boldsymbol\Sigma_b\right|+ \operatorname{tr}\left(\boldsymbol\Sigma_b^{-1}\widehat{\boldsymbol\Sigma}_b\right)\right)\\
		&\operatorname{subject\,to} &      &a_{bn} \in \{0,1\}, \,\forall \, b,n, \label{eq:mle-constraint-binary}
	\end{alignat}
\end{subequations}
where $\widehat {\boldsymbol\Sigma}_b = \mathbf{Y}_b\mathbf{Y}_b^H / M$ is the sample covariance matrix of the received signals at BS $b.$
Problem~\eqref{eq:mle-binary} is a computationally intractable combinatorial problem due to its binary constraints~\eqref{eq:mle-constraint-binary}.
To obtain a more computationally tractable problem, we relax the constraint $a_{bn} \in \{0,1\}$ to $a_{bn} \in [0,1]$ and obtain the following relaxed problem~\cite{chen2021sparse}:
\begin{subequations}\label{eq:mle}
	\begin{alignat}{2}
		&\underset{\mathbf{a}}{\operatorname{minimize}}    &\quad& \sum_{b=1}^B\left(\log\left|\boldsymbol\Sigma_b\right|+ \operatorname{tr}\left(\boldsymbol\Sigma_b^{-1}\widehat{\boldsymbol\Sigma}_b\right)\right)\\
		&\operatorname{subject\,to} &      &a_{bn} \in [0,1], \,\forall \, b,n. \label{eq:mle-constraint}
	\end{alignat}
\end{subequations}
Let $\hat{\mathbf{a}}^{(M)}$ denote the solution to problem \eqref{eq:mle} when the number of antennas $M$ is given.
Once $\hat{\mathbf{a}}^{(M)}$ is obtained, we apply element-wise thresholding to determine the binary activity vector $\mathbf{a} \in \{0,1\}^{BN}$ from $\hat{\mathbf{a}}^{(M)}.$ Specifically, we set a threshold $\ell_{th} \in (0,1)$ and let $a_{bn} = 1$ if $\hat{a}_{bn}^{(M)} \ge \ell_{th}$ and $a_{bn} = 0$ otherwise. 
The probabilities of missed detection~(PM) and false alarm~(PF) are defined as
\begin{align}
	\mathrm{PM} & = \frac{|\{ (b,n) \mid a_{bn} = 0 \text{ and } a_{bn}^{\circ} = 1 \}|}{BK}, \\
	\mathrm{PF} & = \frac{|\{ (b,n) \mid a_{bn} = 1 \text{ and } a_{bn}^{\circ} = 0 \}|}{B(N-K)},
\end{align}
which can be traded off by choosing different values for the threshold $\ell_{th}$ as done in~\cite{chen2022phase}.
As the device activity detection problem is solved by optimizing problem~\eqref{eq:mle}, which depends on the received signals only through the sample covariance matrices, this method is known as the covariance-based approach.

In this paper, we focus on the relaxed MLE problem formulated in~\eqref{eq:mle} and call it the MLE problem for simplicity throughout this paper.
Specifically, in Section~\ref{sec:main-theory}, we investigate the detection performance limit of problem~\eqref{eq:mle} as the number of antennas $M$ approaches infinity. In Section~\ref{sec:algorithm}, we propose efficient accelerated CD algorithms for solving problem~\eqref{eq:mle} to achieve accurate and fast activity detection.

\section{Asymptotic Detection Performance Analysis}
\label{sec:main-theory}

In this section, our focus is on characterizing the detection performance limit of problem~\eqref{eq:mle} as the number of antennas $M$ tends to infinity. We address three questions: (i) What are the statistical properties of the Type~\ref{item:qam} signature sequences, which are generated from the discrete symbol set $\qam^L,$ that would allow the detection performance guarantee of the MLE? (ii) Given the system parameters $L, N,$ and $B,$ how many active devices can be correctly detected by solving the MLE problem~\eqref{eq:mle} as $M$ approaches infinity? (iii) What is the asymptotic distribution of the MLE error?
Answering the first question provides theoretical guarantees for the practical generation of signature sequences. The answer to the second question provides an analytic scaling law that reveals how the inter-cell interference affects the detection performance of the MLE. Finally, the answer to the third question characterizes the error probabilities and estimates the decay rate of the MLE error as $M$ approaches infinity.

\subsection{Consistency of the MLE}

Our analysis is closely related to and based on the consistency result of the MLE problem~\eqref{eq:mle}, derived in \cite[Theorem~3]{chen2021sparse}, in a multi-cell setup.
\begin{lemma}[Consistency of the MLE~\cite{chen2021sparse}]\label{lemma:consistency}
	Consider the MLE problem~\eqref{eq:mle} with a given signature sequence matrix $\mathbf{S},$ large-scale fading component matrices $\mathbf{G}_b$ for all $b,$ and noise variance $\sigma_w^2.$ 
	Let matrix $\widetilde{\mathbf{S}}$ be defined as the matrix whose columns are the Kronecker product of $\mathbf{s}_{bn}^*$ and $ \mathbf{s}_{bn},$ i.e.,
	\begin{equation}\label{eq:s-tilde}
		\widetilde{\mathbf{S}}\triangleq[\mathbf{s}_{11}^*\otimes \mathbf{s}_{11},\mathbf{s}_{12}^*\otimes \mathbf{s}_{12},\ldots ,\mathbf{s}_{BN}^*\otimes \mathbf{s}_{BN}]\in\mathbb{C}^{L^2\times BN}.
	\end{equation}
	Let $\hat{\mathbf{a}}^{(M)}$ be the solution to \eqref{eq:mle} when the number of antennas $M$ is given, and let $\mathbf{a}^{\circ}$ be the true activity indicator vector whose $B(N-K)$ zero entries are indexed by $ \mathcal{I},$ i.e.,
	\begin{equation}\label{eq:def-I}
		\mathcal{I}\triangleq\{i\mid a_i^{\circ}=0\}.
	\end{equation}
	Define
	\begin{align}
		\mathcal{N}&\triangleq\{\mathbf{x}\in \mathbb{R}^{BN}\mid \widetilde{\mathbf{S}}\mathbf{G}_b\mathbf{x}= \mathbf{0}, \, \forall\, b\},\label{eq:subspace} \\
		\mathcal{C}&\triangleq\{\mathbf{x}\in\mathbb{R}^{BN}\mid x_i\geq 0~\text{if}~i\in \mathcal{I},\, x_i\leq 0 ~\text{if}~i \notin \mathcal{I}\}.\label{eq:cone}
	\end{align}
	Then a necessary and sufficient condition for $\hat{\mathbf{a}}^{(M)}\to \mathbf{a}^{\circ}$ as $M\to\infty$ is that the intersection of $\mathcal{N}$ and $\mathcal{C}$ is the zero vector, i.e., $\mathcal{N}\cap\mathcal{C}=\{\mathbf{0}\}.$
\end{lemma}

Lemma~\ref{lemma:consistency} can be explained as follows. The subspace $\mathcal{N}$ contains all directions from $\mathbf{a}^{\circ}$ along which the likelihood function remains unchanged, i.e., 
\begin{equation}
	p(\mathbf{Y}_1,\mathbf{Y}_2,\ldots,\mathbf{Y}_B\mid\mathbf{a}^{\circ}) = p(\mathbf{Y}_1,\mathbf{Y}_2,\ldots,\mathbf{Y}_B\mid\mathbf{a}^{\circ} + t\, \mathbf{x})
\end{equation}
holds for any small positive $t$ and any $ \mathbf{x} \in \mathcal{N}.$
The cone $\mathcal{C}$ contains all directions starting from $\mathbf{a}^{\circ}$ towards the feasible region, i.e., $\mathbf{a}^{\circ} + t\, \mathbf{x} \in [0,1]^{BN}$ holds for any small positive $t$ and any $ \mathbf{x} \in \mathcal{C}.$
The condition $\mathcal{N}\cap\mathcal{C}=\{\mathbf{0}\}$ implies that the likelihood function $p(\mathbf{Y}_1,\mathbf{Y}_2,\ldots,\mathbf{Y}_B\,|\,\mathbf{a}^{\circ})$ is uniquely identifiable in the feasible neighborhood of $\mathbf{a}^{\circ}.$

Since there is generally no closed-form characterization of $\mathcal{N}\cap\mathcal{C},$ it is difficult to analyze its scaling law, i.e., to characterize the feasible set of system parameters under which $\mathcal{N}\cap\mathcal{C}=\{\mathbf{0}\}$ holds true.
The signature sequences and the large-scale fading coefficients are critical in the scaling law analysis because they are involved in the definition of $\mathcal{N}$ in \eqref{eq:subspace}.
Therefore, taking these factors into consideration is essential for accurate scaling law analysis.

\subsection{Statistical Properties of Signature Sequences}

It turns out that the way signature sequences are generated plays a crucial role in the scaling law analysis of the MLE problem~\eqref{eq:mle}.
In this subsection, we consider the following two ways of generating signature sequences, formally state them in the following Assumption~\ref{assu:sequence}, and specify their statistical properties.
\begin{assumption}\label{assu:sequence}
	The signature sequence matrix $\mathbf{S}$ is generated from one of the following two ways, and the corresponding signature sequences are called Type~\ref{item:qam} and Type~\ref{item:sphere}, respectively:
	\begin{enumerate}[Type I:]
		\item \label{item:qam} Drawing the components of $\mathbf{S}$ uniformly and independently from the set $\qam,$ i.e., drawing the columns of $\mathbf{S}$ randomly and uniformly from the discrete set $\qam^L;$ 
		\item \label{item:sphere} Drawing the columns of $\mathbf{S}$ uniformly and independently from the complex sphere of radius $\sqrt{L}.$ 
	\end{enumerate}
\end{assumption}

It is worth mentioning that Type~\ref{item:qam} signature sequences are more cost effective for system implementation than Type~\ref{item:sphere} signature sequences.
In contrast to Type~\ref{item:qam} signature sequences, to generate Type~\ref{item:sphere} sequences, we first need to generate an i.i.d. Gaussian sequence of length $L,$ and then normalize it to obtain the sequence on the sphere of radius $\sqrt{L}.$
For storage, Type~\ref{item:qam} sequences can be stored with $2L$ bits per sequence (only $2$ bits per component), whereas each component of Type~\ref{item:sphere} sequences needs to be stored as a high-precision floating-point number. Therefore, Type~\ref{item:qam} is better than Type~\ref{item:sphere} in terms of the complexity of generating and storing the signature sequences. 
It is also worth stating that the Type~\ref{item:sphere} sequences have superior detection performance compared to i.i.d. Gaussian sequences at a fixed variance, as verified by simulation later in the paper.

In the following theorem, we show that the two types of signature sequences in Assumption~\ref{assu:sequence} share the same statistical properties.

\begin{theorem}\label{theorem:nsp}
	For both the Type~\ref{item:qam} and Type~\ref{item:sphere} sequences stated in Assumption~\ref{assu:sequence}, the following holds. For any given parameter $\bar{\rho}\in (0,1),$ there exist constants $c_1$ and $c_2$ depending only on $\bar{\rho}$ such that if
	\begin{equation}
		s \le c_1L^2/ \log^2(eBN/L^2),
	\end{equation}
	then with probability at least $1 - \exp(-c_2L),$ the matrix $\widetilde{\mathbf{S}}$ defined in \eqref{eq:s-tilde} has the stable NSP of order $s$ with parameters $\rho \in (0,\bar{\rho}).$ 
	More precisely, for any $\mathbf{v} \in \mathbb{R}^{BN}$ that satisfies $\widetilde{\mathbf{S}} \mathbf{v} = \mathbf{0},$ the following inequality holds for any index set $\mathcal{S}\subseteq\{1,2,\ldots,BN\}$ with $|\mathcal{S}|\leq s$:
	\begin{equation}\label{eq:nsp}
		\left\|\mathbf{v}_{\mathcal{S}}\right\|_1 \leq \rho \left\|\mathbf{v}_{\mathcal{S}^c}\right\|_1,
	\end{equation}
	where $\mathbf{v}_{\mathcal{S}}$ is a sub-vector of $\mathbf{v}$ with entries from $\mathcal{S},$ and $\mathcal{S}^{c}$ is the complementary set of $\mathcal{S}$ with respect to $\{1,2,\ldots,BN\}.$
\end{theorem}
\begin{IEEEproof}
	Please see Appendix~\ref{sec:proof-nsp-bernoulli} in the supplementary material.
\end{IEEEproof}

The main contribution of Theorem~\ref{theorem:nsp} is to establish the stable NSP of $\widetilde{\mathbf{S}}$ when Type~\ref{item:qam} signature sequences are used.
For the case where Type~\ref{item:sphere} signature sequences are used, this conclusion has already been proved in \cite{fengler2021non}, where the proof relies on the concentration properties of the signature sequences on the sphere of radius $\sqrt{L}.$ In our proof, we build on the techniques used in \cite{fengler2021non} and leverage the concentration properties of Type~\ref{item:qam} signature sequences due to their independent sub-Gaussian coordinates.
Therefore, the stable NSP of $\widetilde{\mathbf{S}}$ holds for two different types of signature sequences in Assumption~\ref{assu:sequence}.

\subsection{Scaling Law Analysis}

In this subsection, we present one of the main results of this paper, which is the scaling law analysis of problem~\eqref{eq:mle} under two types of signature sequences as specified in Assumption~\ref{assu:sequence}.
It turns out that the scaling law analysis in the single-cell and multi-cell scenarios is fundamentally different due to the presence of the large-scale fading coefficients. To gain a better understanding of the impact of the large-scale fading coefficients on the scaling law analysis, we present the scaling law results in the single-cell and multi-cell scenarios separately.

\subsubsection{Single-Cell Scenario}

First, we focus on the single-cell scenario where $B = 1.$
To simplify the notation, we remove all the cell and BS indices $b$ and index the devices by $n$ only.

The following consistency result of the MLE in the single-cell scenario is a special case of Lemma~\ref{lemma:consistency}.
\begin{proposition}\label{prop:consistency-single}
	Consider the MLE problem~\eqref{eq:mle} in the single-cell scenario with given $\mathbf{S},$ $\mathbf{G},$ and $\sigma_w^2.$ 
	Let
	\begin{equation}
		\widetilde{\mathbf{S}}=[\mathbf{s}_{1}^*\otimes \mathbf{s}_{1},\mathbf{s}_{2}^*\otimes \mathbf{s}_{2},\ldots ,\mathbf{s}_{N}^*\otimes \mathbf{s}_{N}]\in\mathbb{C}^{L^2\times N}.
	\end{equation}
	Let
	$\hat{\mathbf{a}}^{(M)}$ be the MLE solution when the number of antennas $M$ is given, and let $\mathbf{a}^{\circ}$ be the true activity indicator, and 
	$
	\mathcal{I}^{\prime}=\{i\mid a_i^{\circ}=0\}.
	$
	Define
	\begin{align}
		\mathcal{N}^{\prime}&=\{\mathbf{x}\in \mathbb{R}^{N}\mid \widetilde{\mathbf{S}}\mathbf{G}\mathbf{x}= \mathbf{0}\}, \label{eq:subspace-single} \\
		\mathcal{C}^{\prime}&=\{\mathbf{x}\in\mathbb{R}^{N}\mid x_i\geq 0~\text{if}~i\in \mathcal{I}^{\prime},\, x_i\leq 0~\text{if}~i \notin \mathcal{I}^{\prime}\}.
	\end{align}
	Then a necessary and sufficient condition for $\hat{\mathbf{a}}^{(M)}\to \mathbf{a}^{\circ}$ as $M\to\infty$ is that $\mathcal{N}^{\prime}\cap\mathcal{C}^{\prime}=\{\mathbf{0}\}.$
\end{proposition}

Notice that the definition of $\mathcal{N}^{\prime}$ in \eqref{eq:subspace-single} contains only one equation, in contrast to the multi-cell scenario where there are $B$ mutually coupled equations.
This structural difference allows for the elimination of the large-scale fading coefficients in $\mathcal{N}^{\prime},$ as demonstrated by the following proposition.
\begin{proposition}\label{prop:NC-equivalent}
	Define $\mathcal{N}^{\prime\prime}$ as the null space of $\widetilde{\mathbf{S}},$ i.e.,
	\begin{equation}
		\mathcal{N}^{\prime\prime}\triangleq\{\mathbf{x}\in \mathbb{R}^{BN}\mid \widetilde{\mathbf{S}}\mathbf{x}= \mathbf{0}\}.
	\end{equation}
	Then the condition $\mathcal{N}^{\prime}\cap\mathcal{C}^{\prime}=\{\mathbf{0}\}$ in Proposition~\ref{prop:consistency-single} holds if and only if $\mathcal{N}^{\prime\prime}\cap\mathcal{C}^{\prime}=\{\mathbf{0}\}.$
\end{proposition}
\begin{IEEEproof}
	Please see Appendix~\ref{sec:NC-equivalent}.
\end{IEEEproof}

Proposition~\ref{prop:NC-equivalent} highlights a key observation that the large-scale fading coefficients have no impact on the consistency of the MLE in the single-cell scenario. This consistency solely depends on the null space of $\widetilde{\mathbf{S}}.$
The stable NSP of $\widetilde{\mathbf{S}}$ in Theorem~\ref{theorem:nsp} leads to the following scaling law result in the single-cell scenario, which extends the result in \cite[Theorem~9]{chen2022phase} from Type~\ref{item:sphere} signature sequences to Type~\ref{item:qam} signature sequences.
\begin{proposition}\label{prop:scaling-law-single}
	For both the Type~\ref{item:qam} and Type~\ref{item:sphere} sequences stated in Assumption~\ref{assu:sequence},	there exist constants $c_1$ and $c_2 > 0,$ independent of system parameters $K,\, L,$ and $N,$ such that if
	\begin{equation}\label{eq:scaling-law-single}
		K \le c_1 L^2 / \log^2(eN/L^2),
	\end{equation}
	then the condition $\mathcal{N}^{\prime\prime}\cap \mathcal{C}^{\prime}=\{\mathbf{0}\}$ holds with probability at least $1-\exp(-c_2L).$
\end{proposition}
\begin{IEEEproof}
	Please see Appendix~\ref{sec:scaling-law-single}.
\end{IEEEproof}

Proposition~\ref{prop:scaling-law-single} demonstrates that the scaling law in the single-cell scenario holds true for both types of signature sequences in Assumption~\ref{assu:sequence}.
Specifically, with a sufficiently large $M,$ the maximum number of active devices that can be accurately detected by solving the MLE is in the order of $L^2.$
Consequently, the detection performance of the MLE remains reliable, even when using the more practical Type~\ref{item:qam} signature sequences.
It is worth mentioning that the scaling law in the single-cell scenario is independent of any assumptions regarding the large-scale fading coefficients. But as demonstrated later, the scaling law in the multi-cell scenario is established based on the assumption that the path-loss exponent $\gamma > 2.$

\subsubsection{Multi-Cell Scenario}

We now consider the multi-cell scenario (i.e., $B \ge 2$).
The scaling law analysis in the multi-cell scenario is significantly different from that in the single-cell scenario. Specifically, in the single-cell scenario, Proposition~\ref{prop:NC-equivalent} implies that the large-scale fading coefficients (i.e., $\mathbf{G}$) have no impact on the consistency of the MLE.
The reason is that there is only one equality in the definition of $\mathcal{N}^{\prime}$ in \eqref{eq:subspace-single}, which allows the large-scale fading coefficients to be eliminated via algebraic operations.
However, in the multi-cell scenario, since there are multiple coupled equations in the definition of $\mathcal{N}$ in \eqref{eq:subspace}, the large-scale fading coefficients (i.e., $\mathbf{G}_1, \mathbf{G}_2, \ldots, \mathbf{G}_B$) cannot be eliminated as in the single-cell scenario.
As a result, the large-scale fading coefficients play a critical role in the scaling law analysis in the multi-cell scenario, i.e., determining the feasible set of system parameters under which the condition $\mathcal{N}\cap\mathcal{C}=\{\mathbf{0}\}$ in Lemma~\ref{lemma:consistency} is satisfied.
To establish the scaling law in the multi-cell scenario, we must first specify the assumption made regarding the large-scale fading coefficients and the properties that result from the assumption.

\begin{assumption}\label{assu:cell}
	The multi-cell system consists of $B$ hexagonal cells with radius $R.$
	In this system, the large-scale fading components are inversely proportional to the distance raised to the power $\gamma$~\cite{rappaport1996wireless}, i.e.,
	\begin{equation}\label{eq:gamma}
		g_{bjn} = P_0 \left(\frac{D_0}{D_{bjn}}\right)^{\gamma},
	\end{equation}
	where $P_0$ is the received power at the point with distance $D_0$ from the transmitting antenna, $D_{bjn}$ is the BS-device distance between device $n$ in cell $j$ and BS $b,$ and $\gamma$ is the path-loss exponent.
\end{assumption}

\begin{lemma}\label{lemma:gamma}
	Suppose that Assumption~\ref{assu:cell} holds true with $\gamma>2.$
	Then, there exists a constant $C > 0$ depending only on $\gamma,$ $P_0,$ $D_0,$ and $R$ defined in Assumption~\ref{assu:cell}, such that for each BS $b,$ the large-scale fading coefficients satisfy
	\begin{equation}\label{eq:less-C}
		\sum_{j=1,\,j\neq b}^{B} \Big(\max_{1\le n \le N} g_{bjn} \Big) \le C.
	\end{equation}
\end{lemma}
\begin{IEEEproof}
	Please see Appendix~\ref{sec:proof-gamma} in the supplementary material.
\end{IEEEproof}

According to Lemma~\ref{lemma:gamma}, if the path-loss exponent $\gamma > 2$ in Assumption~\ref{assu:cell}, then the summation in the left-hand side of \eqref{eq:less-C} is upper bounded by a constant $C$ that is independent of $B.$
The intuitive explanation for the validity of Lemma~\ref{lemma:gamma} is as follows.
For a particular BS $b,$ since most of the interfering cells are far away from it, the distances $D_{bjn}$'s for all devices in these cells are large.
Consequently, most of the terms $\{\max_{n} g_{bjn}\}$ in the summation in the left-hand side of~\eqref{eq:less-C} are sufficiently small to make \eqref{eq:less-C} hold true.
Notably, $\gamma > 2$ is a sufficient condition for \eqref{eq:less-C}, and it typically holds for most channel models and application scenarios~\cite{rappaport1996wireless}. Finally, it is worth mentioning that the hexagonal structure is not an essential aspect of Lemma~\ref{lemma:gamma}, and similar results can be obtained for other structured systems, such as square cells.

Building on the results presented in Theorem~\ref{theorem:nsp} and Lemma~\ref{lemma:gamma}, we can now state the scaling law for the MLE, which establishes a sufficient condition for $\mathcal{N}\cap \mathcal{C}=\{\mathbf{0}\}.$

\begin{theorem}\label{theorem:scaling-law}
	For both Type~\ref{item:qam} and Type~\ref{item:sphere} sequences in Assumption~\ref{assu:sequence},	and under Assumption~\ref{assu:cell} with $\gamma>2,$ there exist constants $c_1$ and $c_2 > 0,$ independent of system parameters $K,$ $L,$ $N,$ and $B,$ such that if
	\begin{equation}\label{eq:scaling-law}
		K \le c_1 L^2 / \log^2(eBN/L^2),
	\end{equation}
	then the condition $\mathcal{N}\cap \mathcal{C}=\{\mathbf{0}\}$ in Lemma~\ref{lemma:consistency} holds with probability at least $1-\exp(-c_2L).$
\end{theorem}
\begin{IEEEproof}
	Please see Appendix~\ref{sec:proof-scaling-law}.
\end{IEEEproof}

Theorem~\ref{theorem:scaling-law} affirms that the MLE estimator $\hat{\mathbf{a}}^{(M)}$ converges to the true value $\mathbf{a}^{\circ}$ with overwhelmingly high probability as $M$ approaches infinity, under the scaling law \eqref{eq:scaling-law}.
As a result, with a sufficiently large $M,$ the MLE can accurately detect up to $L^2$ active devices by solving problem~\eqref{eq:mle}. Additionally, we observe that the scaling law \eqref{eq:scaling-law} in the multi-cell scenario is almost identical to the scaling law \eqref{eq:scaling-law-single} in the single-cell scenario. This implies that solving the MLE problem~\eqref{eq:mle} can detect almost as many active devices in each cell in the multi-cell scenario as it does in the single-cell scenario. Furthermore, it indicates that the inter-cell interference is not a limiting factor of the detection performance since the total number of cells $B$ only affects $K$ through $\log B$ in \eqref{eq:scaling-law}.

It is important to note that Theorem~\ref{theorem:scaling-law} applies not only to two types of signature sequences outlined in Assumption~\ref{assu:sequence} but also to any other types of signature sequences that have the following two properties: (i) the stable NSP of $\widetilde{\mathbf{S}}$ stated in Theorem~\ref{theorem:nsp} and (ii) normalized norms, i.e., $\| \mathbf{s}_{bn} \|_2^2 = L$ for all $b$ and $n.$
The proof of the scaling law is established based only on the properties (i) and (ii) of the signature sequences.

\subsection{Asymptotic Distribution of Estimation Error}

In this subsection, assuming the consistency of the MLE, we characterize the asymptotic distribution of the estimation error $\hat{\mathbf{a}}^{(M)} - \mathbf{a}^{\circ}$ as $M$ tends to infinity.
The following result is a generalization of \cite[Theorem~4]{chen2022phase} from the single-cell case to the multi-cell case.
\begin{theorem}\label{theorem:error}
	Consider the MLE problem~\eqref{eq:mle} with given $\mathbf{S},$ $\{\mathbf{G}_b\},$ $\sigma_w^2,$ and $M.$
	Assume that $\mathcal{N}\cap\mathcal{C}=\{\mathbf{0}\}$ as in Lemma~\ref{lemma:consistency} holds.
	The Fisher information matrix of problem~\eqref{eq:mle} is given by
	\begin{equation}
		\mathbf{J}(\mathbf{a}) = M \sum_{b=1}^{B} \big( \mathbf{Q}_b \odot \mathbf{Q}_b^* \big),
	\end{equation}
	where
	$\mathbf{Q}_b = \mathbf{G}_b^{\frac{1}{2}} \mathbf{S}^H \big(\mathbf{S} \mathbf{G}_b\mathbf{A}\mathbf{S}^H + \sigma_w^2 \mathbf{I}  \big)^{-1} \mathbf{S} \mathbf{G}_b^{\frac{1}{2}}.$
	Let $\mathbf{x} \in \mathbb{R}^{BN}$ be a random vector sampled from $\mathcal{N} \big(\mathbf{0}, M \mathbf{J}(\mathbf{a}^{\circ})^{\dagger}\big).$
	Then, for each realization of $\mathbf{x},$ there exists a vector $\hat{\boldsymbol{\eta}} \in \mathbb{R}^{BN}$ which is the solution to the following QP:
	\begin{subequations}\label{eq:QP}
		\begin{alignat}{2}
			&\underset{\boldsymbol\eta}{\operatorname{minimize}}    &\quad& \frac{1}{M} (\mathbf{x}-\boldsymbol\eta)^T \mathbf{J}(\mathbf{a}^{\circ})  (\mathbf{x}-\boldsymbol\eta) \label{eq:quad-fisher} \\
			&\operatorname{subject\,to} &      & \boldsymbol{\eta} \in \mathcal{C},
		\end{alignat}
	\end{subequations}
	where $\mathcal{C}$ is defined in \eqref{eq:cone}, such that $\sqrt{M} \big(\hat{\mathbf{a}}^{(M)} - \mathbf{a}^{\circ}\big)$ converges in distribution to the collection of $\hat{\boldsymbol{\eta}}$'s as $M \to \infty.$ 
\end{theorem}
\begin{IEEEproof}
	Please see Appendix~\ref{sec:proof-error}.
\end{IEEEproof}

Theorem~\ref{theorem:error} can be explained as follows. Consider a sample drawn from $\mathbf{x} \sim \mathcal{N} \big(\mathbf{0}, M \mathbf{J}(\mathbf{a}^{\circ})^{\dagger}\big).$ Let $\hat{\boldsymbol{\eta}}$ be the projection of $\mathbf{x}$ onto the cone $\mathcal{C}$ using the distance metric defined by the quadratic function \eqref{eq:quad-fisher}. Then, the MLE error $\sqrt{M} \big(\hat{\mathbf{a}}^{(M)} - \mathbf{a}^{\circ}\big)$ converges in distribution to the same distribution as the projected sample $\hat{\boldsymbol{\eta}}.$ This result is consistent with the fact that the MLE error is confined within the cone $\mathcal{C},$ since the true vector $\mathbf{a}^{\circ}$ lies on the boundary of the feasible set $[0,1]^{BN}.$

It is not possible to analytically characterize the distribution of the MLE error using Theorem~\ref{theorem:error} since the QP~\eqref{eq:QP} generally does not have a closed-form solution. Nevertheless, this result provides insight into the decay rate of the MLE error, which can be approximated by $\frac{1}{\sqrt{M}}\hat{\boldsymbol{\eta}}$ for a sufficiently large $M.$ Furthermore, it shows that we can numerically compute the distribution of the estimation error for large values of $M$ by solving the QP~\eqref{eq:QP}.

\section{Efficient Accelerated CD Algorithms} 
\label{sec:algorithm}

In this section, we develop efficient algorithms for solving the MLE problem~\eqref{eq:mle}.
While the optimization problem~\eqref{eq:mle} is not convex due to the convexity of $\op{tr}\big(\boldsymbol\Sigma_b^{-1}\widehat{\boldsymbol\Sigma}_b\big)$ and the concavity of $\log\left|\boldsymbol\Sigma_b\right|,$ the CD algorithm that iteratively updates each coordinate of the unknown variable $\mathbf{a}$ while holding the others fixed has demonstrated good computational performance and detection error probability.
However, when $B$ and $N$ are large, the CD algorithm (e.g., as proposed in \cite{chen2021sparse}) can become inefficient and involve unnecessary computations, as discussed in detail in Section~\ref{subsec:cd}. To address these issues, we propose the use of inexact and active set strategies to accelerate the CD algorithm. The proposed algorithms are based on the following principles:
\begin{itemize}
	\item \textit{Inexactly solve the subproblem in CD.}
	The CD algorithm in \cite{chen2021sparse} updates coordinates by solving the one-dimensional subproblem exactly, which can result in high complexity in the cooperative multi-cell system. To address this issue, the inexact CD algorithm uses a low-complexity algorithm to approximately solve the one-dimensional subproblem to update the coordinates. As a result, the overall complexity of inexact CD is much lower than that of CD.
	\item \textit{Exploit the special structure of the solution.}
	The consistency of the MLE in Section~\ref{sec:main-theory} implies that the solution to problem~\eqref{eq:mle} is close to the binary vector $\mathbf{a}^{\circ}.$ The proposed active set CD algorithm fully exploits this special structure of the solution, wherein most components of the solution are located on the boundary of the box constraint. The active set idea can be applied to accelerate any algorithm for solving problem~\eqref{eq:mle} that does not exploit the special structure of its solution.
\end{itemize}

Our goal in designing algorithms in this section is to quickly find a feasible point that satisfies the first-order optimality condition of problem~\eqref{eq:mle}, which is a necessary condition for its global solution.
Let $F(\mathbf{a})$ denote the objective function of problem~\eqref{eq:mle}.
For any $b=1,2,\ldots,B,$ $n=1,2,\ldots,N,$ the gradient of $F(\mathbf{a})$ with respect to $a_{bn}$ is given by
\begin{equation}
	\left[\nabla F(\mathbf{a})\right]_{bn} = \sum_{j=1}^{B} g_{jbn} \left( \mathbf{s}_{bn}^H \boldsymbol\Sigma_j^{-1} \mathbf{s}_{bn} - \mathbf{s}_{bn}^H\boldsymbol\Sigma_j^{-1}\widehat{\boldsymbol\Sigma}_j\boldsymbol\Sigma_j^{-1}\mathbf{s}_{bn} \right).
\end{equation}
The first-order optimality condition of problem~\eqref{eq:mle} is then given by
\begin{equation}\label{eq:kkt}
	\left[\nabla F(\mathbf{a})\right]_{bn}
	\begin{cases}
		\geq 0, & \text{if}~a_{bn} = 0;   \\
		\leq 0, & \text{if}~a_{bn} = 1; \\
		= 0, &  \text{if}~ 0 < a_{bn} < 1,
	\end{cases}\ \forall~b,n ,
\end{equation}
which is a necessary condition for the optimality of $\mathbf{a}.$
To quantify the degree to which each coordinate of $\mathbf{a}$ violates condition~\eqref{eq:kkt}, we define a non-negative vector as follows:
\begin{equation}\label{eq:def-va}
	\mathbb{V}(\mathbf{a}) \triangleq \left | \op{Proj}(\mathbf{a} - \nabla F(\mathbf{a})) - \mathbf{a} \right | \in \mathbb{R}_{+}^{BN},
\end{equation}
where $\op{Proj}(\cdot)$ is the (coordinate-wise) projection operator onto the interval $[0, 1],$ and $|\cdot|$ is the (coordinate-wise) absolute operator.
Notice that the first-order optimality condition~\eqref{eq:kkt} is equivalent to $\mathbb{V}(\mathbf{a}) = \mathbf{0}.$
The vector $\mathbb{V}(\mathbf{a})$ plays a central role in this section; see the discussion below Proposition~\ref{prop:icd}.

\subsection{CD Algorithm}
\label{subsec:cd}

Randomly permuted CD~\cite{chen2019covariance} is one of the most efficient variants of the CD algorithm for solving problem~\eqref{eq:mle}. 
At each iteration, the algorithm randomly permutes the indices of all coordinates and then updates the coordinates one by one according to the order in the permutation.
For any given coordinate $(b,n)$ of device $n$ in cell $b,$ the algorithm solves the following one-dimensional problem \cite{chen2021sparse}:
\begin{multline}\label{eq:one-dim}
	\underset{d \in [-a_{bn},\,1-a_{bn}]}{\operatorname{minimize}} \quad  \sum_{j=1}^B\Bigg(\log\left(1+d\,g_{jbn}\,\mathbf{s}_{bn}^H\boldsymbol\Sigma_j^{-1}\mathbf{s}_{bn}\right) \\
	-\frac{d\,g_{jbn}\,\mathbf{s}_{bn}^H\boldsymbol\Sigma_j^{-1}\widehat{\boldsymbol\Sigma}_j\boldsymbol\Sigma_j^{-1}\mathbf{s}_{bn}}{1+d\,g_{jbn}\,\mathbf{s}_{bn}^H\boldsymbol\Sigma_j^{-1}\mathbf{s}_{bn}} \Bigg)
\end{multline}
to possibly update $a_{bn}.$
Problem~\eqref{eq:one-dim} does not have a closed-form solution, which is different from the single-cell case.
However, it can be transformed into a polynomial root-finding problem of degree $2B - 1,$ which can be solved exactly by computing all the eigenvalues of the companion matrix formed by the coefficients of the corresponding polynomial function~\cite[Chap.~6]{mcnamee2007numerical}.
The computational complexity of this approach to solving problem~\eqref{eq:one-dim} is $\mathcal{O}(B^3).$

\begin{algorithm}[t]
	\caption{CD and Inexact CD Algorithms for Solving the MLE Problem~\eqref{eq:mle}}
	\label{alg:cd}
	\begin{algorithmic}[1]
		\STATE Initialize $\mathbf{a} = \mathbf{0},$ $\boldsymbol\Sigma_b^{-1} = \sigma_w^{-2}\mathbf{I},$ $1 \le b \le B,$ and $\epsilon > 0;$ 
		\REPEAT
		\STATE Randomly select a permutation $\{ (b,n)_1, (b,n)_2, \ldots, (b,n)_{BN} \}$ of the coordinate indices $\{(1,1), (1,2), \ldots, (B,N)\}$ of $\mathbf{a};$
		\FOR{$(b,n) = (b,n)_1$ to $(b,n)_{BN}$}
		\STATE \label{alg:cd-one-dim} \textbf{If CD:} Apply the root-finding algorithm~\cite{mcnamee2007numerical} to solve subproblem~\eqref{eq:one-dim} \textit{exactly} to obtain $\hat{d},$ and set $d = \hat{d};$
		\STATE \textbf{If inexact CD:} Apply Algorithm~\ref{alg:icd} to solve subproblem~\eqref{eq:one-dim} \textit{inexactly} to obtain $\bar{d},$ and set $d = \bar{d};$
		\STATE $a_{bn} \leftarrow a_{bn} + d;$
		\STATE $\boldsymbol\Sigma_j^{-1} \leftarrow \boldsymbol\Sigma_j^{-1} -  \frac{d\, g_{jbn}\,\boldsymbol\Sigma_j^{-1} \mathbf{s}_{bn} \mathbf{s}_{bn}^H \boldsymbol\Sigma_j^{-1}}{1\, +\, d\, g_{jbn}\, \mathbf{s}_{bn}^H \boldsymbol\Sigma_j^{-1} \mathbf{s}_{bn}},$ $ 1 \le j \le B;$  \label{alg:cd-update-sigma}
		\ENDFOR
		\UNTIL $\big\|\mathbb{V}(\mathbf{a})\big\|_{\infty} \le \epsilon;$
		\STATE Output $\mathbf{a}.$
	\end{algorithmic}
\end{algorithm}

The termination criterion for the CD algorithm is the first-order optimality condition, which indicates that the algorithm stops when the following condition is satisfied:
\begin{equation}\label{eq:terminate}
	\big\|\mathbb{V}(\mathbf{a})\big\|_{\infty} \le \epsilon,
\end{equation}
where $\epsilon \ge 0$ is the error tolerance.
We call the feasible point $\mathbf{a}$ satisfying \eqref{eq:terminate} an $\epsilon$-stationary point.
When $\epsilon=0,$ the $\epsilon$-stationary point is the first-order stationary point.
The (randomly permuted) CD algorithm is summarized in Algorithm~\ref{alg:cd}.

The dominant computational complexity of CD Algorithm~\ref{alg:cd} arises from the polynomial root-finding and the matrix-vector multiplications in lines~\ref{alg:cd-one-dim} and~\ref{alg:cd-update-sigma}, respectively. The complexity of the polynomial root-finding is $\mathcal{O}(B^3),$ whereas the complexity of the matrix-vector multiplications is $\mathcal{O}(BL^2).$ Therefore, the overall complexity of one iteration in CD Algorithm~\ref{alg:cd} is $\mathcal{O}(BN(B^3 + BL^2)).$

\subsection{Inexact CD Algorithm}
\label{subsec:icd}

The CD algorithm (e.g., Algorithm~\ref{alg:cd}) always solves problem~\eqref{eq:one-dim} exactly to update the coordinates, which can significantly complicate the algorithm.
The purpose of the inexact CD algorithm is to solve problem~\eqref{eq:one-dim} inexactly in a controllable fashion, which enables the coordinates to be updated with a substantially lower complexity.

Notice that the large-scale fading coefficient $g_{jbn}$ appears in the $j$-th summation term in \eqref{eq:one-dim} by multiplying with $d.$
Moreover, because BS $b$ is the closest BS to device $(b,n),$ the path-loss model in Assumption~\ref{assu:cell} implies that $g_{bbn}$ is significantly larger than $g_{jbn}$ for all $j \neq b.$
As a result, for the variable $a_{bn},$ the $b$-th term in problem~\eqref{eq:one-dim} dominates all the other terms, i.e., it contains most of the information of the whole objective function. The underlying idea of constructing an approximate solution to \eqref{eq:one-dim} is to retain the $b$-th term and approximate the other terms in order to efficiently update the variable $a_{bn}.$

To demonstrate the inexact algorithm for solving problem~\eqref{eq:one-dim}, we can first rewrite it as follows:
\begin{equation}\label{eq:one-dim-2}
	\underset{d \in [-a_{bn},\,1-a_{bn}]}{\op{minimize}} \quad \sum_{j=1}^{B} f_j(\mathbf{a} + d\,\mathbf{e}_{bn}),
\end{equation}
where $f_j(\mathbf{a}) = \log\left|\boldsymbol\Sigma_j\right|+ \op{tr}\big(\boldsymbol\Sigma_j^{-1}\widehat{\boldsymbol\Sigma}_j\big)$ is the $j$-th summation term in the objective function $F(\mathbf{a}),$
and $\mathbf{e}_{bn}$ is a vector whose $(b,n)$-th component is $1$ and the other components are $0.$
Instead of directly solving problem~\eqref{eq:one-dim-2} exactly, we solve an approximate problem, given by
\begin{multline}\label{eq:icd}
	d^{(i)} \, = \underset{d \in [-a_{bn},\,1-a_{bn}]}{\op{argmin}}
	\quad \bigg \{ f_b(\mathbf{a} + d\,\mathbf{e}_{bn}) \\
	+\sum_{j=1,\,j\neq b}^{B} \left(f_j(\mathbf{a}) + \left[ \nabla f_j(\mathbf{a})\right]_{bn}d \right) + \frac{\mu^{(i)}}{2} d^2\bigg\}.
\end{multline}
In problem~\eqref{eq:icd}, we retain the dominant term $f_b(\mathbf{a} + d\,\mathbf{e}_{bn})$ because it dominates the objective function in \eqref{eq:one-dim-2}, and approximate the other terms by a quadratic polynomial, where $\mu^{(i)}$ is a properly selected parameter, and
\begin{equation}
	\left[ \nabla f_j(\mathbf{a})\right]_{bn} = g_{jbn} \left( \mathbf{s}_{bn}^H \boldsymbol\Sigma_j^{-1} \mathbf{s}_{bn} - \mathbf{s}_{bn}^H\boldsymbol\Sigma_j^{-1}\widehat{\boldsymbol\Sigma}_j\boldsymbol\Sigma_j^{-1}\mathbf{s}_{bn} \right)
\end{equation}
is the gradient of $f_j(\mathbf{a})$ with respect to $a_{bn}.$

In sharp contrast to problem~\eqref{eq:one-dim-2} (equivalent to problem~\eqref{eq:one-dim}), the approximate problem~\eqref{eq:icd} can be solved with complexity $\mathcal{O}(1).$
Specifically, by setting the derivative of \eqref{eq:icd} to zero, we get the following cubic polynomial equation:
\begin{multline}\label{eq:cubic}
	\left( 1+d\,\xi_{bn} \right)^2 \Bigg(\sum_{j=1,\,j\neq b}^{B} \left[ \nabla f_j(\mathbf{a})\right]_{bn} + \mu^{(i)} d\Bigg) \\ 
	+ \xi_{bn}\left( 1+d\, \xi_{bn} \right) - \zeta_{bn}
	= 0,
\end{multline}
where $ \xi_{bn} = g_{bbn}\,\mathbf{s}_{bn}^H\boldsymbol\Sigma_b^{-1}\mathbf{s}_{bn},\,\zeta_{bn} = g_{bbn}\,\mathbf{s}_{bn}^H\boldsymbol\Sigma_b^{-1}\widehat{\boldsymbol\Sigma}_b\boldsymbol\Sigma_b^{-1}\mathbf{s}_{bn}.$
All roots of \eqref{eq:cubic} can be obtained from the cubic formula,
and the optimal solution to \eqref{eq:icd} can be obtained by choosing the one with the smallest objective among the roots of \eqref{eq:cubic} and the two boundary points $-a_{bn}$ and $1-a_{bn}.$

\begin{algorithm}[t]
	\caption{An Inexact Algorithm for Solving Problem~\eqref{eq:one-dim}}
	\label{alg:icd}
	\renewcommand{\algorithmicrepeat}{\textbf{Repeat} for $i = 0,1,\ldots$}
	\begin{algorithmic}[1]
		\STATE Initialize $\mu^{(0)} > 0,$ and $\beta > 1;$
		\REPEAT
		\STATE $\mu^{(i)} \leftarrow \beta^i \mu^{(0)};$
		\STATE Solve problem~\eqref{eq:icd} with $\mu^{(i)}$ to obtain $d^{(i)};$
		\UNTIL{condition~\eqref{eq:decrease-condition} is satisfied;}
		\STATE Output $\bar{d}$ as $d^{(i)}.$
	\end{algorithmic}
\end{algorithm}

Now we discuss the choice of the parameter $\mu^{(i)}$ in \eqref{eq:icd}. In particular, the parameter $\mu^{(i)}$ in problem~\eqref{eq:icd} should be chosen such that its solution $d^{(i)}$ satisfies the following sufficient decrease condition:
\begin{multline}\label{eq:decrease-condition}
	\sum_{j=1,\,j\neq b}^{B} f_j\left(\mathbf{a} + d^{(i)}\mathbf{e}_{bn}\right) \\
	\le \sum_{j=1,\,j\neq b}^{B} \left(f_j(\mathbf{a}) + \left[ \nabla f_j(\mathbf{a})\right]_{bn}d^{(i)} \right) + \frac{\mu^{(i)}}{2} \left(d^{(i)}\right)^2.
\end{multline}
This condition guarantees that for each coordinate update, the inexact CD update decreases the objective function; see Proposition~\ref{prop:icd} further ahead.
A way of finding the desirable parameter $\mu^{(i)}$ is summarized in Algorithm~\ref{alg:icd}, which involves gradually increasing the value of $\mu^{(i)}$ starting from a sufficiently small initial value of $\mu^{(0)}.$
In practice, we can exploit the second-order information of the objective function to efficiently determine $\mu^{(0)}.$
To be specific, we set $\mu^{(0)}$ as
\begin{equation}\label{eq:initialize-mu}
	\mu^{(0)} = \sum_{j=1,\,j\neq b}^{B} \left[\nabla^2 f_j(\mathbf{a}) \right]_{bn,bn}
\end{equation}
if the right-hand side of \eqref{eq:initialize-mu} is positive, where $\nabla^2 f_j(\mathbf{a})$ is the Hessian matrix of $f_j(\mathbf{a}),$ and
\begin{multline}
	\left[\nabla^2 f_j(\mathbf{a}) \right]_{bn,bn} = g_{jbn}^2 \, \mathbf{s}_{bn}^H \boldsymbol\Sigma_j^{-1} \mathbf{s}_{bn} \\
	\times \left( 2\,\mathbf{s}_{bn}^H\boldsymbol\Sigma_j^{-1}\widehat{\boldsymbol\Sigma}_j\boldsymbol\Sigma_j^{-1}\mathbf{s}_{bn} - \mathbf{s}_{bn}^H \boldsymbol\Sigma_j^{-1} \mathbf{s}_{bn} \right) .
\end{multline}
Otherwise, we set $\mu^{(0)}$ to a small positive number, e.g., $10^{-2}.$
The complete inexact CD algorithm is summarized in Algorithm~\ref{alg:cd}.

The following proposition demonstrates that Algorithm~\ref{alg:icd} is guaranteed to return a value of $\mu^{(i)}$ such that the solution to problem~\eqref{eq:icd} satisfies the sufficient decrease condition in \eqref{eq:decrease-condition}.

\begin{proposition}\label{prop:icd}
	Suppose that the parameters in Algorithm~\ref{alg:icd} satisfy $\mu^{(0)} > 0$ and $\beta > 1.$
	Then, the algorithm will terminate at a certain $i\le \log_{\beta} \big(\lip / \mu^{(0)}\big) +1,$ where $\lip > 0$ is the Lipschitz constant of $\nabla F(\mathbf{a}).$ Furthermore, the output $\bar{d}$ satisfies
	\begin{equation}\label{eq:sufficiently-decrease}
		F(\mathbf{a} + \bar{d}\,\mathbf{e}_{bn}) \le F(\mathbf{a}) - \frac{\left[\mathbb{V}(\mathbf{a})\right]_{bn}^2}{2\lip(\beta+1)},
	\end{equation}
	where $\mathbb{V}(\mathbf{a})$ is defined in \eqref{eq:def-va}.
\end{proposition}
\begin{IEEEproof}
	Please see Appendix~\ref{sec:proof-icd}.
\end{IEEEproof}

Proposition~\ref{prop:icd} provides important insights into the behavior of Algorithm~\ref{alg:icd}. First, \eqref{eq:sufficiently-decrease} demonstrates that the output of Algorithm~\ref{alg:icd} decreases the objective function of the MLE problem~\eqref{eq:mle}. Second, the degree to which the objective function is decreased depends on $\left[\mathbb{V}(\mathbf{a})\right]_{bn},$ which measures the degree to which the coordinate $a_{bn}$ violates the first-order optimality condition. Therefore, the vector $\mathbb{V}(\mathbf{a})$ characterizes the potential of each coordinate for decreasing the objective function. In particular, updating the coordinate $a_{bn}$ with a larger $\left[\mathbb{V}(\mathbf{a})\right]_{bn}$ is expected to yield a larger decrease in the objective function of problem~\eqref{eq:mle}. Therefore, $\mathbb{V}(\mathbf{a})$ can guide the coordinate selection process in Algorithm~\ref{alg:cd} and improve its convergence speed, which will be exploited in the next subsection.

The computational complexity of inexact CD Algorithm~\ref{alg:cd} is primarily determined by matrix-vector multiplications, where updating a single coordinate requires $\mathcal{O}(BL^2)$ operations.
As a result, the overall complexity of one iteration in the inexact CD Algorithm~\ref{alg:cd} is $\mathcal{O}(BN\cdot BL^2).$

\begin{table*}[t]
	\centering
	\caption{Complexity Comparison Between the CD Algorithm~\cite{chen2021sparse} and the Proposed Accelerated CD Algorithms.}
	\label{table:complexity}
	\resizebox{\linewidth}{!}{
	\begin{tabular}{|c|cc|cc|}
		\hline
		& \multicolumn{1}{c|}{\parbox[t]{81pt}{\centering CD~\cite{chen2021sparse}}}  & \parbox[t]{81pt}{\centering Inexact CD} & \multicolumn{1}{c|}{\parbox[t]{81pt}{\centering Active set CD}} & \parbox[t]{81pt}{\centering Active set inexact CD}  \\ \hline
		Total number of coordinates that need &  \multicolumn{2}{c|}{\multirow{2}{*}{$BN$}} &   \multicolumn{2}{c|}{\multirow{2}{*}{$\left|\mathcal{A}^{(k)}\right|$}}   \\
		to be updated at the $k$-th iteration &&  &&\\
		\hline
		Computational complexity of updating & \multicolumn{1}{c|}{\multirow{2}{*}{$\mathcal{O}(BL^2+B^3)$}}  & \multirow{2}{*}{$\mathcal{O}(BL^2)$}  & \multicolumn{1}{c|}{\multirow{2}{*}{$\mathcal{O}(BL^2+B^3)$}} & \multirow{2}{*}{\centering{$\mathcal{O}(BL^2)$}} \\
		one coordinate& \multicolumn{1}{c|}{} && \multicolumn{1}{c|}{} &\\
		\hline
	\end{tabular}}
\end{table*}

\subsection{Active Set CD Algorithm}

In Algorithm~\ref{alg:cd}, both CD and inexact CD treat all coordinates equally and attempt to update all of them at each iteration. However, due to the special structure of the solution to problem~\eqref{eq:mle}, many coordinates $a_{bn}$ would have their corresponding subproblem~\eqref{eq:one-dim} solved (exactly or inexactly), but the coordinate update step $d$ can be zero, and as a result, $a_{bn}$ does not change, which leads to unnecessary computations that slow down the algorithm. To address this issue, we propose an active set CD algorithm to accelerate Algorithm~\ref{alg:cd} by reducing the number of unnecessary coordinate updates.

The active set CD algorithm differs from Algorithm~\ref{alg:cd} in that, at each iteration, it first judiciously selects an active set of coordinates and then applies lines~\ref{alg:cd-one-dim}--\ref{alg:cd-update-sigma} in Algorithm~\ref{alg:cd} to update the coordinates in the active set once. Due to the special structure of the solution to problem~\eqref{eq:mle}, it is expected that the cardinality of the selected active set would be significantly less than the total number of devices (i.e., $BN$), and more importantly, gradually decrease as the iteration goes on. Therefore, the active set CD algorithm reduces unnecessary computational cost and significantly accelerates Algorithm~\ref{alg:cd}.

To select an active set that improves the objective function as much as possible, the desirable active set should contain coordinates that contribute the most to the deviation from the optimality condition in \eqref{eq:kkt}, corresponding to coordinates with a sufficiently large $\left[\mathbb{V}(\mathbf{a})\right]_{bn}$ in \eqref{eq:def-va}, as discussed in the last second paragraph of Section~\ref{subsec:icd}. However, the cardinality of the active set should be as small as possible to avoid unnecessary computation and improve computational efficiency. Therefore, there is a trade-off in selecting the coordinates with large $\left[\mathbb{V}(\mathbf{a})\right]_{bn}$ into the active set. Mathematically, we propose a selection strategy for the active set $\mathcal{A}^{(k)}$ expressed as
\begin{equation}\label{eq:select-active-set}
	\mathcal{A}^{(k)} = \big\{ (b,n) \mid \big[ \mathbb{V}(\mathbf{a}^{(k)})\big]_{bn} \ge \omega^{(k)}  \big\},
\end{equation}
where $\omega^{(k)} \ge 0$ is a properly selected threshold parameter at each iteration.
As a result, at each iteration, the total number of coordinates that need to be updated in the active set CD algorithm is only $\left|\mathcal{A}^{(k)}\right|,$ which could be much less than $BN$ in Algorithm~\ref{alg:cd}.
The active set selection strategy in \eqref{eq:select-active-set} is different from that proposed in our previous work~\cite{wang2021accelerating}.
The selection of the active set in \cite{wang2021accelerating} is based on the variable $a_{bn}$ and the gradient $\big[\nabla F(\mathbf{a})\big]_{bn},$ but the selection strategy \eqref{eq:select-active-set} depends only on $\big[ \mathbb{V}(\mathbf{a}^{(k)})\big]_{bn},$ which is more concise and easier to implement.

The choice of the threshold parameter $\omega^{(k)}$ in \eqref{eq:select-active-set} is crucial in achieving the balance between decreasing the objective function and reducing the cardinality of the active set (and hence the computational cost) at the $k$-th iteration. If the parameter $\omega^{(k)}$ is set to be a large value, the cardinality of the active set at the $k$-th iteration would be small, and the iteration would require fewer computations. However, the decrease in the objective function would also be small. Conversely, if the parameter $\omega^{(k)}$ is set to a small value, more coordinates would be selected into the active set, resulting in a larger decrease in the objective function, but at the cost of increased computational complexity.
In practice, one way of selecting the parameter $\omega^{(k)}$ is to set a relatively large initial value $\omega^{(0)}$ and then gradually decrease $\omega^{(k)}$ at each iteration.

\begin{algorithm}[t]
	\caption{Active Set CD Algorithm for Solving the MLE Problem~\eqref{eq:mle}}
	\label{alg:active-set}
	\renewcommand{\algorithmicrepeat}{\textbf{Repeat} for $k = 0,1,\ldots$}
	\begin{algorithmic}[1]
		\STATE Initialize $\mathbf{a}^{(0)} = \mathbf{0},$ and $\epsilon > 0;$
		\REPEAT
		\STATE Update $\omega^{(k)};$
		\STATE Select the active set $\mathcal{A}^{(k)}$ according to \eqref{eq:select-active-set};
		\STATE Apply lines~\ref{alg:cd-one-dim}--\ref{alg:cd-update-sigma} in Algorithm~\ref{alg:cd} to update all coordinates in $\mathcal{A}^{(k)}$ \textit{only once} in the order of a random permutation;
		\UNTIL{$\big\|\mathbb{V}(\mathbf{a}^{(k)})\big\|_{\infty} \le \epsilon;$} \label{alg:termination-active-set}
		\STATE Output $\mathbf{a}^{(k)}.$
	\end{algorithmic}
\end{algorithm}

Based on the above discussion, we propose the active set CD algorithm for solving problem~\eqref{eq:mle}, which is detailed in Algorithm~\ref{alg:active-set}. It is worth mentioning that the active set strategy can accelerate both CD and inexact CD in Algorithm~\ref{alg:cd}. Furthermore, the idea of the active set can be applied to accelerate any algorithm for solving problem~\eqref{eq:mle} that does not properly exploit the special structure of the solution. A comparison of the per-iteration computational complexity of the CD algorithm and the proposed algorithms is summarized in Table~\ref{table:complexity}.
Notice that among all compared algorithms in Table~\ref{table:complexity}, the active set inexact CD algorithm has the lowest per-iteration complexity.

The convergence of the proposed active set CD Algorithm~\ref{alg:active-set} is presented in Theorem~\ref{theorem:active-set}.
It is important to note that a poor choice of the active set may lead to oscillations and even divergence of the algorithm.
The convergence and the iteration complexity properties of Algorithm~\ref{alg:active-set} are mainly due to the active set selection strategy in \eqref{eq:select-active-set} (and the careful choices of parameters $\omega^{(k)}$), and the convergence property of Algorithm~\ref{alg:cd}.

\begin{theorem}\label{theorem:active-set}
	For any given error tolerance $\epsilon > 0,$ 
	let $\omega^{(k)}$ satisfy the condition
	\begin{equation}\label{eq:active-set-parameters}
		\epsilon \le \omega^{(k)} \le \max\left\{\big\| \mathbb{V}(\mathbf{a}^{(k)}) \big\|_{\infty}, \, \epsilon \right\}.
	\end{equation}
	Then, the proposed active set CD Algorithm~\ref{alg:active-set} would return an $\epsilon$-stationary point	within $\mathcal{O}(\lip\beta/\epsilon^{2})$ iterations.
\end{theorem}
\begin{IEEEproof}
	Please see Appendix~\ref{sec:proof-active-set}.
\end{IEEEproof}

Some remarks on the parameter setting in Theorem~\ref{theorem:active-set} are in order. 
First, setting $\epsilon$ as a lower bound on $\omega^{(k)}$ ensures that the active set always excludes coordinates with $\left[ \mathbb{V}(\mathbf{a}^{(k)}) \right]_{bn} < \epsilon.$
The reason for setting the lower bound is that updating the coordinate $a_{bn}$ with a sufficiently small $\left[ \mathbb{V}(\mathbf{a}^{(k)})\right]_{bn}$ has little potential to decrease the objective function of problem~\eqref{eq:mle}.
Second, the second inequality in \eqref{eq:active-set-parameters} guarantees that the active set $\mathcal{A}^{(k)}$ is always non-empty unless the algorithm terminates.
Specifically, if $\| \mathbb{V}(\mathbf{a}^{(k)}) \|_{\infty} > \epsilon,$ then the last inequality in \eqref{eq:active-set-parameters} implies that $\omega^{(k)} \le \big\| \mathbb{V}(\mathbf{a}^{(k)}) \big\|_{\infty},$ and $\mathcal{A}^{(k)}$ is non-empty due to \eqref{eq:select-active-set}. Otherwise, if $\| \mathbb{V}(\mathbf{a}^{(k)}) \|_{\infty} \le \epsilon,$ then the termination condition in line~\ref{alg:termination-active-set} will be satisfied.

\section{Simulation Results}
\label{sec:simulation}

In this section, we verify the theoretical results and demonstrate the efficiency of the proposed accelerated CD algorithms through simulations.
We consider a multi-cell system where all potential devices within each cell are uniformly distributed.
Unless otherwise specified, we consider the cells to be hexagonal with a radius of $500$\,m and assume a homogeneous setup where the total number of devices $N$ and the number of active devices $K$ are the same among different cells.
In the simulations, we model the channel path-loss as $128.1 + 37.6 \log_{10}(d)$ (satisfying Assumption~\ref{assu:cell} with $\gamma=3.76$), where $d$ is the corresponding BS-device distance in km. We set the transmit power of each device as $23$\,dBm and the background noise power as $-169$\,dBm/Hz over $10$\,MHz.

\subsection{Numerical Validation of Theoretical Results}

\subsubsection{Scaling Law}

\begin{figure}[t]
	\centering
	\includegraphics[width=0.9\columnwidth,clip]{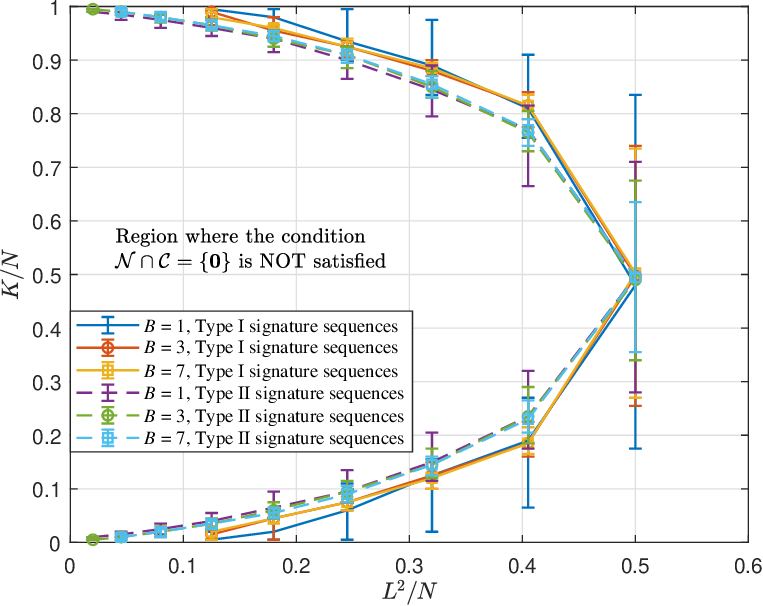}
	\caption{Scaling law comparison of the covariance-based activity detection with Type~\ref{item:qam} and Type~\ref{item:sphere} signature sequences for different $B$'s.}
	\label{fig:scaling-law-both}
\end{figure}

We consider the covariance-based device activity detection problem and solve the linear program proposed in \cite{chen2021sparse} to numerically test the condition $\mathcal{N} \cap \mathcal{C} = \{\mathbf{0}\}$ in Lemma~\ref{lemma:consistency} under a variety of choices of $L$ and $K,$ given $N = 200$ devices and $B = 1, 3, 7$ cells.
The signature sequences of length $L$ used are Type~\ref{item:qam} or Type~\ref{item:sphere} in Assumption~\ref{assu:sequence}.
Fig.~\ref{fig:scaling-law-both} plots the region of $(L^2/N, K/N)$ in which the condition is satisfied or not under the two types of signature sequences.
The result is obtained based on $100$ random realizations of $\mathbf{S}$ and $\mathbf{a}^{\circ}$ for each given $K$ and $L.$
The error bars indicate the range beyond which either all realizations or no realization satisfy the condition.
We observe from Fig.~\ref{fig:scaling-law-both} that the curves with different $B$'s overlap with each other for a given type of signature sequences, which implies that the scaling law for $\mathcal{N}\cap \mathcal{C}=\{\mathbf{0}\}$ is almost independent of $B.$
This is consistent with our analysis in Theorem~\ref{theorem:scaling-law} that the number of active devices $K$ which can be correctly detected in each cell in the multi-cell scenario depends on $B$ only through $\log B.$
We can also observe from Fig.~\ref{fig:scaling-law-both} that $K$ is approximately proportional to $L^2,$ which verifies the scaling law in \eqref{eq:scaling-law}.
It is worth mentioning that the curves in Fig.~\ref{fig:scaling-law-both} are symmetric. Further interpretations and discussions on the symmetry of the curves can be found in \cite{chen2021sparse}.

Another key observation from Fig.~\ref{fig:scaling-law-both} is that the curves of the two types of signature sequences almost overlap exactly, and Type~\ref{item:sphere} signature sequences seem slightly better than Type~\ref{item:qam} signature sequences.
This verifies Theorem~\ref{theorem:scaling-law}, which states that the (same) scaling law holds for both types of signature sequences.
From a practical point of view, Type~\ref{item:qam} signature sequences are easier to generate and store, although slightly more active devices can be correctly detected using Type~\ref{item:sphere} signature sequences.

\subsubsection{Asymptotic Distribution of Estimation Error}

\begin{figure}[t]
	\centering
	\includegraphics[width=0.9\columnwidth,clip]{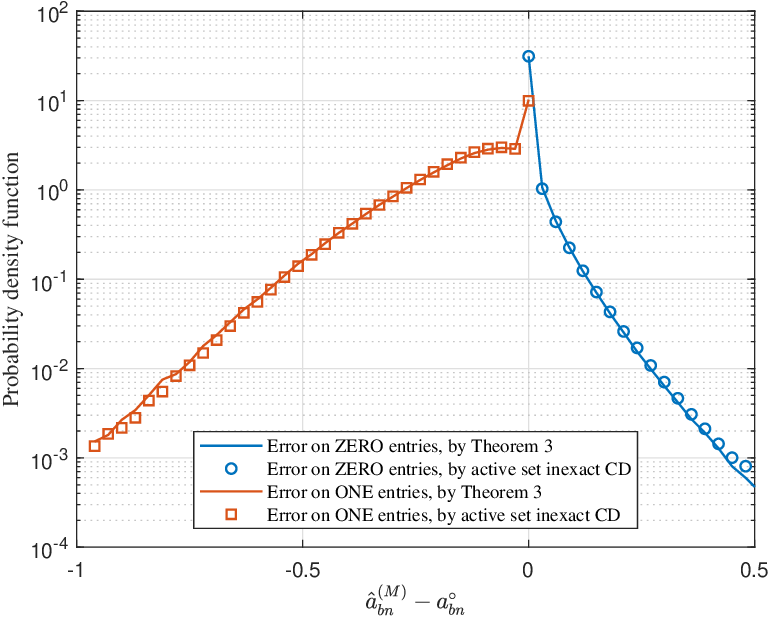}
	\caption{Probability density functions (PDFs) of the error on the zero entries and on the one entries.}
	\label{fig:QP-PDF}
\end{figure}

We validate the distribution of the estimation error $\hat{\mathbf{a}}^{(M)}-\mathbf{a}^{\circ}$ in Fig.~\ref{fig:QP-PDF} with $B = 7$ cells, $N=200$ devices, $K = 20$ active devices, $L = 20$ symbols, and $M = 128$ antennas.
For simplicity, we only use Type~\ref{item:qam} signature sequences, and similar simulation results can be obtained using Type~\ref{item:sphere} signature sequences.
We use the active set inexact CD Algorithm~\ref{alg:active-set} to solve the MLE problem~\eqref{eq:mle} and compare its error distribution with the theoretical distribution from solving the QP \eqref{eq:QP}.
For simplicity, we treat each coordinate of $\hat{\mathbf{a}}^{(M)}-\mathbf{a}^{\circ}$ as independent and plot the empirical distribution of the coordinate-wise error.
We consider two types of coordinates corresponding to inactive and active devices, i.e., entry $a^{\circ}_{bn}$ is zero or one, and plot their corresponding distributions separately.
We observe that the curves obtained from solving problem~\eqref{eq:mle} by active set inexact CD match well with the curves predicted by Theorem~\ref{theorem:error} for both types of coordinates,
and their error distribution is Gaussian distributed (except near point zero).
Another observation is that there is a point mass in the distribution of the error for both types of coordinates, which represents the probabilities of correctly identifying an inactive or active device in the current setup.

\begin{figure}[t]
	\centering
	\includegraphics[width=0.9\columnwidth,clip]{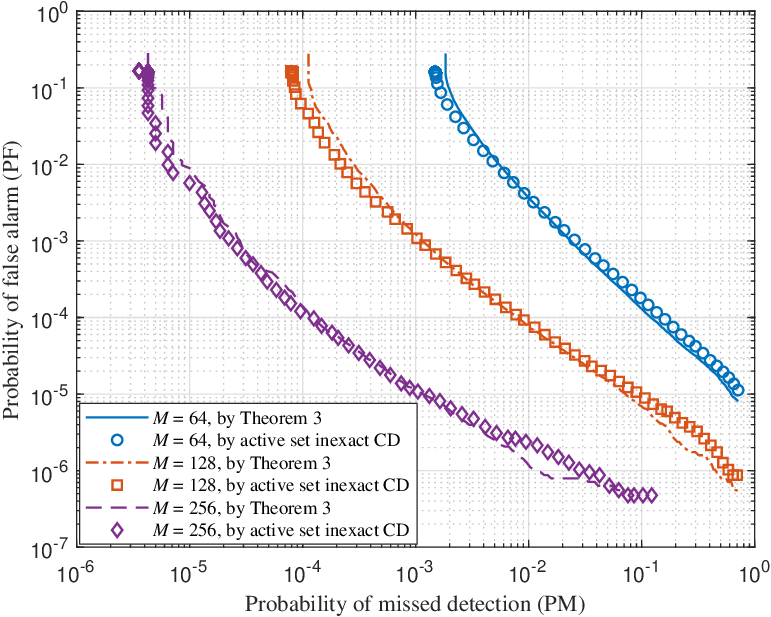}
	\caption{Comparison of the simulated results and the analysis in terms of probability of missed detection (PM) and probability of false alarm (PF) for the covariance-based activity detection for different $M$'s.}
	\label{fig:QP-error}
\end{figure}

In Fig.~\ref{fig:QP-error}, we present the detection performance with $B = 7$ cells, $N=200$ devices, $K = 20$ active devices, $L = 20$ symbols, and $M = 64,$ $128,$ or $256$ antennas.
We compare the detection performance obtained from solving problem~\eqref{eq:mle} using the active set inexact CD Algorithm~\ref{alg:active-set} with the result from solving the QP \eqref{eq:QP}.
The probability of missed detection (PM) and probability of false alarm (PF) are traded off by choosing different values for the threshold.
We observe from Fig.~\ref{fig:QP-error} that the curves obtained from the active set inexact CD algorithm match well with the theoretical ones.

\subsection{Detection Performance Comparison for Different Types of Signature Sequences}

In this subsection, we perform the covariance-based activity detection and compare the detection performance using different types of signature sequences.
We consider three types of signature sequences including Type~\ref{item:qam} and Type~\ref{item:sphere} in Assumption~\ref{assu:sequence} and the following Type~\ref{item:Gaussian}:
\begin{enumerate}[Type I:]\setcounter{enumi}{2}
	\item \label{item:Gaussian} Drawing the signature sequences from an i.i.d. complex Gaussian distribution $\mathcal{CN}(\mathbf{0},\mathbf{I}).$
\end{enumerate}
We fix $B = 7$ cells, $N = 200$ devices, $K = 20$ active devices, and consider various values of $L$ and $M.$

\begin{figure}[t]
	\centering
	\includegraphics[width=0.9\columnwidth,clip]{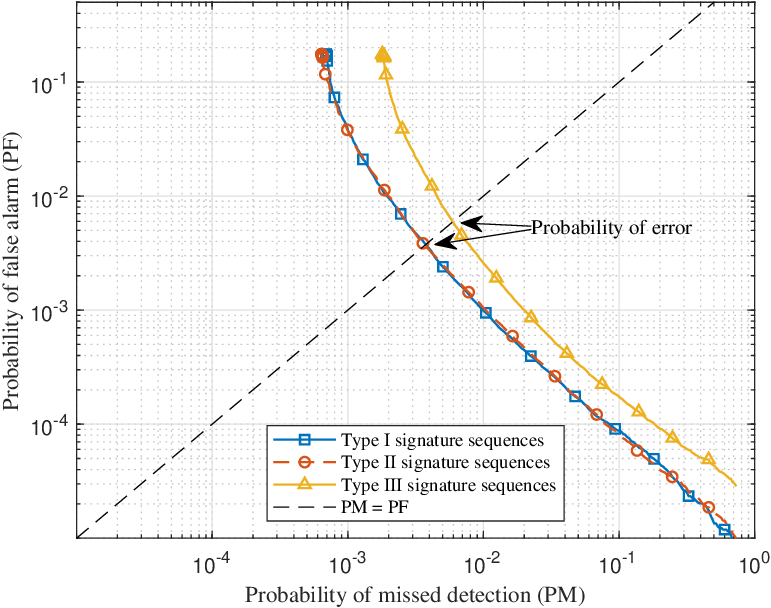}
	\caption{Detection performance comparison of the covariance-based activity detection with three different types of signature sequences.}
	\label{fig:sequence-PM-PF}
\end{figure}

In Fig.~\ref{fig:sequence-PM-PF}, we show the detection performance when using these three types of signature sequences with $L = 16$ symbols and $M = 256$ antennas.
We observe that almost the same detection performance can be obtained by using Type~\ref{item:qam} and Type~\ref{item:sphere} signature sequences.
Considering the hardware cost, it is more practical to use Type~\ref{item:qam} signature sequences without degrading the detection performance.
Additionally, we observe from Fig.~\ref{fig:sequence-PM-PF} that the detection performance of Type~\ref{item:Gaussian} signature sequences is worse than the other two types of sequences.
It turns out that the normalization of the signature sequences (assigned to different devices) is critical to the detection performance.

\begin{table*}[t]
	\centering
	\caption{Comparison of the Estimated Activity Indicator for the Signature Sequences With Different Norms.}
	\label{table:norm-sequence}
	\resizebox{\linewidth}{!}{
	\begin{tabular}{|c|ccccc|ccccc|}
		\hline
		& \multicolumn{5}{c|}{Device $(b,n)$ is inactive: $a_{bn}^{\circ} = 0$}                      & \multicolumn{5}{c|}{Device $(b,n)$ is active: $a_{bn}^{\circ} = 1$}                     \\ \hline
		Estimated $a_{bn}$ with & \multicolumn{1}{c|}{\multirow{2}{*}{$0.0206$}} & \multicolumn{1}{c|}{\multirow{2}{*}{$0.0309$}} & \multicolumn{1}{c|}{\multirow{2}{*}{$0.0101$}}  & \multicolumn{1}{c|}{\multirow{2}{*}{$0.0547$}} & \multicolumn{1}{c|}{\multirow{2}{*}{$0.0000$}} & \multicolumn{1}{c|}{\multirow{2}{*}{$0.9091$}} &\multicolumn{1}{c|}{\multirow{2}{*}{$0.9035$}}&\multicolumn{1}{c|}{\multirow{2}{*}{$0.9806$}}&\multicolumn{1}{c|}{\multirow{2}{*}{$1.0000$}}& \multicolumn{1}{c|}{\multirow{2}{*}{$1.0000$}} \\ 
		sequence $\s_{bn}$&\multicolumn{1}{c|}{}&\multicolumn{1}{c|}{}&\multicolumn{1}{c|}{}&\multicolumn{1}{c|}{}&\multicolumn{1}{c|}{}&\multicolumn{1}{c|}{}&\multicolumn{1}{c|}{}&\multicolumn{1}{c|}{}&\multicolumn{1}{c|}{}& \\ \hline
		Estimated $a_{bn}$ with & \multicolumn{1}{c|}{\multirow{2}{*}{$0.0821$}} & \multicolumn{1}{c|}{\multirow{2}{*}{$0.1248$}} & \multicolumn{1}{c|}{\multirow{2}{*}{$0.0404$}}  & \multicolumn{1}{c|}{\multirow{2}{*}{$0.2188$}} & \multicolumn{1}{c|}{\multirow{2}{*}{$0.0000$}} & \multicolumn{1}{c|}{\multirow{2}{*}{$0.8143$}} &\multicolumn{1}{c|}{\multirow{2}{*}{$0.9091$}}&\multicolumn{1}{c|}{\multirow{2}{*}{$0.9626$}}&\multicolumn{1}{c|}{\multirow{2}{*}{$1.0000$}}& \multicolumn{1}{c|}{\multirow{2}{*}{$0.7341$}} \\ 
		sequence $\frac{1}{2}\,\s_{bn}$&\multicolumn{1}{c|}{}&\multicolumn{1}{c|}{}&\multicolumn{1}{c|}{}&\multicolumn{1}{c|}{}&\multicolumn{1}{c|}{}&\multicolumn{1}{c|}{}&\multicolumn{1}{c|}{}&\multicolumn{1}{c|}{}&\multicolumn{1}{c|}{}& \\ \hline
		Estimated $a_{bn}$ with & \multicolumn{1}{c|}{\multirow{2}{*}{$0.0052$}} & \multicolumn{1}{c|}{\multirow{2}{*}{$0.0077$}} & \multicolumn{1}{c|}{\multirow{2}{*}{$0.0025$}}  & \multicolumn{1}{c|}{\multirow{2}{*}{$0.0137$}} & \multicolumn{1}{c|}{\multirow{2}{*}{$0.0000$}} & \multicolumn{1}{c|}{\multirow{2}{*}{$0.9036$}} &\multicolumn{1}{c|}{\multirow{2}{*}{$0.9165$}}&\multicolumn{1}{c|}{\multirow{2}{*}{$1.0000$}}&\multicolumn{1}{c|}{\multirow{2}{*}{$0.9877$}}& \multicolumn{1}{c|}{\multirow{2}{*}{$1.0000$}} \\ 
		sequence $2\,\s_{bn}$&\multicolumn{1}{c|}{}&\multicolumn{1}{c|}{}&\multicolumn{1}{c|}{}&\multicolumn{1}{c|}{}&\multicolumn{1}{c|}{}&\multicolumn{1}{c|}{}&\multicolumn{1}{c|}{}&\multicolumn{1}{c|}{}&\multicolumn{1}{c|}{}& \\ \hline
	\end{tabular}}
\end{table*}

To understand why this is the case, in Table~\ref{table:norm-sequence}, we demonstrate the impact of the signature sequences' norm on the device activity detection performance.
We assign all devices Type~\ref{item:sphere} signature sequences.
For each column of Table~\ref{table:norm-sequence}, we first generate a realization of the channel and the noise and fix them.
Then, we randomly select a device $(b,n)$ and rescale its signature sequence $\s_{bn}.$
We record the estimated activity indicator $a_{bn}$ by solving problem~\eqref{eq:mle} both before and after the rescaling process.
In Table~\ref{table:norm-sequence}, we present the results for inactive and active devices of $5$ different realizations. We observe that if the device is inactive, the estimated $a_{bn}$ becomes four times as large if the norm of the signature sequence is halved, and vice versa.
Therefore, for an inactive device whose sequence has a significantly smaller norm than other sequences, the device is more likely to be identified as active and cause a false alarm.
For active devices, we also observe that rescaling the signature sequence (slightly) affects the estimated $a_{bn}.$ This result explains why sequences without the normalization, such as Type~\ref{item:Gaussian}, are more likely to cause detection errors.

\begin{figure}[t]
	\centering
	\includegraphics[width=0.9\columnwidth,clip]{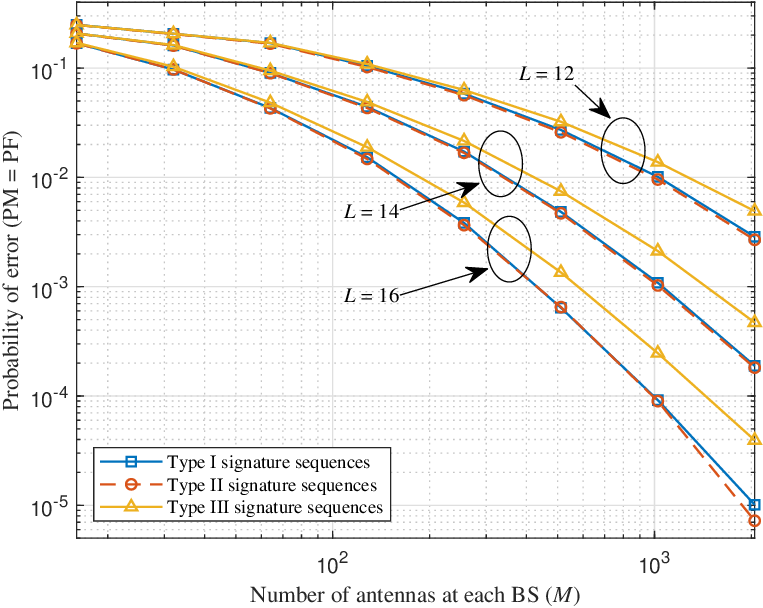}
	\caption{Detection performance comparison of the covariance-based activity detection with three different types of signature sequences for different $M$'s.}
	\label{fig:sequence-PE-M}
\end{figure}

\begin{figure}[t]
	\centering
	\includegraphics[width=0.9\columnwidth,clip]{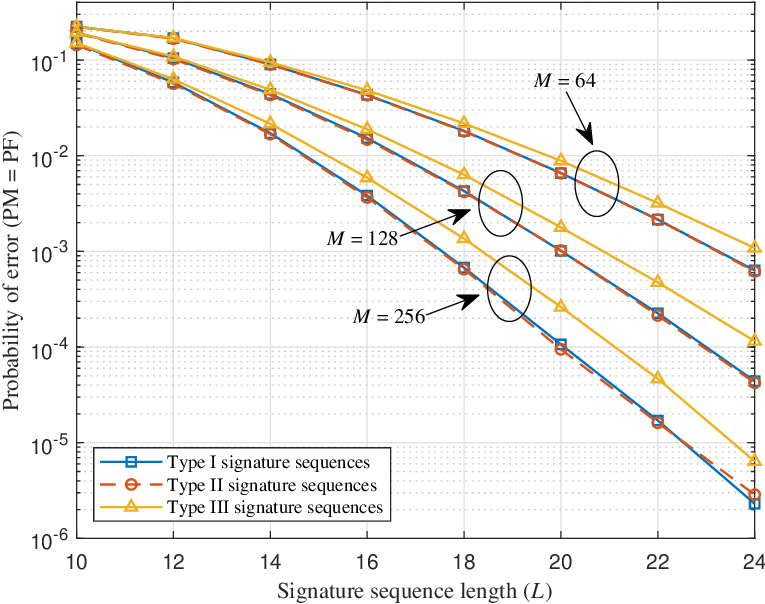}
	\caption{Detection performance comparison of the covariance-based activity detection with three different types of signature sequences for different $L$'s.}
	\label{fig:sequence-PE-L}
\end{figure}

To conveniently illustrate how the estimation error varies with increasing $M$ and $L,$ we select the threshold such that PM and PF are equal and denote the error at this point as ``probability of error''.
Fig.~\ref{fig:sequence-PE-M} and Fig.~\ref{fig:sequence-PE-L} plot the probability of error for different values of $M$ and $L.$
We observe that increasing $M$ or $L$ can improve the detection performance when using the three types of signature sequences.
Furthermore, we can see from Fig.~\ref{fig:sequence-PE-M} and Fig.~\ref{fig:sequence-PE-L} that the detection performance of Type~\ref{item:qam} and Type~\ref{item:sphere} signature sequences is almost the same for different values of $M$ and $L.$ 
Additionally, their detection performance is increasingly better than that of Type~\ref{item:Gaussian} signature sequences as $M$ or $L$ increases.
It is interesting to note that while Fig.~\ref{fig:scaling-law-both} shows slightly different scaling law curves for Type~\ref{item:qam} and Type~\ref{item:sphere} signature sequences, Fig.~\ref{fig:sequence-PE-M} and Fig.~\ref{fig:sequence-PE-L} demonstrate that when the system parameters are set such that the consistency of the MLE in Lemma~\ref{lemma:consistency} holds, the detection performance of the two types of sequences is the same.

\subsection{Computational Efficiency of the Proposed Algorithms}
\label{subsec:algorithms}

In this subsection, we showcase the efficiency of the proposed accelerated CD algorithms for solving the device activity detection problem~\eqref{eq:mle}.
The signature sequences used are Type~\ref{item:qam}.

The algorithms use the following parameters:
the error tolerance is set to $\epsilon = 10^{-3}.$
For the proposed Algorithm~\ref{alg:icd}, we set $\beta = 2$ and the choice of parameter $\mu^{(0)}$ is discussed in Section~\ref{subsec:icd}.
In Algorithm~\ref{alg:active-set}, we set $\omega^{(k)}$ as
\begin{equation}
	\omega^{(k)} = \max\left\{ 5^{-k-1}\big\| \mathbb{V}(\mathbf{a}^{(k)}) \big\|_{\infty}, \, \epsilon \right\}.
\end{equation}

We consider the following two algorithms as benchmarks:
\begin{itemize}
	\item \textit{Vanilla CD~\cite{chen2021sparse}:} At each iteration, this algorithm randomly permutes the indices of all coordinates and then applies the root-finding algorithm~\cite{mcnamee2007numerical} to solve subproblem~\eqref{eq:one-dim} in order to update the coordinates, as shown in Algorithm~\ref{alg:cd}.
	\item \textit{Clustering-based CD~\cite{ganesan2021clustering}:} 
	This algorithm accelerates the vanilla CD algorithm by approximately solving subproblem~\eqref{eq:one-dim}. In order to more efficiently update a variable $a_{bn}$ in cell $b,$ instead of considering all $B$ cells in \eqref{eq:one-dim}, it only considers a cluster of cells that are close to cell $b$ and neglects those that are far away from cell $b.$ In this case, the degree of the polynomial function associated with the derivative of the objective function of \eqref{eq:one-dim} would be much smaller, which improves the efficiency of solving the corresponding subproblem.
	In the simulation, the number of clusters $T$ is chosen to be $1,2,3$ as in \cite{ganesan2021clustering}.
\end{itemize}
To demonstrate the efficiency of the two proposed acceleration strategies, i.e., the inexact and active set strategies, we consider the following three algorithms:
\begin{itemize}
	\item \textit{Inexact CD:} 
	This algorithm uses the inexact Algorithm~\ref{alg:icd} to solve subproblem~\eqref{eq:one-dim} in the vanilla CD algorithm, as shown in Algorithm~\ref{alg:cd}.
	\item \textit{Active set CD:} 
	This algorithm reduces the number of coordinate updates in the vanilla CD algorithm by using the active set selection strategy, as shown in Algorithm~\ref{alg:active-set}.
	\item \textit{Active set inexact CD:} 
	This algorithm is a combination of the above two algorithms. Specifically, it uses the active set selection strategy in Algorithm~\ref{alg:active-set} to select coordinates and updates each coordinate by the inexact Algorithm~\ref{alg:icd}.
\end{itemize}

\begin{figure*}[t]
	\centering
	\begin{subfigure}[t]{0.32\textwidth}
		\includegraphics[width=\textwidth]{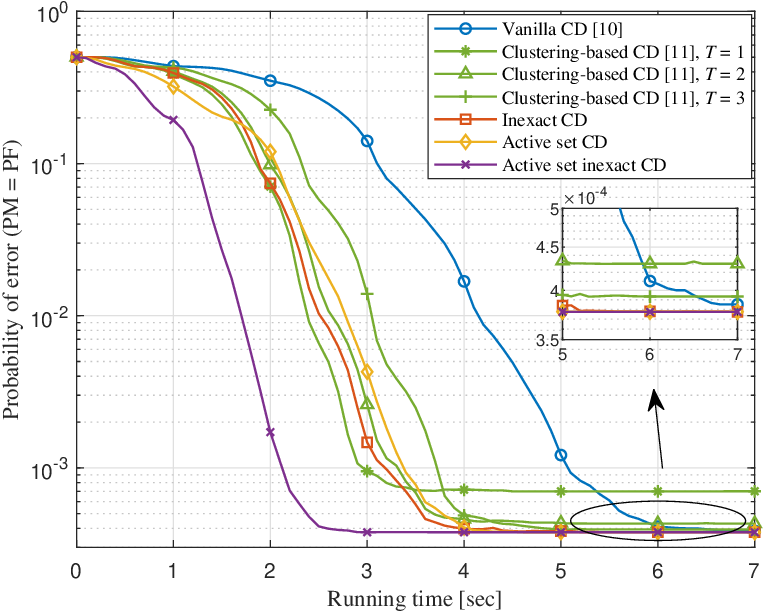}
		\caption{Homogeneous scenario: in each cell $N = 1000$ and $K = 50.$}
	\end{subfigure}
	\hfill
	\begin{subfigure}[t]{0.32\textwidth}
		\includegraphics[width=\textwidth]{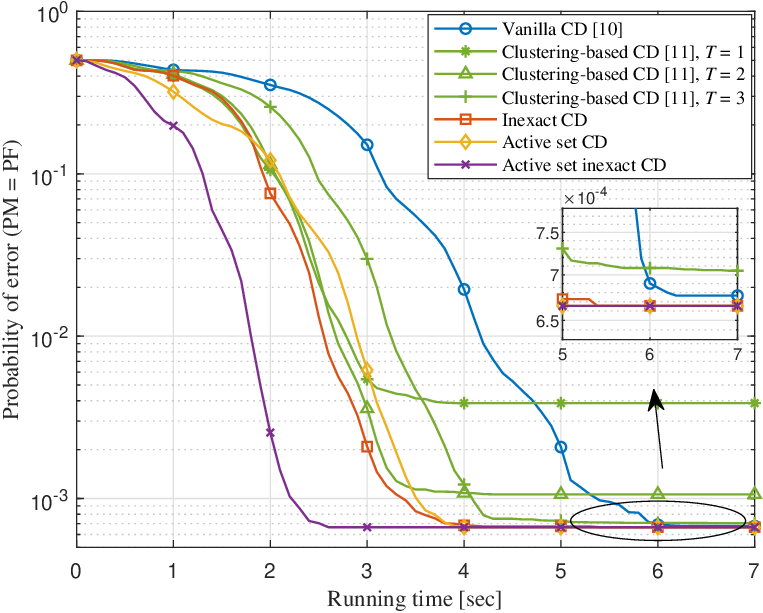}
		\caption{Heterogeneous scenario: in the central cell $N = 1600$ and $K = 80,$ and in other edge cells $N = 900$ and $K = 45.$}
	\end{subfigure}
	\hfill
	\begin{subfigure}[t]{0.32\textwidth}
		\includegraphics[width=\textwidth]{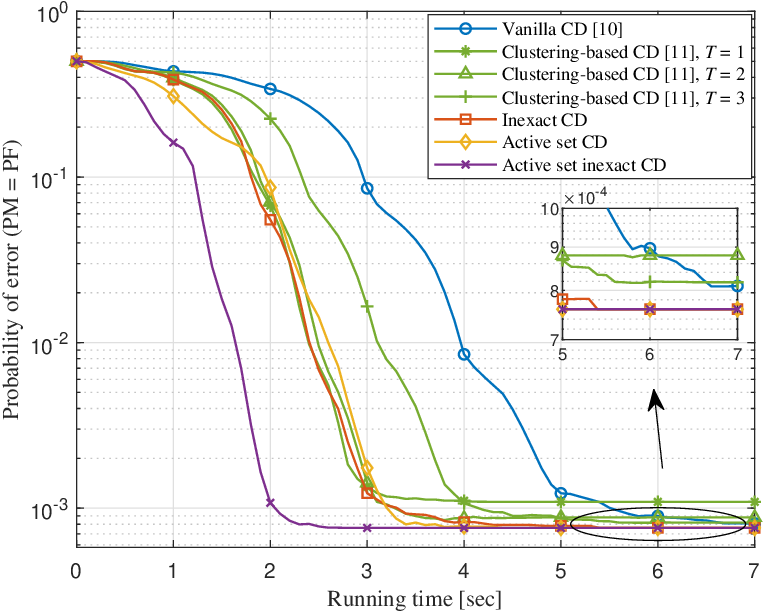}
		\caption{Heterogeneous scenario: in the central cell $N = 400$ and $K = 20,$ and in other edge cells $N = 1100$ and $K = 55.$}
	\end{subfigure}
	\caption{Comparison of the probability of error of the proposed algorithms and the benchmark algorithms versus the running time in both homogeneous and heterogeneous scenarios involving $7$ hexagonal cells.}
	\label{fig:alg-running-7hexagonal}
\end{figure*}

Notice that the compared algorithms are not limited to the homogeneous scenario, where the total number of devices $N$ and the number of active devices $K$ are the same among different cells.
They can also be applied to the heterogeneous scenario, where $N$ and $K$ are different among different cells.
Fig.~\ref{fig:alg-running-7hexagonal} plots the decrease in the probability of error with respect to the running time of the algorithms in both homogeneous and heterogeneous scenarios involving $B = 7$ hexagonal cells with $L = 50$ symbols and $M = 128$ antennas.
The results are obtained by averaging over $500$ Monte-Carlo runs.
For each realization, we record the variable $\mathbf{a}$ and calculate the corresponding probability of error when the algorithms run to a fixed moment.
As shown in Fig.~\ref{fig:alg-running-7hexagonal}, in both homogeneous and heterogeneous scenarios, the clustering-based CD algorithm~\cite{ganesan2021clustering} and the proposed algorithms are significantly more efficient than vanilla CD~\cite{chen2021sparse}, with active set inexact CD being the most efficient among all.
We also observe that all the algorithms achieve the same low probability of error if their running time is allowed to be sufficiently long, except for the clustering-based CD algorithm, whose probability of error is worse than that of the other algorithms due to the coarse approximation.
In particular, when $T = 1,$ the clustering-based CD algorithm demonstrates high efficiency but has the worst probability of error among all compared algorithms. As $T$ increases, the probability of error of the clustering-based CD algorithm improves, but its computational efficiency deteriorates.
The simulation results show that our proposed active set inexact CD has the best efficiency and detection performance among the compared algorithms in both homogeneous and heterogeneous scenarios.

\begin{figure*}[t]
	\centering
	\begin{subfigure}[t]{0.32\textwidth}
		\includegraphics[width=\textwidth]{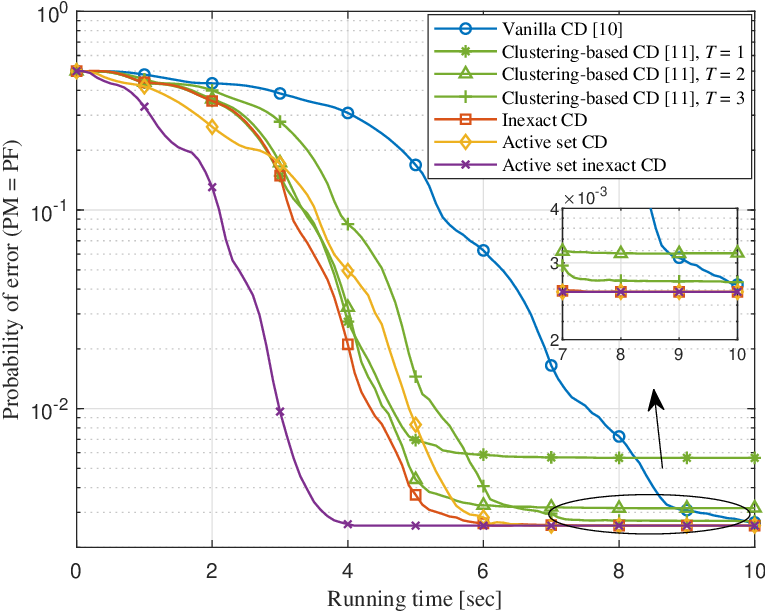}
		\caption{Homogeneous scenario: in each cell $N = 1000$ and $K = 50.$}
		\label{fig:alg-running-9square-a}
	\end{subfigure}
	\hfill
	\begin{subfigure}[t]{0.32\textwidth}
		\includegraphics[width=\textwidth]{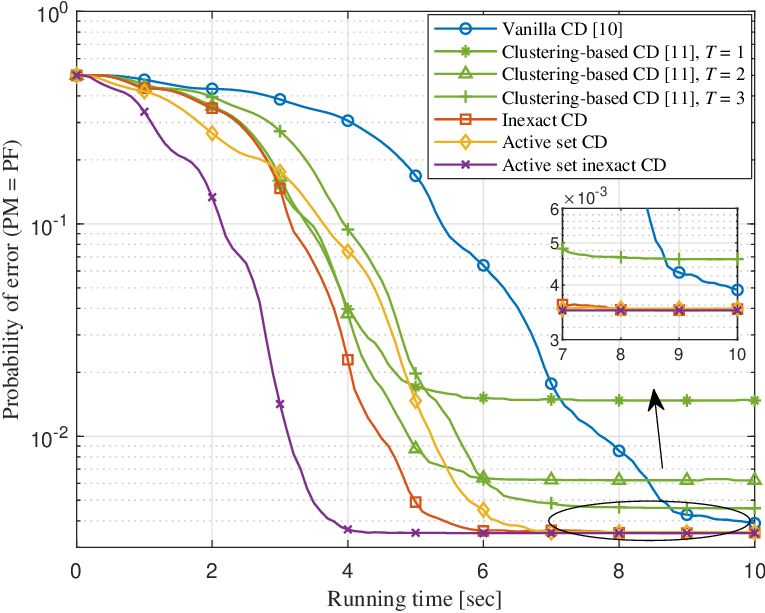}
		\caption{Heterogeneous scenario: in the central cell $N = 1800$ and $K = 90,$ and in other edge cells $N = 900$ and $K = 45.$}
		\label{fig:alg-running-9square-b}
	\end{subfigure}
	\hfill
	\begin{subfigure}[t]{0.32\textwidth}
		\includegraphics[width=\textwidth]{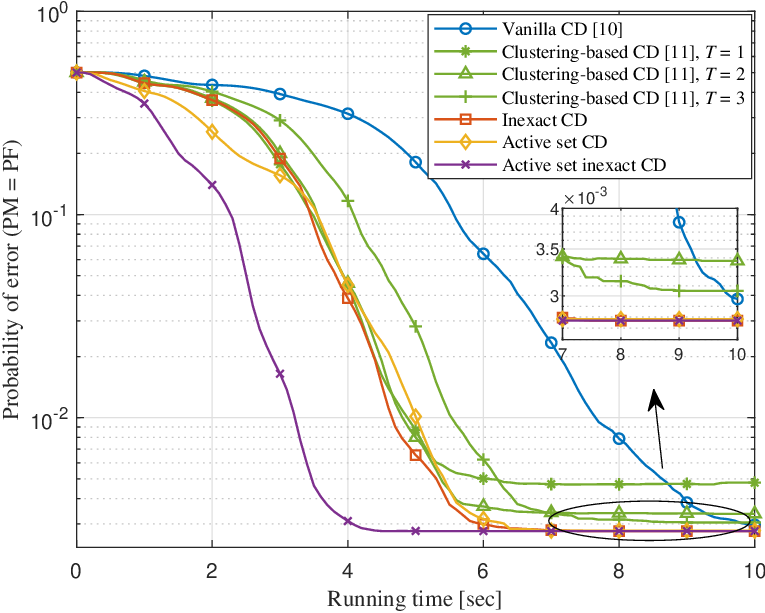}
		\caption{Heterogeneous scenario: in the central cell $N = 200$ and $K = 10,$ and in other edge cells $N = 1100$ and $K = 55.$}
	\end{subfigure}
	\caption{Comparison of the probability of error of the proposed algorithms and the benchmark algorithms versus the running time in both homogeneous and heterogeneous scenarios involving $9$ square cells.}
	\label{fig:alg-running-9square}
\end{figure*}

Fig.~\ref{fig:alg-running-9square} plots the decrease in the probability of error with respect to the running time of the algorithms in both homogeneous and heterogeneous scenarios involving $B = 9$ square cells arranged in a $3 \times 3$ grid with $L = 50$ symbols and $M = 128$ antennas.
As shown in Fig.~\ref{fig:alg-running-9square}, in both homogeneous and heterogeneous scenarios, the proposed algorithms (inexact CD, active set CD, and active set inexact CD) are significantly more efficient than vanilla CD~\cite{chen2021sparse} and achieve a lower probability of error than clustering-based CD~\cite{ganesan2021clustering}. In particular, the proposed active set inexact CD algorithm is the most efficient among all compared algorithms.
Moreover, by comparing Fig.~\ref{fig:alg-running-9square-a} and Fig.~\ref{fig:alg-running-9square-b}, we observe that, as the number of devices in the central cell increases and the number of devices in the edge cells decreases, the gap in the error probability between the clustering-based CD algorithm with $T = 1,2,3$ and the proposed algorithm (e.g., active set inexact CD) widens.
This observation shows that in heterogeneous scenarios, detecting devices solely through a cluster of cells may result in a more significant detection performance loss compared to homogeneous scenarios.

\begin{figure}[t]
	\centering
	\includegraphics[width=0.9\columnwidth,clip]{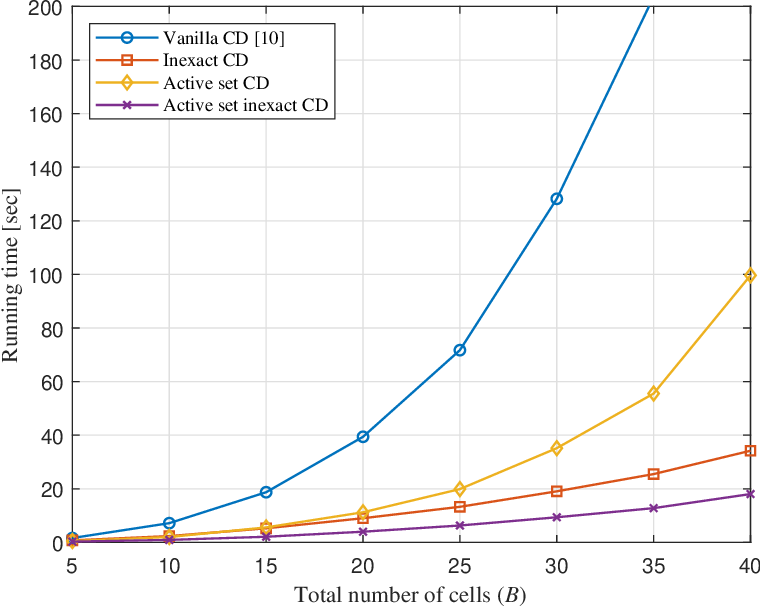}
	\caption{Average running time comparison of the proposed algorithms and the vanilla CD algorithm~\cite{chen2021sparse} for different $B$'s.}
	\label{fig:alg-time-B}
\end{figure}

To demonstrate the efficiency of the proposed inexact Algorithm~\ref{alg:icd} (i.e., the inexact acceleration strategy), Fig.~\ref{fig:alg-time-B} plots the running time comparison of the proposed algorithms with the vanilla CD algorithm~\cite{chen2021sparse} versus the total number of cells (i.e., $B$).
The simulation is conducted in the homogeneous scenario involving hexagonal cells with $N = 200$ devices, $K = 20$ active devices, $L = 20$ symbols, and $M = 128$ antennas.
As shown in Fig.~\ref{fig:alg-time-B}, when $B$ increases, the running time of the vanilla CD algorithm increases drastically, and the algorithms that solve subproblem~\eqref{eq:one-dim} inexactly (inexact CD, and active set inexact CD) are much more efficient than the other algorithms.
This observation highlights that solving subproblem~\eqref{eq:one-dim} exactly can be computationally expensive, especially when $B$ is large, and the proposed inexact acceleration strategy (i.e., Algorithm~\ref{alg:icd}) can significantly reduce the complexity of vanilla CD by solving subproblem~\eqref{eq:one-dim} inexactly in a controllable fashion.

\begin{figure}[t]
	\centering
	\includegraphics[width=0.9\columnwidth,clip]{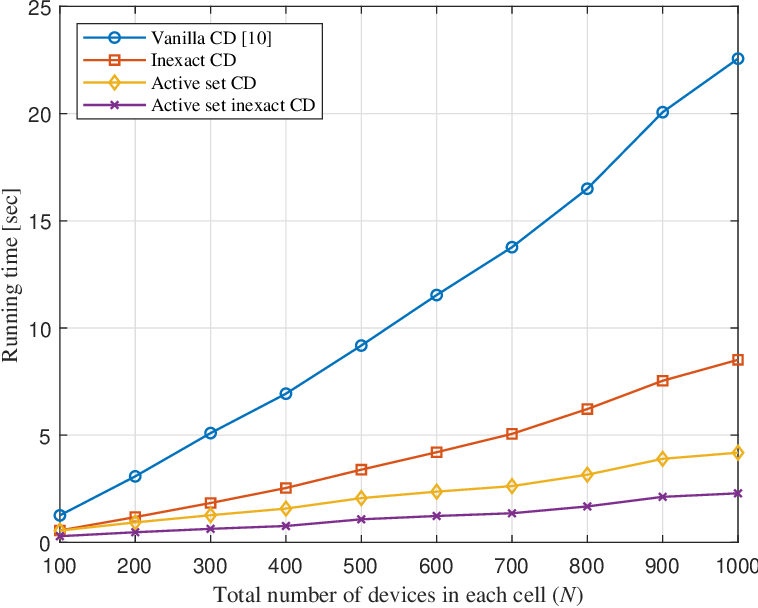}
	\caption{Average running time comparison of the proposed algorithms and the vanilla CD algorithm~\cite{chen2021sparse} for different $N$'s.}
	\label{fig:alg-time-N}
\end{figure}

To demonstrate the efficiency of the proposed active set CD algorithm~\ref{alg:active-set} (i.e., active set selection strategy), Fig.~\ref{fig:alg-time-N} plots the running time of the proposed algorithms with the vanilla CD algorithm~\cite{chen2021sparse} versus the total number of devices in each cell (i.e., $N$).
The simulation is conducted in the homogeneous scenario involving $B = 7$ hexagonal cells with $K = 20$ active devices, $L = 20$ symbols, and $M = 128$ antennas.
As shown in Fig.~\ref{fig:alg-time-N}, for larger $N,$ the algorithms that use the active set selection strategy (active set CD and active set inexact CD) are significantly more efficient than the other algorithms.
It is worth mentioning that as $N$ increases, the dimension of problem~\eqref{eq:mle} becomes larger and the proportion of the components of the solution are zero becomes larger.
This observation illustrates that the active set selection strategy can significantly accelerate the vanilla CD algorithm by exploiting the special structure of the solution.

\begin{figure}[t]
	\centering
	\includegraphics[width=0.9\columnwidth,clip]{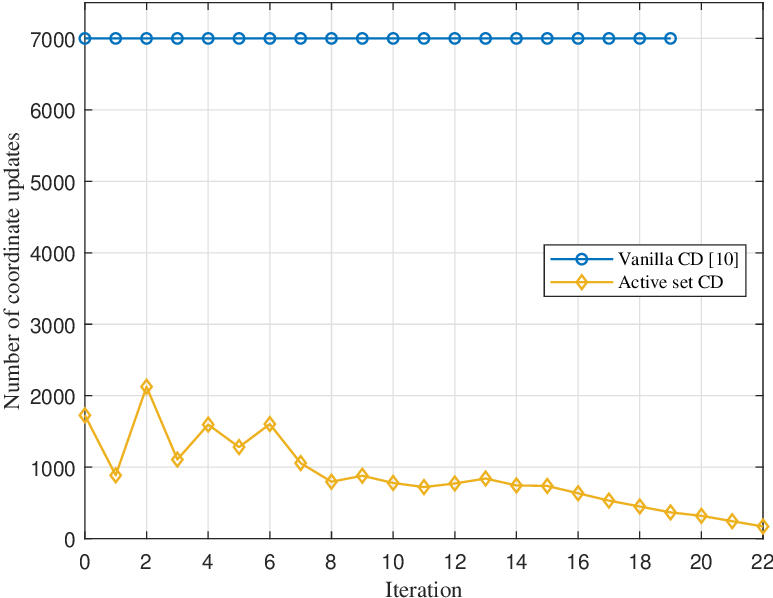}
	\caption{Comparison of the number of coordinate updates of the vanilla CD algorithm~\cite{chen2021sparse} and the active set CD algorithm over the iteration.}
	\label{fig:alg-iter}
\end{figure}

To provide a detailed performance of the active set selection strategy, we run vanilla CD~\cite{chen2021sparse} and active set CD once and plot the number of coordinate updates at each iteration in Fig.~\ref{fig:alg-iter}.
The simulation is conducted in the homogeneous scenario involving $B = 7$ hexagonal cells with $N = 1000$ devices, $K = 50$ active devices, $L = 50$ symbols, and $M = 128$ antennas.
As shown in Fig.~\ref{fig:alg-iter}, the number of coordinate updates of active set CD is significantly less than that of vanilla CD at each iteration, and the overall number of iterations of active set CD is slightly more than that of vanilla CD.
Therefore, the total number of coordinate updates in active set CD is significantly less than that in vanilla CD, which explains why active set CD is more computationally efficient.

A summary of the numerical comparisons is as follows.
First, the vanilla CD algorithm~\cite{chen2021sparse} achieves good error probability but has poor computational efficiency, especially when $B$ or $N$ is large.
Second, the clustering-based CD algorithm~\cite{ganesan2021clustering} improves the computational efficiency of vanilla CD at the cost of its error probability.
Third, the proposed two accelerated CD algorithms (inexact CD and active set CD) are significantly more efficient than vanilla CD and achieve the same low error probability as vanilla CD.
Due to the different features of these acceleration algorithms, inexact CD is more efficient when $B$ is large, while active set CD performs better when $N$ is large.
Finally, the above two acceleration strategies can be combined, and active set inexact CD performs the best in terms of computational efficiency and error probability.

\section{Conclusion}
\label{sec:conclusion}

This paper addresses the covariance-based activity detection problem in a cooperative multi-cell massive MIMO system.
The device activity detection problem is formulated as the MLE problem, and a necessary and sufficient condition for the consistency of the MLE is provided in the literature.
By analyzing the statistical properties of the signature sequences generated from the finite alphabet (i.e., Type~\ref{item:qam} signature sequences) and making a reasonable assumption regarding the large-scale fading coefficients, the quadratic scaling law for the MLE in the multi-cell scenario is derived for the first time.
This result also provides theoretical guarantees for the use of the more practical Type~\ref{item:qam} signature sequences.
Moreover, this paper characterizes the distribution of the estimation error in the multi-cell scenario.
Finally, two efficient accelerated CD algorithms with convergence and iteration complexity guarantees are proposed.
Simulation results demonstrate that the proposed algorithms significantly outperform the vanilla CD algorithm and its accelerated version in the literature.

We conclude this paper with a short discussion on open problems.
First, this paper assumes that the large-scale fading coefficients are known. It would be interesting to formulate the corresponding detection problem in the case of unknown large-scale fading coefficients and design algorithms for solving the problem.
Some recent progress along this direction has been made in \cite{zhang2024activity}.
Second, the scaling law analysis in this paper is based on the assumption that we can globally solve the MLE problem~\eqref{eq:mle}. However, most practical algorithms can only guarantee local optimality. It would be desirable to analyze the detection performance of specific algorithms for solving the MLE problem~\eqref{eq:mle}.

\appendices

\section{Proof of Proposition~\ref{prop:NC-equivalent}} \label{sec:NC-equivalent}

To prove the proposition, we need to show that $\mathcal{N}^{\prime}\cap\mathcal{C}^{\prime}\neq\{\mathbf{0}\}$ if and only if $\mathcal{N}^{\prime\prime}\cap\mathcal{C}^{\prime}\neq\{\mathbf{0}\}.$
Since $\mathbf{G}$ is a diagonal matrix with positive diagonal elements, if there exists a non-zero vector $\mathbf{x} \in \mathcal{N}^{\prime}\cap\mathcal{C}^{\prime},$ then we can find another non-zero vector $\hat{\mathbf{x}} = \mathbf{G}\mathbf{x} \in \mathcal{N}^{\prime\prime}\cap\mathcal{C}^{\prime},$ and vice versa.
This completes the proof of Proposition~\ref{prop:NC-equivalent}.

\section{Proof of Proposition~\ref{prop:scaling-law-single}} \label{sec:scaling-law-single}

The scaling law for Type~\ref{item:sphere} signature sequences has been proved in \cite[Theorem~9]{chen2022phase}.
The proof in \cite{chen2022phase} depends only on the stable NSP of $\widetilde{\mathbf{S}}$ and the normalization of the signature sequences, i.e., $\| \mathbf{s}_n \|_2^2 = L.$
For Type~\ref{item:qam} signature sequences, it is simple to verify that the same normalization property holds, and the stable NSP of $\widetilde{\mathbf{S}}$ is established in Theorem~\ref{theorem:nsp}. Therefore, the same techniques used in \cite{chen2022phase} can be applied to prove the scaling law result for this case as well.
This completes the proof of Proposition~\ref{prop:scaling-law-single}.

\section{Proof of Theorem~\ref{theorem:scaling-law}} \label{sec:proof-scaling-law}

Theorem~\ref{theorem:scaling-law} can be proved by contradiction. The proof can be outlined as follows: first, we use the stable NSP of the matrix $\widetilde{\mathbf{S}}$ (in Theorem~\ref{theorem:nsp}) to derive some useful inequalities. Then, we carefully examine these derived inequalities by exploiting the properties of sets in \eqref{eq:subspace} and \eqref{eq:cone}. Finally, we use the property of the path-loss model (in Lemma~\ref{lemma:gamma}) to derive the contradiction. 
It is worth mentioning that the proof is applicable to both types of signature sequences in Assumption~\ref{assu:sequence}.

We now prove Theorem~\ref{theorem:scaling-law} by contradiction.
Let us assume that there exists a non-zero vector $\mathbf{x} \in \mathcal{N}\cap\mathcal{C}.$
Using Theorem~\ref{theorem:nsp} repeatedly, we can obtain $B$ inequalities.
More specifically, we define $\mathcal{K}_b$ as the index set of $\mathbf{a}^{\circ}$ corresponding to all active devices in cell $b,$ i.e., $|\mathcal{K}_b| = K$ and
\begin{equation}
	\mathcal{K}_b = \{ i \mid a_i^{\circ}=1,\, (b-1)N+1 \le i \le bN \}, \, 1 \le b \le B.
\end{equation}
From \eqref{eq:gamma}, we can see that
\begin{equation}\label{eq:large-D}
	\min_{1 \le n \le N} g_{bbn} \ge D, \quad   1 \le b \le B,
\end{equation}
where $D\triangleq P_0\left(\frac{D_0}{R}\right)^{\gamma}.$
We set
\begin{equation}
	\bar{\rho} = \frac{D}{D+2C}
\end{equation}
in Theorem~\ref{theorem:nsp}, where $C$ is defined in \eqref{eq:less-C}.
Recall the definition of $\mathcal{N}$ in \eqref{eq:subspace}. For each cell $b,$ we set the index set $\mathcal{S} = \mathcal{K}_b$ and $\mathbf{v} = \mathbf{G}_b\mathbf{x}$ in \eqref{eq:nsp}. This gives us $B$ inequalities:
\begin{equation}\label{eq:nsp-main}
	\big\|\left[ \mathbf{G}_b\mathbf{x} \right]_{\mathcal{K}_b}\big\|_1 \leq \rho\big\|\left[ \mathbf{G}_b\mathbf{x} \right]_{\mathcal{K}_b^c}\big\|_1, \, 1 \le b \le B.
\end{equation}

We can derive an equivalent form of \eqref{eq:nsp-main} by studying its right-hand side.
Notice that $\mathcal{K}_b^c = \big( \cup_{j=1,\,j\neq b}^B \mathcal{K}_j \big) \cup \mathcal{I},$ where $\mathcal{I}$ is defined in \eqref{eq:def-I}. Then, we get
\begin{equation}\label{eq:l1-eq-com-sum}
	\big\|\left[ \mathbf{G}_b\mathbf{x} \right]_{\mathcal{K}_b^c}\big\|_1 = \big\|\left[ \mathbf{G}_b\mathbf{x}\right]_{\mathcal{I}}\big\|_1 + \sum_{j=1,\,j\neq b}^{B} \big\|\left[ \mathbf{G}_b\mathbf{x}\right]_{\mathcal{K}_j}\big\|_1.
\end{equation}
Since $\mathbf{x} \in \mathcal{C}$ and $\mathbf{G}_b$ is a positive definite diagonal matrix, it follows that the two sub-vectors of $\mathbf{G}_b\mathbf{x}$ with entries from $\mathcal{I}$ and $\mathcal{I}^c$ are non-negative and non-positive, respectively, i.e.,
$\left[ \mathbf{G}_b\mathbf{x}\right]_{\mathcal{I}} \ge \mathbf{0}$ and $\left[ \mathbf{G}_b\mathbf{x}\right]_{\mathcal{I}^c} \le \mathbf{0}.$
Now let $\mathbf{1} \in \mathbb{R}^{BN}$ denote the all-one vector, then
\begin{equation}\label{eq:minus-l1norm}
	\mathbf{1}^T\mathbf{G}_b\mathbf{x} = \big\|\left[ \mathbf{G}_b\mathbf{x}\right]_{\mathcal{I}}\big\|_1 - \big\|\left[ \mathbf{G}_b\mathbf{x}\right]_{\mathcal{I}^c}\big\|_1.
\end{equation}
Let $\mathbf{u}$ denote the vectorization of the $L \times L$ identity matrix, i.e., $\mathbf{u} = \operatorname{vec}\left(\mathbf{I}\right) \in \mathbb{R}^{L^2}.$ Then, for each column of $\widetilde{\mathbf{S}},$ it holds that
\begin{equation}
	\mathbf{u}^T \left( \mathbf{s}_{bn}^*\otimes \mathbf{s}_{bn} \right) = \operatorname{tr} \big( \mathbf{I} \cdot \mathbf{s}_{bn} \mathbf{s}_{bn}^H \big) = \|\mathbf{s}_{bn}\|_2^2.
\end{equation}
Using $\mathbf{x} \in \mathcal{N}$ and the normalization of the sequence $\|\mathbf{s}_{bn}\|_2^2 = L,$ we get
\begin{equation}\label{eq:sum-zero}
	\mathbf{u}^T \widetilde{\mathbf{S}} \mathbf{G}_b \mathbf{x} = L\mathbf{1}^T\mathbf{G}_b\mathbf{x} = 0.
\end{equation}
Combining \eqref{eq:minus-l1norm} and \eqref{eq:sum-zero}, and noting that $\mathcal{I}^c = \cup_{j=1}^{B} \mathcal{K}_j,$ we obtain
\begin{equation}\label{eq:l1-eq-sum}
	\big\|\left[ \mathbf{G}_b\mathbf{x}\right]_{\mathcal{I}}\big\|_1 = \big\|\left[ \mathbf{G}_b\mathbf{x}\right]_{\mathcal{I}^c}\big\|_1 = \sum_{j=1}^{B} \big\|\left[ \mathbf{G}_b\mathbf{x}\right]_{\mathcal{K}_j}\big\|_1.
\end{equation}
By substituting \eqref{eq:l1-eq-sum} into \eqref{eq:l1-eq-com-sum}, we can equivalently rewrite \eqref{eq:nsp-main} as
\begin{multline}\label{eq:ineq-main}
	\big\|\left[ \mathbf{G}_b\mathbf{x} \right]_{\mathcal{K}_b}\big\|_1 \leq \rho \Big( \big\|\left[ \mathbf{G}_b\mathbf{x} \right]_{\mathcal{K}_b}\big\|_1 +  2\sum_{j=1,\,j\neq b}^{B} \big\|\left[ \mathbf{G}_b\mathbf{x}\right]_{\mathcal{K}_j}\big\|_1 \Big), \\
	1 \le b \le B .
\end{multline}

Finally, we will derive the contradiction by putting all pieces together. Since $\rho<\bar{\rho} <1,$ it follows from \eqref{eq:ineq-main} that
\begin{equation}\label{eq:reduced-main}
	\big\|\left[ \mathbf{G}_b\mathbf{x} \right]_{\mathcal{K}_b}\big\|_1 \le \frac{2\rho}{1-\rho}\sum_{j=1,\,j\neq b}^{B} \big\|\left[ \mathbf{G}_b\mathbf{x}\right]_{\mathcal{K}_j}\big\|_1, \, 1 \le b \le B.
\end{equation}
Since $\mathbf{G}_b = \operatorname{diag}(\mathbf{G}_{b1},\mathbf{G}_{b2},\ldots,\mathbf{G}_{bB})$ is a diagonal matrix, its diagonal elements with entries from $\mathcal{K}_b$ are larger than $\min_{n} g_{bbn},$ and those with entries from $\mathcal{K}_j$ are smaller than $\max_{n} g_{bjn},$ which further implies
\begin{align}
	\big\|\left[ \mathbf{G}_b\mathbf{x} \right]_{\mathcal{K}_b}\big\|_1 & \ge \big(\min_{n} g_{bbn}\big) \big\|\mathbf{x}_{\mathcal{K}_b}\big\|_1 \ge D \big\|\mathbf{x}_{\mathcal{K}_b}\big\|_1, \label{eq:l1-lower} \\
	\big\|\left[ \mathbf{G}_b\mathbf{x}\right]_{\mathcal{K}_j}\big\|_1 & \le \big(\max_{n} g_{bjn}\big) \big\|\mathbf{x}_{\mathcal{K}_j}\big\|_1, \, \forall \, j \neq b, \label{eq:l1-upper}
\end{align}
where the last inequality in \eqref{eq:l1-lower} is from \eqref{eq:large-D}.
Substituting \eqref{eq:l1-lower} and \eqref{eq:l1-upper} into \eqref{eq:reduced-main}, we get
\begin{equation}\label{eq:upper-lower-main}
	D \big\|\mathbf{x}_{\mathcal{K}_b}\big\|_1 \le \frac{2\rho}{1-\rho}  \sum_{j=1,\,j\neq b}^{B} \big(\max_{n} g_{bjn}\big) \big\|\mathbf{x}_{\mathcal{K}_j}\big\|_1, \, 1 \le b \le B. 
\end{equation}
In particular, consider $\bar{b} = \operatorname{argmax}_{b}\big\|\mathbf{x}_{\mathcal{K}_b}\big\|_1.$
Then, we get the following contradiction:
\begin{align}\label{eq:contradictory-main}
	D \big\|\mathbf{x}_{\mathcal{K}_{\bar{b}}}\big\|_1 & \le \frac{2\rho}{1-\rho} \sum_{j=1,\,j\neq \bar{b}}^{B} (\max_{n} g_{\bar{b}jn}) \big\|\mathbf{x}_{\mathcal{K}_j}\big\|_1 \nonumber \\
	& \le \frac{2\rho}{1-\rho} \sum_{j=1,\,j\neq \bar{b}}^{B} (\max_{n} g_{\bar{b}jn}) \big\|\mathbf{x}_{\mathcal{K}_{\bar{b}}}\big\|_1 \nonumber \\
	& \overset{(a)}{\le} \frac{2\rho}{1-\rho} C \big\|\mathbf{x}_{\mathcal{K}_{\bar{b}}}\big\|_1 \overset{(b)}{<} D \big\|\mathbf{x}_{\mathcal{K}_{\bar{b}}}\big\|_1,
\end{align}
where $(a)$ follows from \eqref{eq:less-C} and $(b)$ results from the fact that $\rho < \bar{\rho} = \frac{D}{D+2C}.$ 
The contradiction in \eqref{eq:contradictory-main} implies our assumption that there exists a non-zero vector $\mathbf{x} \in \mathcal{N}\cap\mathcal{C}$ must be false, hence $\mathcal{N}\cap\mathcal{C} = \{ \mathbf{0} \}.$ This completes the proof of Theorem~\ref{theorem:scaling-law}.

\section{Proof of Theorem~\ref{theorem:error}} \label{sec:proof-error}

The proof follows a similar approach to that of \cite[Theorem~4]{chen2022phase}, with the main difference being the dimensions and construction of the cone $\mathcal{C}.$
Specifically, the constraints $a_{bn} \in [0,1]$ for all $b$ and $n$ in \eqref{eq:mle-constraint} ensure that for small positive $t$ and any $\mathbf{x}^{\prime} \in \mathcal{C},$ $\mathbf{a}^{\circ} + t\, \mathbf{x}^{\prime}$ is still feasible, i.e., $\mathbf{a}^{\circ} + t\, \mathbf{x}^{\prime} \in [0,1]^{BN}.$
This completes the proof of Theorem~\ref{theorem:error}.

\section{Proof of Proposition~\ref{prop:icd}} \label{sec:proof-icd}

The proof relies on the fact that every function $f_j(\mathbf{a})$ for $1 \le j \le B$ has a quadratic upper bound. The outline of the proof is as follows. First, we utilize the quadratic upper bounds on $f_j(\mathbf{a})$ for all $j \neq b$ to demonstrate that Algorithm~\ref{alg:icd} will terminate at a finite $i.$ Then, we employ the quadratic upper bound on $f_b(\mathbf{a})$ to establish the sufficient decrease of the whole objective function with the output $\bar{d}$ of Algorithm~\ref{alg:icd}.
Finally, we combine the properties of the quadratic upper bound to derive the desired inequality~\eqref{eq:sufficiently-decrease}.

First, we prove that Algorithm~\ref{alg:icd} will terminate at a finite $i.$
Since the gradient of $f_j(\mathbf{a})$ is Lipschitz continuous~\cite{shao2020cooperative}, we denote $\lip_j > 0$ as the Lipschitz constant of $\nabla f_j(\mathbf{a}),$ and denote $\lip = \sum_{j=1}^{B} \lip_j$ as the Lipschitz constant of $\nabla F(\mathbf{a}).$
Then, we have a quadratic upper bound on $f_j(\mathbf{a})$ \cite[Proposition~A.24]{bertsekas1999nonlinear} i.e., for any $\mathbf{a}, \mathbf{a}^{\prime} \in [0,1]^{BN},$ the following inequalities hold:
\begin{equation}\label{eq:quadratic-upper-bound}
	f_j(\mathbf{a}^{\prime}) \le f_j(\mathbf{a}) + \nabla f_j(\mathbf{a})^T(\mathbf{a}^{\prime}-\mathbf{a}) + \frac{\lip_j}{2} \| \mathbf{a}^{\prime} - \mathbf{a} \|_2^2, \, 1 \le j \le B.
\end{equation}
Setting $\mathbf{a}^{\prime} = \mathbf{a} + d\, \mathbf{e}_{bn}$ in \eqref{eq:quadratic-upper-bound}, we obtain
\begin{equation}\label{eq:quadratic-upper-bound-one-dim}
	f_j\left(\mathbf{a} + d\,\mathbf{e}_{bn}\right) \le f_j(\mathbf{a}) + \left[ \nabla f_j(\mathbf{a})\right]_{bn}d + \frac{\lip_j}{2} d^2, \, 1 \le j \le B.
\end{equation}
Summing all inequalities in \eqref{eq:quadratic-upper-bound-one-dim} for all $j \neq b,$ we have
\begin{multline}\label{eq:upper-bound}
	\sum_{j=1,\,j\neq b}^{B} f_j\left(\mathbf{a} + d\,\mathbf{e}_{bn}\right) \\
	\le \sum_{j=1,\,j\neq b}^{B} \left(f_j(\mathbf{a}) + \left[ \nabla f_j(\mathbf{a})\right]_{bn}d \right) + \frac{ \lip }{2} d^2
\end{multline}
holds for all $d \in [-a_{bn},\,1-a_{bn}].$
We observe from \eqref{eq:upper-bound} that the sufficient decrease condition~\eqref{eq:decrease-condition} always holds true when $\mu^{(i)} = \beta^i \mu^{(0)} \ge \lip.$
Therefore, there exists an integer $i_0 \le \log_{\beta} \big(\lip / \mu^{(0)}\big) +1,$ such that Algorithm~\ref{alg:icd} will terminate at $i = i_0.$

Next, we utilize the quadratic upper bound on $f_b(\mathbf{a})$ to establish an upper bound on the objective function value $F(\mathbf{a}+d^{(i_0)} \mathbf{e}_{bn}).$
Since the sufficient decrease condition~\eqref{eq:decrease-condition} holds at $i = i_0,$ we add $f_b(\mathbf{a}+d^{(i_0)} \mathbf{e}_{bn} )$ to both sides of \eqref{eq:decrease-condition}, use the fact that $F(\mathbf{a})=\sum_{j=1}^{B}f_j(\mathbf{a}),$ and obtain
\begin{multline}\label{eq:Fa-less}
	F(\mathbf{a}+d^{(i_0)} \mathbf{e}_{bn} ) \le f_b(\mathbf{a}+d^{(i_0)} \mathbf{e}_{bn} ) \\
	+ \sum_{j=1,\,j\neq b}^{B} \left(f_j(\mathbf{a}) 
	+ \left[ \nabla f_j(\mathbf{a})\right]_{bn}d^{(i_0)} \right)
	+ \frac{\mu^{(i_0)}}{2} \left(d^{(i_0)}\right)^2.
\end{multline}
Since $d^{(i_0)}$ is the optimal solution to problem~\eqref{eq:icd}, inequality~\eqref{eq:Fa-less} still holds when $d^{(i_0)}$ in its right-hand side is replaced with any $d\in [-a_{bn},\,1-a_{bn}],$ i.e.,
\begin{multline}\label{eq:upper-bound-one-dim}
	F(\mathbf{a}+d^{(i_0)} \mathbf{e}_{bn} ) \le f_b(\mathbf{a}+d\, \mathbf{e}_{bn} ) \\
	+ \sum_{j=1,\,j\neq b}^{B} \left(f_j(\mathbf{a}) + \left[ \nabla f_j(\mathbf{a})\right]_{bn}d \right) + \frac{\mu^{(i_0)}}{2} d^2.
\end{multline}
By substituting \eqref{eq:quadratic-upper-bound-one-dim} for $j = b$ into \eqref{eq:upper-bound-one-dim}, we get the quadratic upper bound:
\begin{align}\label{eq:Fa-upper-bound}
	F(\mathbf{a}+d^{(i_0)} \mathbf{e}_{bn} ) & \le F(\mathbf{a}) + \left[\nabla F(\mathbf{a})\right]_{bn}d +\frac{\lip_b + \mu^{(i_0)}}{2} d^2 \nonumber \\
	& \le F(\mathbf{a}) + \left[\nabla F(\mathbf{a})\right]_{bn}d +\frac{\lip(\beta+1)}{2} d^2,
\end{align}
where the last inequality holds because $\lip_b \le \lip$ and $\mu^{(i_0)} \le \beta \lip.$

Finally, we study the optimal value of the right-hand side of~\eqref{eq:Fa-upper-bound} in order to obtain the desired inequality~\eqref{eq:sufficiently-decrease}.
The minimum of the right-hand side of \eqref{eq:Fa-upper-bound} over the interval $[-a_{bn},\,1-a_{bn}]$ is reached at
\begin{equation}\label{eq:d-prime}
	d^{\prime} = \op{Proj} \left(a_{bn}- \frac{\left[\nabla F(\mathbf{a})\right]_{bn}}{\lip(\beta+1)}\right) - a_{bn}.
\end{equation}
Based on \eqref{eq:d-prime}, it is easy to verify that $\left[\nabla F(\mathbf{a})\right]_{bn}$ has the opposite sign to $d^{\prime},$ and $\lip(\beta+1)|d^{\prime}| \le \left |\left[\nabla F(\mathbf{a})\right]_{bn}\right|$ due to the nonexpansiveness of the projection operator.
Hence, we have 
\begin{equation}\label{eq:gradF-d-prime}
	\left[\nabla F(\mathbf{a})\right]_{bn}d^{\prime} \le - \lip(\beta + 1) (d^{\prime})^2.
\end{equation}
Substituting $d = d^{\prime}$ into \eqref{eq:Fa-upper-bound} and using \eqref{eq:d-prime} and \eqref{eq:gradF-d-prime}, we obtain
\begin{multline}\label{eq:Fa-less-proj}
	F(\mathbf{a}+d^{(i_0)} \mathbf{e}_{bn}) \le F(\mathbf{a})  \\
	- \frac{\lip(\beta+1)}{2} \left( \op{Proj} \left(a_{bn}- \frac{\left[\nabla F(\mathbf{a})\right]_{bn}}{\lip(\beta+1)}\right) - a_{bn} \right)^2.
\end{multline}
Next, we analyze the right-hand side of \eqref{eq:Fa-less-proj}.
We define a function with respect to $t> 0$ as
\begin{equation}\label{eq:chi-t}
	\chi(t) \triangleq t^2 \left(\op{Proj} \left(a_{bn}- \frac{\left[\nabla F(\mathbf{a})\right]_{bn}}{t}\right)- a_{bn} \right)^2,
\end{equation}
which allows us to rewrite \eqref{eq:Fa-less-proj} as
\begin{equation}\label{eq:Fa-less-chi-L}
	F(\mathbf{a}+d^{(i_0)} \mathbf{e}_{bn}) \le F(\mathbf{a}) - \frac{\chi\big(\lip(\beta + 1)\big)}{2 \lip(\beta + 1)}.
\end{equation}
Notice that $\chi(t)$ is an increasing function of $t$ \cite[Lemma~2.3.1]{bertsekas1999nonlinear}.
Without loss of generality, we assume that $\lip(\beta + 1) \ge 1.$ Applying the inequality $\chi\big(\lip(\beta + 1)\big) \ge \chi(1)$ into \eqref{eq:Fa-less-chi-L}, we obtain
\begin{equation}
	F(\mathbf{a}+d^{(i_0)} \mathbf{e}_{bn}) \le F(\mathbf{a}) - \frac{\chi(1)}{2 \lip(\beta+1)},
\end{equation}
where $\chi(1) = \left[ \mathbb{V}(\mathbf{a}) \right]_{bn}^2$ by the definitions of $\mathbb{V}(\mathbf{a})$ and $\chi(t)$ in \eqref{eq:def-va} and \eqref{eq:chi-t}, respectively.
Notice that the output $\bar{d} = d^{(i_0)}.$
This completes the proof of Proposition~\ref{prop:icd}.

\section{Proof of Theorem~\ref{theorem:active-set}} \label{sec:proof-active-set}

We first analyze the convergence of Algorithm~\ref{alg:active-set} when the inexact Algorithm~\ref{alg:icd} is used to update each coordinate in the selected active set. We shall show that Algorithm~\ref{alg:active-set} will terminate within $\mathcal{O}(\lip \beta/\epsilon^{2})$ iterations in this case.  
In particular, we shall first show that the active set $\mathcal{A}^{(k)}$ is non-empty when Algorithm~\ref{alg:active-set} does not terminate at the $k$-th iteration; then we shall derive an upper bound on $k$ after which the algorithm will terminates.  

Assuming that the algorithm does not terminate at the $k$-th iteration, we prove that $\mathcal{A}^{(k)}$ is non-empty by contradiction.
If $\mathcal{A}^{(k)}$ is empty, then \eqref{eq:select-active-set} implies that
\begin{equation}\label{eq:omega-ge-infnorm}
	\omega^{(k)} > \big\| \mathbb{V}(\mathbf{a}^{(k)}) \big\|_{\infty}.
\end{equation}
From \eqref{eq:active-set-parameters}, we can conclude that \eqref{eq:omega-ge-infnorm} holds only if
\begin{equation}\label{eq:termination}
	\big\| \mathbb{V}(\mathbf{a}^{(k)}) \big\|_{\infty} < \epsilon.
\end{equation}
This contradicts the assumption that Algorithm~\ref{alg:active-set} does not terminate at the $k$-th iteration (since the algorithm would have terminated if \eqref{eq:termination} holds). Therefore, $\mathcal{A}^{(k)}$ must be non-empty.

Next, we derive an upper bound on $k$ based on the result in Proposition~\ref{prop:icd}.
Let $T_k \ge 1$ be the number of coordinates in $\mathcal{A}^{(k)},$ and let $\mathbf{a}^{(k,t)}$ denote the value of $\mathbf{a}^{(k)}$ after $t$ coordinate updates, where $0 \le t \le T_k.$
Notice that $\mathbf{a}^{(k,0)} = \mathbf{a}^{(k)}$ and $\mathbf{a}^{(k,T_k)} = \mathbf{a}^{(k+1)}.$
Suppose that $(b,n)$ is the first coordinate to be updated, then using \eqref{eq:sufficiently-decrease} in Proposition~\ref{prop:icd}, we have
\begin{align}\label{eq:decrease-first-update}
	F(\mathbf{a}^{(k,1)}) & \le F(\mathbf{a}^{(k)}) - \frac{\left[\mathbb{V}(\mathbf{a}^{(k)})\right]_{bn}^2}{2 \lip(\beta+1)} \nonumber \\
	& \le F(\mathbf{a}^{(k)}) - \frac{(\omega^{(k)})^2}{2 \lip(\beta+1)} \nonumber \\
	& \le F(\mathbf{a}^{(k)}) - \frac{\epsilon^2}{2 \lip(\beta+1)},
\end{align}
where the second inequality is due to the active set selection strategy, i.e., $\left[\mathbb{V}(\mathbf{a}^{(k)})\right]_{bn} \ge \omega^{(k)}$ in \eqref{eq:select-active-set}, and the last inequality follows from $\omega^{(k)} \ge \epsilon$ in \eqref{eq:active-set-parameters}.
It follows from Proposition~\ref{prop:icd} that the objective function does not increase after each coordinate update by Algorithm~\ref{alg:icd}, i.e.,
\begin{equation}\label{eq:non-increase}
	F(\mathbf{a}^{(k+1)}) = F(\mathbf{a}^{(k,T_k)}) \le F(\mathbf{a}^{(k,T_k-1)}) \le \cdots \le F(\mathbf{a}^{(k,1)}).
\end{equation}
Combining \eqref{eq:decrease-first-update} and \eqref{eq:non-increase}, we obtain
\begin{equation}\label{eq:decrease-iteration}
	F(\mathbf{a}^{(k+1)}) \le F(\mathbf{a}^{(k)}) - \frac{\epsilon^2}{2 \lip(\beta+1)}.
\end{equation}
Using \eqref{eq:decrease-iteration} recursively, we get
\begin{equation}\label{eq:decrease-recursion}
	F^* \le F(\mathbf{a}^{(k+1)}) \le F(\mathbf{a}^{(0)}) - (k+1)\frac{\epsilon^2}{2 \lip(\beta+1)},
\end{equation}
where $F^*$ denotes the optimal value of $F(\mathbf{a}).$ 
Then it follows from \eqref{eq:decrease-recursion} that $k$ has the upper bound:
\begin{equation}\label{eq:upper-bound-k}
	k+1 \le \frac{2 \lip(\beta+1)}{\epsilon^2} \left( F(\mathbf{a}^{(0)}) - F^* \right),
\end{equation}
which shows that Algorithm~\ref{alg:active-set} will terminate within $\mathcal{O}(\lip \beta/\epsilon^{2})$ iterations.

Finally, we consider the convergence of Algorithm~\ref{alg:active-set} when the root-finding algorithm~\cite{mcnamee2007numerical} is employed to update the coordinates in the selected active set.
Since the root-finding algorithm solves problem~\eqref{eq:one-dim} exactly, the decrease in the objective function achieved by the root-finding algorithm when updating the first coordinate is not less than that achieved by the inexact Algorithm~\ref{alg:icd}.
Specifically, for the first updated coordinate $(b,n)$ at the $k$-th iteration, the root-finding algorithm and Algorithm~\ref{alg:icd} output $\hat{d}$ and $\bar{d},$ respectively, such that
\begin{equation}\label{eq:decrease-root-alg2}
	F(\mathbf{a} + \hat{d}\,\mathbf{e}_{bn}) \le F(\mathbf{a} + \bar{d}\,\mathbf{e}_{bn}) \le F(\mathbf{a}) - \frac{\left[\mathbb{V}(\mathbf{a})\right]_{bn}^2}{2 \lip(\beta+1)}.
\end{equation}
Additionally, the objective function does not increase after updating the second to $T_k$-th coordinates at the $k$-th iteration.
Therefore, Eqs.~\eqref{eq:decrease-first-update}--\eqref{eq:upper-bound-k} also hold if the root-finding algorithm is used to update the coordinates.
This completes the proof of Theorem~\ref{theorem:active-set}.

\bibliographystyle{IEEEtran}

\begin{IEEEbiographynophoto}{Ziyue Wang}
	(Graduate Student Member, IEEE) received the B.Sc. degree in information and computing science from Beijing Institute of Technology, Beijing, China, in 2020. He is currently pursuing the Ph.D. degree with the Institute of Computational Mathematics and Scientific/Engineering Computing, Academy of Mathematics and Systems Science, Chinese Academy of Sciences, Beijing. His research interests include optimization algorithms and their applications to signal processing and wireless communications.
\end{IEEEbiographynophoto}

\begin{IEEEbiographynophoto}{Ya-Feng Liu}
	(Senior Member, IEEE) received the B.Sc. degree in applied mathematics from Xidian University, Xi'an, China, in 2007, and the Ph.D. degree in computational mathematics from Chinese Academy of Sciences (CAS), Beijing, China, in 2012. During his Ph.D. study, he was supported by the Academy of Mathematics and Systems Science (AMSS), CAS, to visit Prof. Zhi-Quan (Tom) Luo at the University of Minnesota, Twins Cities, from 2011 to 2012. After his graduation, he joined the Institute of Computational Mathematics and Scientific/Engineering Computing, AMSS, CAS, in 2012, where he became an Associate Professor in 2018. His research interests include nonlinear optimization and its applications to signal processing, wireless communications, and machine learning. 
	He is an Elected Member of the Signal Processing for Communications and Networking Technical Committee (SPCOM-TC) of the IEEE Signal Processing Society from 2020 to 2022 and 2023 to 2025. He received the Best Paper Award from the IEEE International Conference on Communications (ICC) in 2011, the Chen Jingrun Star Award from the AMSS in 2018, the Science and Technology Award for Young Scholars from the Operations Research Society of China in 2018, the 15th IEEE ComSoc Asia-Pacific Outstanding Young Researcher Award in 2020, and the Science and Technology Award for Young Scholars from the China Society for Industrial and Applied Mathematics in 2022. Students supervised and co-supervised by him won the Best Student Paper Award from the International Symposium on Modeling and Optimization in Mobile, Ad Hoc and Wireless Networks (WiOpt), in 2015; and the Best Student Paper Award of IEEE International Conference on Acoustics, Speech and Signal Processing (ICASSP), in 2022.
	He currently serves as an Associate Editor for \textsc{IEEE Transactions on Signal Processing} and \textit{Journal of Global Optimization} and a Lead Guest Editor for \textsc{IEEE Journal on Selected Areas in Communications} Special Issue on ``Advanced Optimization Theory and Algorithms for Next-Generation Wireless Communication Networks.'' He served as an Editor for \textsc{IEEE Transactions on Wireless Communications} from 2019 to 2022 and an Associate Editor for \textsc{IEEE Signal Processing Letters} from 2019 to 2023. 
\end{IEEEbiographynophoto}

\begin{IEEEbiographynophoto}{Zhaorui Wang}
	(Member, IEEE) received the B.S. degree from the University of Electronic Science and Technology of China (UESTC) in 2015 and the Ph.D. degree from The Chinese University of Hong Kong (CUHK) in 2019.
	He was a Post-Doctoral Research Associate with The Hong Kong Polytechnic University from 2019 to 2020 and a Post-Doctoral Research Associate with CUHK from 2021 to 2022.
	He is currently a Research Assistant Professor with the School of Science and Engineering, CUHK, Shenzhen.
	His research interests include system-level design on massive machine-type communications (mMTC) and semantic communications.
	He was a recipient of the Hong Kong Ph.D. Fellowship from 2015 to 2018. He has been selected in the Post of ``Pengcheng Peacock Plan'' (Type~C) since 2022. 
\end{IEEEbiographynophoto}

\begin{IEEEbiographynophoto}{Wei Yu}
	(Fellow, IEEE) received the B.A.Sc. degree in computer engineering and mathematics from the University of Waterloo, Waterloo, ON, Canada, and the M.S. and Ph.D. degrees in electrical engineering from Stanford University, Stanford, CA, USA. He is currently a Professor with the Electrical and Computer Engineering Department, University of Toronto, Toronto, ON, Canada, where he holds the Canada Research Chair (Tier~1) in Information Theory and Wireless Communications. He is a Fellow of Canadian Academy of Engineering and a member of the College of New Scholars, Artists, and Scientists, Royal Society of Canada. He was the President of the IEEE Information Theory Society in 2021 and served on its board of governors from 2015 to 2023.
	He served as the Chair of the Signal Processing for Communications and Networking Technical Committee, IEEE Signal Processing Society, from 2017 to 2018. He was an IEEE Communications Society Distinguished Lecturer from 2015 to 2016.
	He received the IEEE Communications Society and the Information Theory Society Joint Paper Award in 2024; the IEEE Signal Processing Society Best Paper Award in 2021, 2017, and 2008; the IEEE Marconi Prize Paper Award in Wireless Communications in 2019; the IEEE Communications Society Award for Advances in Communication in 2019; the \textit{Journal of Communications and Networks} Best Paper Award in 2017; the IEEE Communications Society Best Tutorial Paper Award in 2015; and the Steacie Memorial Fellowship in 2015. He is a Clarivate Highly Cited Researcher.
	He served as an Area Editor for \textsc{IEEE Transactions on Wireless Communications}, an Associate Editor for \textsc{IEEE Transactions on Information Theory}, and an Editor for \textsc{IEEE Transactions on Communications} and \textsc{IEEE Transactions on Wireless Communications}.
\end{IEEEbiographynophoto}

\clearpage
\begin{center}
	\huge \textbf{Supplementary Material}
\end{center}

\appendices

\setcounter{section}{6}

\section{Proof of Theorem~\ref{theorem:nsp}}
\label{sec:proof-nsp-bernoulli}

In this proof, we only consider Type~\ref{item:qam} signature sequences, since the result for Type~\ref{item:sphere} signature sequences has been proven in \cite{fengler2021non}.
The proof techniques here are similar to those in \cite{fengler2021non}.
We first introduce some basic definitions that will be needed in the proof.

\begin{definition}[Sub-exponential random variables]\label{def:sub-exp}
	A random variable $x$ is called sub-exponential if 
	\begin{equation}\label{eq:moment-sub-exp}
		\left(\mathbb{E} [ \left|x\right|^p ]\right)^{\frac{1}{p}} \le C_1 p \quad \text{for all } p \ge 1.
	\end{equation}
	The sub-exponential norm of $x$ is denoted by $\left\|x\right\|_{\psi_1},$ which is defined as the smallest constant $C_1>0$ that satisfies \eqref{eq:moment-sub-exp}.
\end{definition}

\begin{definition}[Sub-exponential random vectors]\label{def:sub-exp-vec}
	A random vector $\mathbf{x} \in \mathbb{R}^n$ is called sub-exponential if the one-dimensional marginals $\left\langle \mathbf{x}, \mathbf{u} \right\rangle$ are sub-exponential random variables for all vectors $\mathbf{u} \in \mathbb{R}^n.$ The sub-exponential norm of $\mathbf{x}$ is defined as
	\begin{equation}
		\left\|\mathbf{x}\right\|_{\psi_1} \triangleq \sup_{\left\|\mathbf{u}\right\|_2 = 1} \left\|\left\langle \mathbf{x}, \mathbf{u} \right\rangle \right\|_{\psi_1}.
	\end{equation}
\end{definition}

\begin{definition}[Sub-Gaussian random variables]
	A random variable $x$ is called sub-Gaussian if 
	\begin{equation}\label{eq:moment-sub-gaussian}
		\left(\mathbb{E} [ \left|x\right|^p ]\right)^{\frac{1}{p}} \le C_2 \sqrt{p} \quad \text{for all } p \ge 1.
	\end{equation}
	The sub-Gaussian norm of $x$ is denoted by $\left\|x\right\|_{\psi_2},$ which is defined as the smallest constant $C_2>0$ that satisfies \eqref{eq:moment-sub-gaussian}.
\end{definition}

\begin{definition}\label{def:rip}
	The $s$-th restricted isometry constant $\delta_{s}(\mathbf{A})$ of a matrix $\mathbf{A} \in \mathbb{C}^{m\times n}$ is the smallest $\delta \ge 0$ such that
	\begin{equation}
		(1-\delta)\left\|\mathbf{x}\right\|_2^2 \le \left\|\mathbf{A}\mathbf{x}\right\|_2^2 \le (1+\delta)\left\|\mathbf{x}\right\|_2^2
	\end{equation}
	holds for all $s$-sparse vectors $\mathbf{x} \in \mathbb{C}^n.$
	The matrix $\mathbf{A}$ is said to satisfy the restricted isometry property (RIP) of order $s$ with parameter $\delta_{s}(\mathbf{A})$ if $0\le \delta_{s}(\mathbf{A}) < 1.$
\end{definition}

The proof consists of the following three key steps: in Step~I, we establish the RIP of the auxiliary matrix
\begin{multline}\label{eq:def-Shat}
	\widehat{\mathbf{S}} \triangleq \left[ \vecn( \mathbf{s}_{11} \mathbf{s}_{11}^H ), \vecn( \mathbf{s}_{12} \mathbf{s}_{12}^H ), \ldots, \right. \\
	\left. \vecn( \mathbf{s}_{BN} \mathbf{s}_{BN}^H ) \right] \in \mathbb{C}^{L(L-1) \times BN},
\end{multline}
where $\vecn(\cdot)$ denotes the vectorization of the non-diagonal elements of the matrix;
in Step~II, we establish the stable NSP of $\widehat{\mathbf{S}};$
and in Step III, we prove that the matrix $\widetilde{\mathbf{S}}$ defined in \eqref{eq:s-tilde} also satisfies the stable NSP because its null space is a subspace of the null space of $\widehat{\mathbf{S}}.$

\textbf{Step I of showing the RIP of $\widehat{\mathbf{S}}.$}
The following lemma derived in \cite[Theorem~5]{fengler2021non} is key to establish the RIP of matrix $\widehat{\mathbf{S}}.$
\begin{lemma}[\hspace{1sp}\cite{fengler2021non}]\label{lemma:rip-sub-exp}
	Let $\mathbf{R} \in \mathbb{R}^{m \times p}$ be a random matrix whose columns $\mathbf{r}_i \in \mathbb{R}^{m}$ that are independent, sub-exponential, and normalized such that $\left\|\mathbf{r}_i\right\|_2^2 = m,$ and $p \ge m.$
	For any parameter $\delta \in (0,1)$ and $\xi > \max_{1\le i\le p} \left\|\mathbf{r}_{i}\right\|_{\psi_1} + 1,$ there exist constants $0<c_{\delta,\xi}\le1$ and $\hat{c}_{\delta,\xi}> 0$ depending only on $\delta$ and $\xi$ such that if 
	\begin{equation}\label{eq:rip-sub-exp}
		2s \le c_{\delta,\xi} \frac{ m }{ \log^2 ( ep/m ) },
	\end{equation}
	then the RIP constant of $\mathbf{R}/\sqrt{m}$ satisfies $\delta_{2s}(\mathbf{R}/\sqrt{m}) < \delta$ with probability at least $1- \exp(-\hat{c}_{\delta,\xi}\sqrt{m}).$
\end{lemma}

To apply Lemma~\ref{lemma:rip-sub-exp}, we need to construct a real matrix associated with $\widehat{\mathbf{S}}$ in \eqref{eq:def-Shat} that satisfies:
(i) its columns are normalized,
(ii) its columns are independent and have bounded sub-exponential norm, 
and (iii) its RIP is equivalent to that of $\widehat{\mathbf{S}}.$
Using a similar technique as in \cite{fengler2021non}, we stack the real and imaginary parts of $\widehat{\S}$ to form
\begin{equation}\label{eq:def-ShatR}
	\widehat{\S}^{R} \triangleq
	\sqrt{2}
	\begin{bmatrix}
		\op{Re}\Big( \widehat{\S} \Big)^T, \,
		\op{Im}\Big( \widehat{\S} \Big)^T
	\end{bmatrix}^T \in \mathbb{R}^{2L(L-1) \times BN},
\end{equation}
and prove that $\widehat{\S}^{R}$ satisfies all of the above desirable conditions, where $\op{Re}(\cdot)$ denotes the real part, and $\operatorname{Im}(\cdot)$ denotes the imaginary part.
Let $\hat{\s}_{bn}^{R}$ denote the $(b,n)$-th column of $\widehat{\S}^{R}.$

We first show that each column $\hat{\s}_{bn}^{R}$ of matrix $\widehat{\S}^{R}$ is normalized. It follows from the definitions of $\widehat{\mathbf{S}}$ and $\widehat{\S}^{R}$ in \eqref{eq:def-Shat} and \eqref{eq:def-ShatR}, respectively, that
\begin{equation}\label{eq:sR}
	\hat{\s}_{bn}^{R} = \sqrt{2}
	\begin{bmatrix}
		\op{Re}( \vecn( \mathbf{s}_{bn} \mathbf{s}_{bn}^H ) ) \\  
		\op{Im}( \vecn( \mathbf{s}_{bn} \mathbf{s}_{bn}^H ))
	\end{bmatrix} \in \mathbb{R}^{2L(L-1)}.
\end{equation}
Since $\s_{bn} \in \qam^L,$ we can obtain from \eqref{eq:sR} that
\begin{align}
	\left\|\hat{\s}_{bn}^{R}\right\|_2^2 & = 2 \big\|\vecn( \mathbf{s}_{bn} \mathbf{s}_{bn}^H )\big\|_2^2 \nonumber \\
	& = 2 \sum_{\ell_2 = 1}^{L} \, \sum_{\ell_1 = 1, \,\ell_1 \neq \ell_2}^{L} \Big|s_{bn}^{(\ell_1)} \, \big(s_{bn}^{(\ell_2)}\big)^*\Big|^2 \nonumber \\
	& = 2L(L-1),
\end{align}
where $s_{bn}^{(\ell_1)}$ and $s_{bn}^{(\ell_2)}$ denote the $\ell_1$-th and $\ell_2$-th components of $\s_{bn},$ respectively.

We can easily see from \eqref{eq:sR} that the columns $\hat{\s}_{bn}^{R}$'s are independent since the sequences $\s_{bn}$'s are drawn independently.
To show that the sub-exponential norm of $\hat{\s}_{bn}^{R}$ is bounded, we first express the signature sequence in terms of uniformly distributed binary sequences $\mathbf{x}_{bn}$ and $\mathbf{y}_{bn}$:
\begin{equation}\label{eq:s=x+y}
	\s_{bn} = \frac{\sqrt{2}}{2} \left( \mathbf{x}_{bn} + \imath\,\mathbf{y}_{bn}  \right),
\end{equation}
where $\mathbf{x}_{bn},\,\mathbf{y}_{bn} \in \mathbb{R}^{L}$ are drawn uniformly i.i.d. from $\{ \pm 1 \}^L.$
Substituting \eqref{eq:s=x+y} into \eqref{eq:sR}, we get
\begin{equation}\label{eq:shatR-xy}
	\hat{\s}_{bn}^{R} = \frac{\sqrt{2}}{2}
	\begin{bmatrix}
		\vecn( \mathbf{x}_{bn} \mathbf{x}_{bn}^T + \mathbf{y}_{bn} \mathbf{y}_{bn}^T )  \\
		\vecn( \mathbf{y}_{bn} \mathbf{x}_{bn}^T - \mathbf{x}_{bn} \mathbf{y}_{bn}^T )
	\end{bmatrix}.
\end{equation}
We need to prove that all marginal distributions of $\hat{\s}_{bn}^{R}$ have bounded sub-exponential norms.
To this end, we consider any unit vector $\mathbf{u} \in \mathbb{R}^{2L(L-1)}$ and define two matrices $\mathbf{U}_1, \, \mathbf{U}_2 \in \mathbb{R}^{L\times L}$ whose diagonal elements are zero, such that
\begin{equation}\label{eq:u-U1U2}
	\mathbf{u} =
	\begin{bmatrix}
		\vecn( \mathbf{U}_1 )^T, \,
		\vecn( \mathbf{U}_2 )^T
	\end{bmatrix}^T.
\end{equation}
Combining \eqref{eq:shatR-xy} and \eqref{eq:u-U1U2}, we can express the marginal distribution of $\hat{\s}_{bn}^{R}$ as
\begin{align}\label{eq:marginal-distribution}
	& \left\langle \hat{\s}_{bn}^{R}, \mathbf{u} \right \rangle \nonumber \\
	&  = \frac{\sqrt{2}}{2} \left(\mathbf{x}_{bn}^T \mathbf{U}_1 \mathbf{x}_{bn} + \mathbf{y}_{bn}^T \mathbf{U}_1 \mathbf{y}_{bn} + \mathbf{y}_{bn}^T \mathbf{U}_2 \mathbf{x}_{bn} - \mathbf{x}_{bn}^T \mathbf{U}_2 \mathbf{y}_{bn}
	\right)\nonumber \\
	& = \frac{\sqrt{2}}{2} \begin{bmatrix} \mathbf{x}_{bn} \\ \mathbf{y}_{bn} \end{bmatrix}^T
	\begin{bmatrix} \mathbf{U}_1 & -\mathbf{U}_2^T \\ \mathbf{U}_2 & \mathbf{U}_1 \end{bmatrix}
	\begin{bmatrix} \mathbf{x}_{bn} \\ \mathbf{y}_{bn} \end{bmatrix} \nonumber \\
	& = \mathbf{z}_{bn}^T \mathbf{Q}_{\mathbf{u}} \mathbf{z}_{bn},
\end{align}
where the matrix
\begin{equation}\label{eq:def-Qu}
	\mathbf{Q}_{\mathbf{u}} =
	\frac{\sqrt{2}}{2}
	\begin{bmatrix}
		\mathbf{U}_1 & -\mathbf{U}_2^T \\ \mathbf{U}_2 & \mathbf{U}_1
	\end{bmatrix}
	\in \mathbb{R}^{2L\times 2L}
\end{equation}
only depends on $\mathbf{u}$ and has zero diagonal elements, and the vector
$\mathbf{z}_{bn} =
\begin{bmatrix}
	\mathbf{x}_{bn}^T,\mathbf{y}_{bn}^T
\end{bmatrix}^T \in \mathbb{R}^{2L}$ is uniformly distributed on $\left\{ \pm 1 \right\}^{2L}.$
We can use the following lemma to obtain an upper bound on the sub-exponential norm of $\mathbf{z}_{bn}^T \mathbf{Q}_{\mathbf{u}} \mathbf{z}_{bn},$
which builds upon the techniques presented in \cite{fengler2021non} and extends the result from random vectors uniformly distributed on a sphere to random vectors with independent sub-Gaussian coordinates.

\begin{lemma}\label{lemma:quad-sub-exp}
	Let $\mathbf{v} = [ v^{(1)}, v^{(2)}, \ldots, v^{(n)} ]^T \in \mathbb{R}^{n}$ be a random vector with independent, mean zero, sub-Gaussian coordinates with $ \kappa = \max_{i} \|v^{(i)}\|_{\psi_2} .$ For any matrix $\mathbf{A} \in \mathbb{R}^{n\times n},$ the following inequality holds
	\begin{equation}\label{eq:quad-sub-exp}
		\left\|\mathbf{v}^T \mathbf{A} \mathbf{v} - \mathbb{E}[\mathbf{v}^T \mathbf{A} \mathbf{v}]\right\|_{\psi_1} \le \bar{C} \kappa^2 \|\mathbf{A}\|_F,
	\end{equation}
	where $\bar{C} > 0$ is a universal constant, and $\|\cdot\|_F$ denotes the Frobenius norm.
\end{lemma}
\begin{IEEEproof}
	The proof of Lemma~\ref{lemma:quad-sub-exp} relies on the Hanson-Wright inequality in the following Lemma~\ref{lemma:HW}, which guarantees that the difference between $\mathbf{v}^T \mathbf{A} \mathbf{v}$ and its expected value $\mathbb{E}[\mathbf{v}^T \mathbf{A} \mathbf{v}]$ exhibits the following concentration properties.
	\begin{lemma}[Theorem~6.2.1 in \cite{vershynin2018high}]\label{lemma:HW}
		In the setting of Lemma~\ref{lemma:quad-sub-exp}, it follows that for any $t \ge 0,$
		\begin{multline}\label{eq:hw}
			\mathbb{P} \left\{ \left| \mathbf{v}^T \mathbf{A} \mathbf{v} - \mathbb{E}[\mathbf{v}^T \mathbf{A} \mathbf{v}] \right| \ge t \right\} \\
			\le 2\exp\left( -c \min \left( \frac{t^2}{\kappa^4\|\mathbf{A}\|_F^2} , \frac{t}{\kappa^2\|\mathbf{A}\|_2} \right) \right).
		\end{multline}
	\end{lemma}
	We can bound the $p$-th absolute moment of $\mathbf{v}^T \mathbf{A} \mathbf{v} - \mathbb{E}[\mathbf{v}^T \mathbf{A} \mathbf{v}]$ for $p \ge 1$ as follows:
	\begin{align}\label{eq:moment}
		& \mathbb{E}[ \left| \mathbf{v}^T \mathbf{A} \mathbf{v}-\mathbb{E}[\mathbf{v}^T \mathbf{A} \mathbf{v}] \right|^p ] \nonumber \\
		& = \int_{0}^{\infty} \mathbb{P} \left\{ \left | \mathbf{v}^T\mathbf{A}\mathbf{v} - \mathbb{E}[ \mathbf{v}^T\mathbf{A}\mathbf{v}] \right |^p > t \right\} \op{d}t \nonumber \\
		& = p\int_{0}^{\infty} \mathbb{P} \left\{ \left | \mathbf{v}^T\mathbf{A}\mathbf{v} - \mathbb{E}[ \mathbf{v}^T\mathbf{A}\mathbf{v} ] \right | > t \right\} t^{p-1} \op{d}t \nonumber \\
		& \le 2\,p \Bigg( \int_{0}^{\infty}  \exp \left( -c \frac{t^2}{\kappa^4\|\mathbf{A}\|_F^2}  \right)t^{p-1} \op{d}t \nonumber \\
		& \qquad \qquad \qquad + \int_{0}^{\infty}  \exp \left( -c  \frac{t}{\kappa^2\|\mathbf{A}\|_2}  \right)t^{p-1} \op{d}t \Bigg) \nonumber \\
		& \le 2\,p \left( \left(\frac{\kappa^4 \|\mathbf{A}\|_F^2}{c} \right)^{\frac{p}{2}} \Gamma\left(\frac{p}{2}\right) + \left(\frac{\kappa^2 \|\mathbf{A}\|_2}{c} \right)^{p} \Gamma(p) \right) \nonumber \\
		& \le \frac{4\kappa^{2p}\|\mathbf{A}\|_F^p \Gamma(p+1)}{\min ( c,1 )^p},
	\end{align}
	where the first equality is due to \cite[Lemma~1.2.1]{vershynin2018high}, $\Gamma(p) = \int_{0}^{+\infty} x^{p-1} e^{-x} \op{d}x$ denotes the Gamma function, and in the last inequality we utilize the fact that $\|\mathbf{A}\|_2 \le \|\mathbf{A}\|_F.$
	An upper bound on $\Gamma(p+1)$ is provided by Stirling’s formula~\cite{karatsuba2001asymptotic}:
	\begin{align}\label{eq:Stirling}
		\Gamma(p+1) & < \sqrt{\pi} \left( \frac{p}{e} \right)^{p} \left( 8p^3 + 4p^2 + p + \frac{1}{30} \right)^{\frac{1}{6}}, \nonumber \\
		& \le \sqrt{\pi} \left( \frac{p}{e} \right)^{p} \left( 2 \sqrt{p} \right)
		\le 2\sqrt{\pi} \, p^{p},
	\end{align}
	where the last inequality holds because $\sqrt{p} \le e^p$ for any $p \ge 1.$
	Combining \eqref{eq:moment} and \eqref{eq:Stirling}, we obtain the inequality:
	\begin{equation}
		\mathbb{E} \left[\left|\mathbf{v}^T\mathbf{A}\mathbf{v} - \mathbb{E} [\mathbf{v}^T\mathbf{A}\mathbf{v}]\right|^p\right]^{\frac{1}{p}} \le 
		\frac{8\sqrt{\pi}\kappa^{2}\|\mathbf{A}\|_F }{\min ( c,1 )} \, p,
	\end{equation}
	which holds for all $p \ge 1.$
	Recalling Definition~\ref{def:sub-exp}, and letting $\bar{C} = 8\sqrt{\pi}/\min(c,1),$ we can see that \eqref{eq:quad-sub-exp} holds true.
	This completes the proof of Lemma~\ref{lemma:quad-sub-exp}.
\end{IEEEproof}

We observe that each coordinate of $\mathbf{z}_{bn}$ is independent, mean zero, sub-Gaussian random variable with $\kappa = \max_{\ell}\|z_{bn}^{(\ell)}\|_{\psi_2} = 1.$
By Lemma~\ref{lemma:quad-sub-exp}, we obtain
\begin{equation}\label{eq:quad-exp-less}
	\big\| \mathbf{z}_{bn}^T \mathbf{Q}_{\mathbf{u}} \mathbf{z}_{bn} - \mathbb{E} [ \mathbf{z}_{bn}^T \mathbf{Q}_{\mathbf{u}} \mathbf{z}_{bn} ] \big\|_{\psi_1} \le \bar{C} \|\mathbf{Q}_{\mathbf{u}}\|_F,
\end{equation}
which, together with the fact that the diagonal elements of $\mathbf{Q}_{\mathbf{u}}$ are zero, further implies
\begin{equation}\label{eq:mean-zero}
	\mathbb{E} [ \mathbf{z}_{bn}^T \mathbf{Q}_{\mathbf{u}} \mathbf{z}_{bn} ] = \op{tr} ( \mathbf{Q}_{\mathbf{u}} ) = 0.
\end{equation}
Additionally, the definition of $\mathbf{Q}_{\mathbf{u}}$ in \eqref{eq:def-Qu} and the definition of $\mathbf{u}$ in \eqref{eq:u-U1U2} show that
\begin{equation}\label{eq:norm=1}
	\|\mathbf{Q}_{\mathbf{u}}\|_F^2 = \|\mathbf{U}_1\|_F^2 + \|\mathbf{U}_2\|_F^2 = \|\mathbf{u}\|_2^2 = 1.
\end{equation}
By substituting \eqref{eq:mean-zero} and \eqref{eq:norm=1} into \eqref{eq:quad-exp-less} and combining it with \eqref{eq:marginal-distribution}, we bound the sub-exponential norm of $\hat{\s}_{bn}^{R}$:
\begin{align}
	\left\|\hat{\s}_{bn}^{R}\right\|_{\psi_1} & = \sup_{\left\|\mathbf{u}\right\|_2 = 1} \left\| \left\langle \hat{\s}_{bn}^{R}, \mathbf{u} \right \rangle \right\|_{\psi_1} \nonumber \\
	& = \sup_{\left\|\mathbf{u}\right\|_2 = 1} \left\| \mathbf{z}_{bn}^T \mathbf{Q}_{\mathbf{u}} \mathbf{z}_{bn} \right\|_{\psi_1} \le \bar{C}.
\end{align}

Based on all of the above discussion, we now can apply Lemma~\ref{lemma:rip-sub-exp} to the matrix $\widehat{\S}^{R}$ to obtain the RIP of our interested matrix $\widehat{\S}.$
Recalling the definition of $\widehat{\S}^{R}$ in \eqref{eq:def-ShatR}, we have
\begin{equation}
	\Big\|\widehat{\S}^{R}\mathbf{x} \Big\|_2^2 = 2\,\Big \| \op{Re}\Big( \widehat{\S} \Big) \mathbf{x} \Big \|_2^2  + 2\,\Big \| \op{Im}\Big( \widehat{\S} \Big) \mathbf{x} \Big \|_2^2 = 2 \, \Big\|\widehat{\S} \mathbf{x} \Big\|_2^2
\end{equation}
holds for any vector $\mathbf{x}.$ 
Based on Definition~\ref{def:rip}, we get
\begin{equation}\label{eq:rip-equation}
	\delta_{2s}\left(\widehat{\S}^{R}/\sqrt{2L(L-1)}\right) = \delta_{2s}\left(\widehat{\S}/\sqrt{L(L-1)}\right).
\end{equation}
In Lemma~\ref{lemma:rip-sub-exp}, let $\mathbf{R} = \widehat{\S}^{R},$ $p = BN,$ $m = 2L(L-1),$ and fix the parameter $\xi = \bar{C} + 2$ be a universal constant.
For any $\delta \in (0,1), $ there exist two constants $0<c_{1}\le 1$ and $c_{2}> 0$ that depend only on $\delta$ such that if
\begin{equation}\label{eq:order-rip}
	2s \le c_{1} \frac{ L^2 }{ \log^2 ( eBN/L^2 ) },
\end{equation}
then the RIP constant
\begin{equation}\label{eq:rip-less-delta}
	\delta_{2s}\Big(\widehat{\S}^{R}/\sqrt{2L(L-1)}\Big) < \delta
\end{equation}
holds with probability at least $1- \exp(-c_{2}L),$ because $L^2 \le 2L(L-1)$ for $L \ge 2.$
Combining \eqref{eq:rip-equation} and \eqref{eq:rip-less-delta}, we get the RIP of $\widehat{\S}.$

\textbf{Step II of establishing the stable NSP of $\widehat{\S}.$} We now establish the stable NSP of $\widehat{\S}$ using the following result.
\begin{lemma}[Theorem~6.13 in \cite{foucart2013mathematical}]\label{lemma:nsp-rip}
	Let $\mathbf{A} \in \mathbb{C}^{m \times n}$ be a matrix with $2s$-th restricted isometry constant $\delta_{2s}(\mathbf{A}) < \frac{4}{\sqrt{41}}.$ 
	Then, $\mathbf{A}$ satisfies the $\ell_2$-robust NSP of order $s$ with constants
	\begin{equation}
		\rho = \frac{\delta_{2s}(\mathbf{A})}{ \sqrt{1-\delta_{2s}(\mathbf{A})^2} - \delta_{2s}(\mathbf{A})/4 } < 1
	\end{equation}
	and $\tau > 0,$ which depends only on $\delta_{2s}(\mathbf{A}).$
	More precisely, for any $\mathbf{x} \in \mathbb{C}^{n}$ and any index set $\mathcal{S}\subseteq\{1,2,\ldots,n\}$ with $|\mathcal{S}|\leq s,$ the following inequality holds:
	\begin{equation}\label{eq:l2-robust-nsp}
		\|\mathbf{x}_{\mathcal{S}}\|_2 \le \frac{\rho}{\sqrt{s}} \|\mathbf{x}_{\mathcal{S}^c}\|_1 + \tau \|\mathbf{A} \mathbf{x}\|_2.
	\end{equation}
\end{lemma}

Based on the RIP of $\widehat{\S},$ we can set $\delta = \frac{4}{\sqrt{41}} \bar{\rho}$ for any given parameter $0< \bar{\rho}<1.$ Then, there exist constants $c_{1}, c_{2} > 0$ depending only on $\bar{\rho}$ such that the inequality
\begin{equation}
	\delta_{2s}\Big(\widehat{\S}/\sqrt{L(L-1)}\Big) < \delta = \frac{4}{\sqrt{41}} \bar{\rho}
\end{equation}
holds with probability at least $1- \exp(-c_{2}L)$ under \eqref{eq:order-rip}.
Then it follows from Lemma~\ref{lemma:nsp-rip} that $\widehat{\S}/\sqrt{L(L-1)}$ satisfies the $\ell_2$-robust NSP of order $s$ with
\begin{equation}\label{eq:rhp-barrho}
	\rho  = \frac{\delta_{2s}}{ \sqrt{1-\delta_{2s}^2} - \delta_{2s}/4 } 
	< \frac{\sqrt{41}}{4}\delta = \bar{\rho},
\end{equation}
where $\delta_{2s}$ is the abbreviation of $\delta_{2s}\Big(\widehat{\S}/\sqrt{L(L-1)}\Big)$ for simplicity.
The inequality in \eqref{eq:rhp-barrho} holds because the denominator $\sqrt{1-\delta_{2s}^2} - \delta_{2s}/4 \ge \frac{4}{\sqrt{41}},$ and the numerator $\delta_{2s} < \delta.$
As shown in \cite[Chap.~4]{foucart2013mathematical}, the $\ell_2$-robust NSP is a sufficient condition for the stable NSP.
In other words, for any vector $\mathbf{x}$ in the null space of $\widehat{\S},$ inequality~\eqref{eq:l2-robust-nsp} implies that
\begin{equation}\label{eq:nsp-Shat}
	\| \mathbf{x}_{\mathcal{S}} \|_{1} \le \sqrt{s} \|\mathbf{x}_{\mathcal{S}}\|_2 \le \rho \|\mathbf{x}_{\mathcal{S}^c}\|_1,
\end{equation}
where the first inequality follows from the Cauchy-Schwartz inequality.
Therefore, we have established the stable NSP of $\widehat{\S}$ as in \eqref{eq:nsp-Shat}.

\textbf{Step III of proving the stable NSP $\widetilde{\mathbf{S}}.$} Finally, we prove the stable NSP of $\widetilde{\mathbf{S}}.$ 
Recalling the definitions of $\widetilde{\mathbf{S}}$ and $\widehat{\S}$ in \eqref{eq:s-tilde} and \eqref{eq:def-Shat}, respectively, we can see that $\widehat{\S}$ is a sub-matrix formed by removing $L$ rows from $\widetilde{\mathbf{S}}.$
Hence, for any $\mathbf{x}$ satisfying $\widetilde{\mathbf{S}}\mathbf{x} = \mathbf{0},$ we have $\widehat{\mathbf{S}}\mathbf{x} = \mathbf{0},$ and therefore $\| \mathbf{x}_{\mathcal{S}} \|_{1} \le \rho \|\mathbf{x}_{\mathcal{S}^c}\|_1$ holds.
This completes the proof of Theorem~\ref{theorem:nsp}.

\section{Proof of Lemma~\ref{lemma:gamma}} \label{sec:proof-gamma}

The proof of Lemma~\ref{lemma:gamma} relies mainly on the path-loss model as in Assumption~\ref{assu:cell}.
Let $D_{bj}$ denote the BS-BS distance between BS $b$ and BS $j$ for $b \neq j.$
Since the radius of hexagonal cells is $R,$ the BS-device distance $D_{bjn}$ from any device $n$ in cell $j$ to BS $b$ is bounded by
\begin{equation}
	D_{bjn} \ge D_{bj} - R, \quad \forall~ n.
\end{equation}
Hence, based on the path-loss model in \eqref{eq:gamma}, we can obtain
\begin{equation}\label{eq:upper-g}
	\max_{1\le n \le N} g_{bjn} \le P_0 \left(\frac{D_0}{D_{bj}-R}\right)^{\gamma}.
\end{equation}
To show the lemma, we need to carefully study the properties of $D_{bj}$ and establish an upper bound on the sum of the terms in the right-hand side of \eqref{eq:upper-g}.

\begin{figure}[t]
	\centering
	\begin{tikzpicture}
		\filldraw[fill=gray!5,densely dashed,draw=white] (1,0)--(0.5,0.866)--(-0.5,0.866)--(-1,0)--(-0.5,-0.866)--(0.5,-0.866)--cycle;
		\node[fill=teal,regular polygon, regular polygon sides=3,inner sep=1.5pt] at (0,0) {};
		\filldraw[fill=gray!5,densely dashed,draw=brown] (1.5+1,0.866)--(1.5+0.5,0.866+0.866)--(1.5-0.5,0.866+0.866)--(1.5-1,0.866)--(1.5-0.5,0)--(1.5+0.5,0)--cycle;
		\node[fill=teal,regular polygon, regular polygon sides=3,inner sep=1.5pt] at (1.5,0.866) {};
		\filldraw[fill=gray!5,densely dashed,draw=brown] (1.5+1,-0.866)--(1.5+0.5,0)--(1.5-0.5,0)--(1.5-1,-0.866)--(1.5-0.5,-2*0.866)--(1.5+0.5,-2*0.866)--cycle;
		\node[fill=teal,regular polygon, regular polygon sides=3,inner sep=1.5pt] at (1.5,-0.866) {};
		\filldraw[fill=gray!5,densely dashed,draw=brown] (1,2*0.866)--(0.5,3*0.866)--(-0.5,3*0.866)--(-1,2*0.866)--(-0.5,0.866)--(0.5,0.866)--cycle;
		\node[fill=teal,regular polygon, regular polygon sides=3,inner sep=1.5pt] at (0,2*0.866) {};
		\filldraw[fill=gray!5,densely dashed,draw=brown] (1,-2*0.866)--(0.5,-0.866)--(-0.5,-0.866)--(-1,-2*0.866)--(-0.5,-3*0.866)--(0.5,-3*0.866)--cycle;
		\node[fill=teal,regular polygon, regular polygon sides=3,inner sep=1.5pt] at (0,-2*0.866) {};
		\filldraw[fill=gray!5,densely dashed,draw=brown] (-1.5+1,0.866)--(-1.5+0.5,2*0.866)--(-1.5-0.5,2*0.866)--(-1.5-1,0.866)--(-1.5-0.5,0)--(-1.5+0.5,0)--cycle;
		\node[fill=teal,regular polygon, regular polygon sides=3,inner sep=1.5pt] at (-1.5,0.866) {};
		\filldraw[fill=gray!5,densely dashed,draw=brown] (-1.5+1,-0.866)--(-1.5+0.5,0)--(-1.5-0.5,0)--(-1.5-1,-0.866)--(-1.5-0.5,-2*0.866)--(-1.5+0.5,-2*0.866)--cycle;
		\node[fill=teal,regular polygon, regular polygon sides=3,inner sep=1.5pt] at (-1.5,-0.866) {};
		\draw [->] (-3.5,0)--(3.5,0) node[below] {$x$};
		\draw [->] (0,-3*0.866-0.15)--(0,3*0.866+0.15) node[right] {$y$};
		\draw (-2, 1pt) -- (-2, -1pt) node[anchor=north] {\scriptsize$-2R$};
		\draw (-1, 1pt) -- (-1, -1pt) node[anchor=north] {\scriptsize$-R$};
		\draw (1, 1pt) -- (1, -1pt) node[anchor=north] {\scriptsize$R$};
		\draw (2, 1pt) -- (2, -1pt) node[anchor=north] {\scriptsize$2R$};
		\draw (-1pt, 0.866) -- (1pt, 0.866) node[anchor=west] {\scriptsize$\tfrac{\sqrt{3}}{2}R$};
		\draw (-1pt, 2*0.866) -- (1pt, 2*0.866) node[anchor=west] {\scriptsize$\sqrt{3}R$};
		\draw (-1pt, -0.866) -- (1pt, -0.866) node[anchor=west] {\scriptsize$-\tfrac{\sqrt{3}}{2}R$};
		\draw (-1pt, -2*0.866) -- (1pt, -2*0.866) node[anchor=west] {\scriptsize$-\sqrt{3}R$};
		\node[fill=teal,regular polygon, regular polygon sides=3,inner sep=1pt] at (1.35,2.47) {};
		\node [draw, rounded rectangle,rounded rectangle arc length=90] at (2,2.5) {\scriptsize~~~~Base station};
		\draw (-1.8,2.3) node {\textbf{.\,.\,.\,.\,.\,.}};
		\draw (2.8,0.2) node {\textbf{.\,.\,.\,.\,.\,.}};
		\draw (1.6,-2.2) node {\textbf{.\,.\,.\,.\,.\,.}};
		\draw (-2.8,-1.6) node {\textbf{.\,.\,.\,.\,.\,.}};
	\end{tikzpicture}
	\caption{Base station locations in the multi-cell system.}
	\label{fig:multi-cell}
\end{figure}
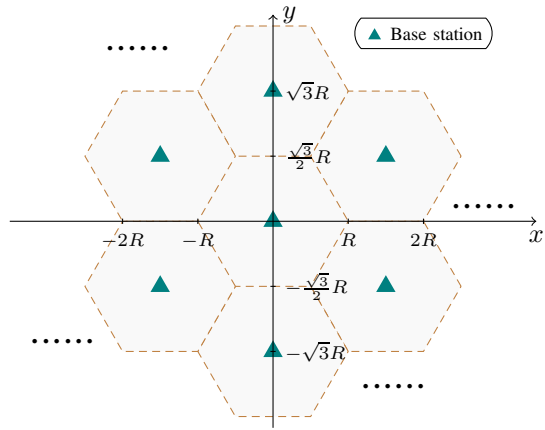

We first study the properties of $D_{bj}.$
We place all BSs in the Cartesian coordinate system as shown in Fig.~\ref{fig:multi-cell}, with BS $b$ located at the coordinate origin.
We observe that the $x$-coordinate of any BS is an integer multiple of $\frac{3}{2}R,$ i.e., its $x$-coordinate is equal to $\frac{3}{2}Rk,$ where $k \in \mathbb{Z}$ with $\mathbb{Z}$ being the set of all integers.
Moreover, if $k$ is even, then the $y$-coordinate of the BS is $\sqrt{3}R\ell$ for some $\ell\in\mathbb{Z};$ if $k$ is odd, then the $y$-coordinate of the BS is $\sqrt{3}R(\ell+\frac{1}{2})$ for some $\ell \in \mathbb{Z}.$
Hence, the $(x,y)$-coordinates of all BSs are in the following set:
\begin{equation}
	\left\{ \left( \tfrac{3}{2}Rk, \sqrt{3}R\left(\ell+\tfrac{1}{2}\op{mod}(k,2)\right) \right) ~ \Big | ~ (k,\ell) \in \mathbb{Z}^2 \right\}.
\end{equation}
Furthermore, the BS-BS distance $D_{bj}$ for $j \neq b$ satisfies:
\begin{multline}\label{eq:structure-D}
	D_{bj} \in \bigg\{ \sqrt{\left(\tfrac{3}{2}Rk\right)^2 + \left( \sqrt{3}R\left(\ell+\tfrac{1}{2}\op{mod}(k,2)\right) \right)^2} \\
	\bigg | ~ (k,\ell) \in \mathbb{Z}^2, \, (k,\ell) \neq (0,0) \bigg\}.
\end{multline}
By combining \eqref{eq:upper-g} and \eqref{eq:structure-D}, we can obtain the following upper bound:
\begin{align}\label{eq:bound-sum-g-long}
	& \sum_{j=1,\,j\neq b}^{B} \Big(\max_{1\le n \le N} g_{bjn} \Big) \nonumber \\
	& \le \sum_{j=1,\,j\neq b}^{B} P_0 \left(\frac{D_0}{D_{bj}-R}\right)^{\gamma} \nonumber \\
	& \le \sum_{\substack{(k,\ell) \in \mathbb{Z}^2\\(k,\ell) \neq (0,0)}} \Bigg( P_0 \nonumber \\
	& \quad \times \Bigg(\frac{D_0}{\sqrt{\left(\frac{3}{2}Rk\right)^2 + \left( \sqrt{3}R\left(\ell+\tfrac{1}{2}\op{mod}(k,2)\right) \right)^2}-R}\Bigg)^{\gamma} \Bigg) \nonumber \\
	& \overset{(a)}{\le} \frac{P_0 \, D_0^{\gamma}}{R^{\gamma}} \sum_{\substack{(k,\ell) \in \mathbb{Z}^2\\(k,\ell) \neq (0,0)}} \left( \tfrac{3}{2}\sqrt{k^2 + \left(\ell+\tfrac{1}{2}\op{mod}(k,2)\right)^2 } - 1 \right)^{-\gamma} \nonumber \\
	& \overset{(b)}{\le} \frac{2 P_0 \, D_0^{\gamma}}{R^{\gamma}} \sum_{\substack{(k,\ell) \in \mathbb{Z}^2\\(k,\ell) \neq (0,0)}} \left( \tfrac{3}{2}\sqrt{k^2 + \ell^2 } - 1 \right)^{-\gamma} \nonumber \\
	& \overset{(c)}{\le} \frac{2 P_0 \, D_0^{\gamma}}{R^{\gamma}} \sum_{\substack{(k,\ell) \in \mathbb{Z}^2\\(k,\ell) \neq (0,0)}} \left( \tfrac{1}{2}\sqrt{k^2 + \ell^2 } \right)^{-\gamma} \nonumber \\
	& = \frac{2^{\gamma+1} P_0 \, D_0^{\gamma}}{R^{\gamma}} \sum_{\substack{(k,\ell) \in \mathbb{Z}^2\\(k,\ell) \neq (0,0)}} \left( k^2 + \ell^2 \right)^{-\frac{\gamma}{2}},
\end{align}
where $(a)$ is due to $\frac{3}{2} < \sqrt{3},$ $(b)$ is due to the fact that the inequality
\begin{equation}
	\sum_{\ell \in \mathbb{Z}} \Big( \tfrac{3}{2}\sqrt{k^2 + \left(\ell+\tfrac{1}{2}\right)^2 } - 1 \Big)^{-\gamma} \le 2 \sum_{\ell \in \mathbb{Z}} \left( \tfrac{3}{2}\sqrt{k^2 + \ell^2 } - 1 \right)^{-\gamma}
\end{equation}
holds true for any odd $k,$ and $(c)$ is due to the fact that $1 \le \sqrt{k^2+\ell^2}.$

Next, we prove that the last line in \eqref{eq:bound-sum-g-long} is bounded when $\gamma >2.$
Since $\left( k^2 + \ell^2 \right)^{-\frac{\gamma}{2}}$ is independent of the signs of $k$ and $\ell,$ we consider $(k,\ell)$ in the first quartile:
\begin{equation}\label{eq:upper-all-non-negative}
	\sum_{\substack{(k,\ell) \in \mathbb{Z}^2\\(k,\ell) \neq (0,0)}} \left( k^2 + \ell^2 \right)^{-\frac{\gamma}{2}} \le 
	4 \sum_{\substack{(k,\ell) \in \mathbb{Z}_+^2\\(k,\ell) \neq (0,0)}} \left( k^2 + \ell^2 \right)^{-\frac{\gamma}{2}},
\end{equation}
where $\mathbb{Z}_+$ is the set of non-negative integers.
Then, we divide all $(k,\ell) \in \mathbb{Z}_+^2,$ $(k,\ell) \neq (0,0)$ into three subsets:
(i) $\{(k,\ell) \mid k,\ell\ge 1,\, (k,\ell) \neq (1,1) \},$ (ii) $\{(k,0) \mid k \ge 1\} \cup \{(0,\ell) \mid \ell \ge 1\},$ and (iii) $\{(1,1)\}.$
Then, we get
\begin{multline}\label{eq:bound-three-sum}
	\sum_{\substack{(k,\ell) \in \mathbb{Z}_+^2\\(k,\ell) \neq (0,0)}} \left( k^2 + \ell^2 \right)^{-\frac{\gamma}{2}} \\
	= \sum_{\substack{k,\ell \ge 1\\(k,\ell) \neq (1,1)}} \left( k^2 + \ell^2 \right)^{-\frac{\gamma}{2}} + 2 \sum_{k\ge 1} k^{-\gamma} + 2^{-\frac{\gamma}{2}} .
\end{multline}

It remains to bound the first sum in \eqref{eq:bound-three-sum} based on the monotonicity of the integral. For all $ k,\ell \ge 1,\, (k,\ell) \neq (1,1),$ we have
\begin{equation}
	\left( k^2 + \ell^2 \right)^{-\frac{\gamma}{2}} \le \int_{k-1}^{k} \int_{\ell-1}^{\ell} \left( x^2 + y^2 \right)^{-\frac{\gamma}{2}} \op{d}x\op{d}y.
\end{equation}
Therefore, we have
\begin{align}\label{eq:upper-klge-1}
	& \sum_{\substack{k,\ell \ge 1\\(k,\ell) \neq (1,1)}} \left( k^2 + \ell^2 \right)^{-\frac{\gamma}{2}} \nonumber \\
	& \le \sum_{\substack{k,\ell \ge 1\\(k,\ell) \neq (1,1)}} \int_{k-1}^{k} \int_{\ell-1}^{\ell} \left( x^2 + y^2 \right)^{-\frac{\gamma}{2}} \op{d}x\op{d}y \nonumber \\
	& = \int_{(x,y)\in \mathbb{R}_+^2, \, \max\{x,y\}\ge 1} \left( x^2 + y^2 \right)^{-\frac{\gamma}{2}} \op{d}x\op{d}y
	\nonumber \\ 
	& \overset{(a)}{\le} \int_{(x,y)\in \mathbb{R}_+^2, \, x^2+y^2\ge 1} \left( x^2 + y^2 \right)^{-\frac{\gamma}{2}} \op{d}x\op{d}y
	\nonumber \\ 
	& \overset{(b)}{=} \int_{0}^{\frac{\pi}{2}} \int_{1}^{\infty} \eta^{1-\gamma}\, \op{d}\eta\,\op{d}\theta \overset{(c)}{=} \frac{\pi}{2(\gamma-2)},
\end{align}
where $(a)$ is because of the reduced integration region, $(b)$ is due to the polar coordinate transformation, i.e., $x = \eta\cos\theta,$ $y = \eta\sin\theta,$ and $\op{d}x\,\op{d}y = \eta\,\op{d}\eta\,\op{d}\theta,$ and $(c)$ holds for all $\gamma > 2.$
Using the same arguments, we can show
\begin{equation}\label{eq:upper-kge-1}
	\sum_{k\ge 1} k^{-\gamma} \le 1 + \sum_{k\ge 2} \int_{k-1}^{k} t^{-\gamma} \op{d}t
	= 1 + \int_{1}^{\infty} t^{-\gamma} \op{d}t = \frac{\gamma}{\gamma-1}.
\end{equation}
As such, the last line in \eqref{eq:bound-sum-g-long} is bounded when $\gamma >2.$

Finally, we can derive the desired conclusion by putting all pieces together.
In particular, substituting \eqref{eq:upper-klge-1} and \eqref{eq:upper-kge-1} into \eqref{eq:bound-three-sum}, and combining \eqref{eq:bound-sum-g-long} and \eqref{eq:upper-all-non-negative}, we obtain
\begin{multline}\label{eq:upper-g-end}
	\sum_{j=1,\,j\neq b}^{B} \Big(\max_{1\le n \le N} g_{bjn} \Big) \\
	\le \frac{2^{\gamma+3} P_0 D_0^{\gamma}}{R^{\gamma}} \left( \frac{\pi}{2(\gamma-2)} + \frac{2\gamma}{\gamma-1} + \frac{1}{2^{\frac{\gamma}{2}}} \right).
\end{multline}
Denote the right-hand side of \eqref{eq:upper-g-end} as the constant $C,$ which is independent of the total number of cells $B.$
This completes the proof of Lemma~\ref{lemma:gamma}.

\end{document}